\RequirePackage[T1]{fontenc}
\documentclass[12pt]{article}
\pdfoutput=1 
\usepackage[height=8.85in,width=6.45in]{geometry}

\usepackage[utf8]{inputenc}
\usepackage{amsmath}
\usepackage{amssymb}
\usepackage{mathtools}
\numberwithin{equation}{section}
\usepackage{slashed}
\usepackage{braket}
\usepackage[svgnames]{xcolor}
\usepackage[colorlinks,linktocpage=true,citecolor=DarkGreen,linkcolor=FireBrick,urlcolor=FireBrick]{hyperref}
\usepackage{cite}
\usepackage{graphicx}
\usepackage{tikz}
\usepackage{tikz-cd}
\usepackage{times}
\usepackage[scaled]{couriers}
\usepackage{bm}
\usepackage{subfig}
\usepackage{mathrsfs}
\usepackage{amsthm}
\usepackage{makecell}
\usepackage{xcolor}
\usepackage{mdframed}

\renewenvironment{figure}[1][]{
  \begin{originalfigure}[#1]
    \begin{mdframed}[linecolor=black!0,backgroundcolor=black!1]
}{
    \end{mdframed}
  \end{originalfigure}
}

\theoremstyle{plain}

\theoremstyle{definition}

\numberwithin{thm}{section}

\newcommand{\lsim}{ \mathop{}_{\textstyle \sim}^{\textstyle <} }
\newcommand{\vev}[1]{ \left\langle {#1} \right\rangle }

\def\d{\text{d}}
\def\i{{\mathsf i}}

\DeclareMathOperator{\tr}{tr}

\usepackage{slashed}

\def\a{\alpha}

\def\cA{{\cal A}}

\def\cD{{\cal D}}

\def\cH{{\cal H}}

\def\cN{{\cal N}}
\def\cO{{\cal O}}

\def\cW{{\cal W}}
\def\cX{{\cal X}}
\def\cY{{\cal Y}}

\def\bA{{\mathbb A}}

\def\bC{{\mathbb C}}

\def\bH{{\mathbb H}}

\def\bL{{\mathbb L}}

\def\bR{{\mathbb R}}

\def\bZ{{\mathbb Z}}

\def\sF{{\mathsf F}}

\def\sL{{\mathsf L}}

\def\sP{{\mathsf P}}
\def\sQ{{\mathsf Q}}
\def\sR{{\mathsf R}}

\def\sX{{\mathsf X}}
\def\sY{{\mathsf Y}}

\def\sa{{\mathsf a}}
\def\sb{{\mathsf b}}
\def\sc{{\mathsf c}}
\def\sd{{\mathsf d}}
\def\se{{\mathsf e}}

\def\si{{\mathsf i}}

\def\so{{\mathsf o}}
\def\sp{{\mathsf p}}
\def\sq{{\mathsf q}}
\def\sr{{\mathsf r}}

\def\su{{\mathsf u}}

\def\sx{{\mathsf x}}

\def\U{\mathrm{U}}
\def\SU{\mathrm{SU}}
\def\O{\mathrm{O}}
\def\SO{\mathrm{SO}}
\def\Sp{\mathrm{Sp}}

\def\Spin{\mathrm{Spin}}

\def\u{\mathfrak{u}}
\def\su{\mathfrak{su}}

\def\so{\mathfrak{so}}
\def\sp{\mathfrak{sp}}

\def\e{\mathfrak{e}}
\def\g{\mathfrak{g}}
\def\h{\mathfrak{h}}

\def\beq#1\eeq{\begin{align}#1\end{align}}

\def\hete{(E_8 \times E_8) \rtimes \bZ_2}
\def\hets{\Spin(32)/\bZ_2}
\def\hetss{\dfrac{\Spin(32)}{\bZ_2}}
\def\sugra{\Omega^\text{SUGRA}}
\def\sqft{\text{SQFT}}
\def\GS{\text{GS}}

\def\TMF{\text{TMF}}
\def\SQFT{\text{SQFT}}

\def\KO{\mathrm{KO}}

\def\Nequals#1{$\mathcal{N}{=}#1$}

\def\mod{\ \text{mod}\ }
\def\Arf{\mathsf{Arf}}
\def\KO{\mathrm{KO}}

\begin{document}

\begin{titlepage}

\begin{flushright}
TU-1248
\\
KYUSHU-HET-294
\end{flushright}

\vskip 3cm

\begin{center}

{\Large \bfseries On non-supersymmetric heterotic branes}

\vskip 2cm
Justin Kaidi$^{1,2}$,
Yuji Tachikawa$^3$, 
and Kazuya Yonekura$^4$
\vskip 2cm

\begin{tabular}{ll}
1 &Institute for Advanced Study, Kyushu University, Fukuoka 812-8581, Japan \\
2 &Department of Physics, Kyushu University, Fukuoka 819-0395, Japan\\
3 & Kavli Institute for the Physics and Mathematics of the Universe (WPI), \\
& University of Tokyo,   Kashiwa, Chiba 277-8583, Japan \\
4 & Department of Physics, Tohoku University, Sendai, Miyagi, 980-8578, Japan \\
\end{tabular}

\vskip 1cm

\end{center}

\noindent\textbf{Abstract:}
A uniform construction of non-supersymmetric 0-, 4-, 6- and 7-branes in heterotic string theory 
was announced and outlined in our letter \cite{Kaidi:2023tqo}.
In this full paper, we provide details on their properties.
Among other things, 
we discuss the charges carried by the branes,
their topological and dynamical stability,
the exact worldsheet descriptions of their near-horizon regions, 
and the relationship of the branes to the mathematical notion of topological modular forms.

\end{titlepage}

\setcounter{tocdepth}{2}

\newpage

\tableofcontents

\section{Introduction}

It is generally believed that any consistent theory of quantum gravity must contain dynamical objects carrying all possible charges. 
The initial motivation for this idea involved creating black holes with a given charge and allow them to Hawking radiate, 
with the charges in question being those for conventional 
symmetries;
see e.g.~\cite{Banks:2010zn} for a summary of the discussions in 2010.
In more recent years, this idea has been generalized to include more
subtle types of charges, stemming often from topological properties of the system;
this idea was most clearly put forward in for example \cite{McNamara:2019rup}. 
These charges may arise from both the gauge and gravitational sectors of the
 theory, but our focus in the current work will be on a class of charges associated with the former.
Our goal in the present work will be to discuss new non-supersymmetric objects in the $E_8 \times E_8$ and $\SO(32)$ heterotic strings, expanding on the results given in our previous work with Ohmori \cite{Kaidi:2023tqo}. 

In more detail, in this work we will be focused on the two supersymmetric heterotic string theories, which, though often referred to as the $E_8 \times E_8$ and $\SO(32)$ heterotic strings, actually have the gauge groups $\hete$ and $\hets$. These groups have a number of non-trivial homotopy groups $\pi_{n-1}(G)$, which can be used to engineer topologically non-trivial gauge configurations on $S^n$. The presence of such non-trivial gauge configurations can be measured by integrating appropriate characteristic classes along $S^n$, and it is natural to think of these characteristic classes as a sort of topological charge. The aforementioned quantum gravity lore then leads one to expect that there exist $(8-n)$-branes which carry these charges, i.e. which act as sources for the non-trivial gauge configurations on the transverse $S^n$. As was first discussed in \cite{Kaidi:2023tqo} and will be reviewed in Sec.~\ref{sec:mainbranecharge}, this results in the prediction of a 4- and 7-brane in the $\hete$ heterotic string, as well as a 6- and 0-brane in the $\hets$ heterotic string.\footnote{Note that the 0-brane had been previously discussed in work by Polchinski \cite{Polchinski:2005bg}, while the existence of the 4-brane was anticipated in \cite{Bergshoeff:2006bs}. }

Seemingly unrelatedly, it has long been known that in addition to the two supersymmetric heterotic strings in 10-dimensions, there exist a number of \textit{non}-supersymmetric heterotic string theories as well \cite{Dixon:1986iz,Kawai:1986vd}. All of these theories possess a closed string tachyon, but in \cite{Hellerman:2007zz,Kaidi:2020jla} it was shown that tachyon condensation could be used to produce (meta)stable vacua for each of them. In particular, upon closed string tachyon condensation one finds a 9-dimensional vacuum with gauge algebra $\mathfrak{e}_8$, an 8-dimensional vacuum with algebra $\mathfrak{su}(16)$, a 6-dimensional vacuum with algebra $\mathfrak{e}_7 \times \mathfrak{e}_7$, and a $2$-dimensional vacuum with algebra $\mathfrak{so}(24)$.\footnote{One also finds a second $2$-dimensional vacua with algebra $\mathfrak{so}(24)$, as well as a $2$-dimensional vacua with algebra $\mathfrak{e}_8 \times \mathfrak{o}(16)$. These vacua will not play a role in the current work.} Based on dimension counting, together with the fact that $\mathfrak{e}_8, \,\,\mathfrak{e}_7 \times \mathfrak{e}_7 \subset \mathfrak{e}_8 \times \mathfrak{e}_8$ while $\mathfrak{su}(16),\,\,\mathfrak{so}(24) \subset \mathfrak{so}(32)$, one is led to suspect that these vacua may serve as the near-horizon limits of the tentative 7-, 6-, 4-, and 0-branes, respectively.

\begin{table}
\centering
\hskip-.5em\begin{tabular}{c|cccc}
& 7-brane & 6-brane & 4-brane & 0-brane \\
\hline
\hline
\small\makecell{parent \\ heterotic theory} & $\hete$ & $\hets$ & $\hete$ & $\hets$ \\
\hline
\small\makecell{relevant\\ homotopy groups} & $\pi_0$ & $\pi_1$ & $\pi_3$ & $\pi_7$\\
\hline
\small\makecell{relevant\\ characteristic classes} & $H^1(-,\bZ_2)$ &$H^2(-,\bZ_2)$ &$H^4(-,\bZ_2)$ & $H^8(-,\bZ)$ \\
\hline
\small\makecell{near-horizon\\ worldsheet theory} & $(E_8)_2$ & $(\SU(16)/\bZ_4)_1$ & $((E_7\times E_7)/\bZ_2)_1$ & $(\Spin(24)/\bZ_2)_1$ 
\end{tabular}
\caption{Summary of the properties of our non-supersymmetric branes.  \label{tab:summary}}
\end{table}

In this work, we give more detail on these and other claims appearing in our previous work \cite{Kaidi:2023tqo}, including a detailed description of the charges carried by the branes, as well as their near-horizon limits. 
We tabulate the basic properties of our branes in Table~\ref{tab:summary}.
There, for each brane, we list
\begin{itemize}
\item the parent supersymmetric heterotic string theory,
distinguished by its gauge group $G$,
\item the nonzero homotopy group $\pi_{n-1}(G)$ responsible for the existence of the brane,
\item the characteristic class which measures the topological charge of the brane,
\item and the component $H_k$ of the near-horizon worldsheet theory $
\mathbb{R}^{p,1} \times \mathbb{R}_\text{linear dilaton} \times H_k,
$ where $H_k$ is the level $k$ current algebra theory with group $H$.
\end{itemize}


In the rest of the introduction, we will give a brief overview of the contents of the paper.
We start in Sec.~\ref{sec:mainbranecharge} by discussing the charges carried by our branes in detail.
We will also provide an explicit gauge field configurations which have these charges,
which are often obtained by embedding the tangent bundle of the $S^n$ surrounding the brane into the gauge bundle.

A key feature of our new branes is that 
they are non-supersymmetric, which leads to the important question of their stability. We address this question in two steps.  First, in Sec.~\ref{sec:bordism} we discuss topological aspects of stability, in particular showing that 
our branes carry non-trivial topological charges, given by the aforementioned characteristic classes defined in Sec.~\ref{sec:mainbranecharge}. As such, the branes
 cannot decay into topologically trivial configurations. We then proceed in Sec.~\ref{sec:dynstability} to discuss the dynamical aspects of stability. 
 This is particularly subtle for the 0-brane, and while we are not able to provide a complete proof of dynamical stability, we will confirm the absence of tachyonic modes in both the near-horizon and asymptotically-far regimes. More concretely, denoting the radial direction to the brane by $r$, we will show that in the region $r \leq \a'^{1/2}$ all tachyonic modes are absent by means of constructing an exact worldsheet description of that region; on the other hand,  in appendix \ref{sec:instability}, by explicitly studying the spectrum of the Laplacian in the presence of the brane we will show the absence of tachyons in the region $r \gg \a'^{1/2}$.

As we have just mentioned, the issue of stability can be partially addressed by understanding the tachyonic modes (or lack thereof) in the throat region 
of the branes. The exact worldsheet descriptions of the throat regions were announced in our previous work \cite{Kaidi:2023tqo},
and in the current paper we describe these results in full detail. In Sec.~\ref{sec:6brane}, we begin the discussion with the simplest case of the 6-brane in the $\hets$ heterotic string, 
for which a supergravity solution exists. After reviewing this solution, we take the near-horizon limit and use the result to motivate the worldsheet description of the throat region.
Equipped with these results, in Sec.~\ref{sec:throats} we are able to write the exact CFTs for the throat regions of not only the 6-brane, but also the 0-, 4-, and 7-branes. In all cases, the near-horizon theory is of the form $\mathbb{R}^{p,1} \times \mathbb{R}_\text{linear dilaton} \times H_k$ with $\mathbb{R}^{p,1}$ denoting the $\cN=(0,1)$ free sigma model with $\mathbb{R}^{p,1}$ target space,  $\mathbb{R}_\text{linear dilaton}$ denoting an $\cN=(0,1)$ linear dilaton CFT, and $H_k$ denoting a level-$k$ current algebra 
theory with $H_k$ as given in Table \ref{tab:summary}. The algebra $\mathfrak{h}$ corresponding to the group $H$ will be seen to match precisely with that predicted via closed string tachyon condensation in the aforementioned non-supersymmetric heterotic strings.

In Sec.~\ref{sec:boundary} we give more detail on the spectrum of fermions living in the throat region of the branes, as well as their anomalies. 
Anomaly cancellation is achieved via the Green-Schwarz mechanism, and since the branes provide boundary conditions 
for the theory in the throat, one encounters an interesting higher-dimensional version of the Callan-Rubakov problem. This leads to the expectation that the worldvolume theories of the branes will be strongly-coupled, non-Lagrangian theories. Finally, in Sec.~\ref{sec:GScoupling}, we offer a more formal understanding of the non-triviality of our branes. In particular, we will first discuss how the Green-Schwarz coupling provides a bilinear pairing between the worldsheet theory and spacetime manifold, such that by computing a non-trivial Green-Schwarz coupling one can prove that the worldsheet theory is non-trivial, in a precise sense to be expounded upon in the main text. For the case of the 0-brane this is rather straightforward, whereas for the 4-, 6-, and 7-branes the proper definition of the Green-Schwarz coupling turns out to be more involved. Finally, we close by drawing a connection between these results and the theory of topological modular forms. 

We have a few appendices.
In Appendix~\ref{sec:instability}, we provide a detailed analysis of the dynamical stability of our branes in the asymptotically-far region.
This will be used in Sec.~\ref{sec:dyn-stab}.
In Appendix~\ref{app:CFTsection},
we present a brief review of some facts about chiral CFTs related to the 
$\hete$ and  $\hets$ current algebra theories
to make the discussion in this paper self-contained.
This will be used in Sec.~\ref{sec:throats}.
In Appendix~\ref{app:mod2},
we discuss a mod-2 index which is used in the analysis of the 7-brane.
This will be used in Sec.~\ref{sec:vacuumstructure-7brane}.
Finally, in Appendix~\ref{app:elliptic},
we compute the elliptic genus of the internal worldsheet theory of our branes.
This will be used in Sec.~\ref{sec:GScoupling}.

Before proceeding, we would like to mention another non-supersymmetric brane recently found in the literature.
The work \cite{McNamara:2019rup} which rekindled recent interest in the search for new branes utilized the language of the bordism groups.
As such, one route towards the identification of topological charges is 
to compute appropriate bordism groups in spacetime.
Recent progress has led to the computation of said bordism groups in F-theory \cite{Debray:2023yrs}, 
the $E_8 \times E_8$ heterotic string \cite{Debray:2023rlx}, 
the $\SO(32)$ heterotic and Type I strings \cite{Kneissl:2024zox}, 
as well as in a number of non-tachyonic non-supersymmetric string theories \cite{Basile:2023knk}. 
The results in F-theory in particular 
motivated the discovery of a new non-supersymmetric object in Type IIB, now known as a \emph{reflection 7-brane} \cite{Dierigl:2022reg,Dierigl:2023jdp}. Codimension-2 branes in some nonsupersymmetric heterotic string theories are studied in \cite{Hamada:2024cdd}. Noncritical heterotic strings related to the throat region of the branes are further investigated in \cite{Yonekura:2024spl,Saxena:2024eil,DeFreitas:2024yzr}. A potential application is discussed in \cite{Alvarez-Garcia:2024vnr}.

In our case, the charges which the branes carry can be understood readily from a more traditional viewpoint using the homotopy groups of the gauge groups involved.
Bordism groups also play a role in the stability of the branes,
as will be discussed in the main text. 
Another difference is that, unlike in the case of the reflection 7-brane, we will describe our branes not only in the spacetime language, but also on the worldsheet, for which we make use of the notion of bordism between two-dimensional (2d) worldsheet \Nequals{(0,1)} supersymmetric  theories.

\section{Brane charges}\label{sec:mainbranecharge}

\subsection{Generalities}\label{sec:generalcharge}
In general, charges in gauge and gravitational theories are determined by the behavior of fields at infinity. 
For instance, electric charge in QED can be measured by the behavior of the electric field $\vec E$ at infinity, and is proportional to
$\int_{S^2} \d S\, \vec n \cdot \vec E$, where the integral is over the sphere at spatial infinity and $\vec n$ is the unit normal vector to the sphere.
Energy in general relativity can likewise be measured by the behavior of the metric field at infinity and is given by the ADM formula. 

In particular, some charges are characterized by the topology of gauge fields (in the sense described below), and we will refer to them as \emph{topological charges.}\footnote{Note that whether or not a charge is topological depends on the duality frame. For example, magnetic charge in QED is a topological charge while electric charge is not. However, they are exchanged under  electric-magnetic duality.} Consider a brane (or more generally some singularity such as an orbifold or orientifold) of codimension $n+1$, and take a plane perpendicular to it. The boundary of the plane at spatial infinity is $S^n$ or possibly a more general $n$-manifold $W_n$ in the case of a singularity. On $W_n$, one has some gauge and gravitational configuration, and if the configuration is topologically nontrivial, we say that the brane carries topological charge. 

For instance, magnetic charge in QED is characterized by the topology of a $\U(1)$ bundle on $W_2=S^2$. Let $F = \frac{1}{2} F_{\mu\nu} \d x^\mu \wedge \d x^\nu$ be the $\U(1)$ field strength. Then the magnetic charge is proportional to $\int_{S^2} F$, which measures the first Chern class of the $\U(1)$ bundle on $S^2$.
In the same way, if a codimension-$(n+1)$ brane has a magnetic charge under an $(n-1)$-form field $A$, that charge can be measured in terms of the $n$-form field strength $F = \d A$ via $\int_{W_n} F$. 

In the language of cohomology, the situation is as follows. (This paragraph is not needed for the main line of discussions of the present paper, and can be skipped.)
We can interpret the above brane charge as an element in the cohomology group $H^n(W_n)$.
At the level of differential forms, the charge is determined by the de~Rham cohomology class $[F|_{W_n}] \in H^n(W_n,\bR)$, where $F|_{W_n}$ is the restriction of $F$ to $W_n$, and $[F|_{W_n}]$ is its de~Rham cohomology class. More precisely, however, charges are classified by the cohomology group $H^n(W_n,\bZ)$ with integer coefficients. Even more precisely, we may need to use generalized cohomology theories. For instance, D-brane charges are classified by the behavior of Ramond-Ramond (RR) fields at infinity, and topologies of RR-fields are classified by K-theory~\cite{Moore:1999gb}. In Type I, IIA, and IIB string theories, the relevant cohomology groups are $KO^{-1}(W_n)$, $K^{0}(W_n)$, and $K^{-1}(W_n)$, respectively.
Compared with the usual description of brane charges in terms of K-theory, the degrees of cohomology groups for RR fields are shifted by $-1$. The heuristic explanation is as follows. Let $J$ be a brane current and $F$ be a field strength. Then we have equations of motion $\d F = J$. Therefore, if $J$ is a $p$-form, $F$ is a $(p-1)$-form. For Type IIA string theory, we have also used the periodicity theorem $K^{-2} \simeq K^0$.
Now let us understand e.g.~$KO^{-1}(W_n)$ in a little more detail. When $W_n=S^n$, we have (by using the suspension isomorphism in generalized cohomology theory) $KO^{-1}(S^n) \simeq KO^{-n-1}(\text{pt}) \oplus KO^{-1}(\text{pt})$, where $\text{pt}$ represents a point. The part 
$KO^{-1}(\text{pt}) \sim \bZ_2$ just classifies a discrete theta angle in the ten-dimensional supergravity.\footnote{
Type IIB string theory has the 0-form RR field $C_0$. After orientifold projection to Type I string theory, it takes values in $\bZ_2$, namely $C_0 = 0, \pi$. In the full string theory (rather than just $KO$-theory), these two values give physically equivalent theories,
since it can be compensated by the sign change of a fermion field.
} 
On the other hand, the part $KO^{-n-1}(\text{pt})$ gives the well-known fact that codimension-$(n+1)$ brane charges in Type~I string theory are classified by $KO^{-n-1}(\text{pt})$. The same is true for Type II string theories.\footnote{
For type IIA, we have $K^{0}(S^n) \simeq K^{-n}(\text{pt}) \oplus K^{0}(\text{pt})$ and the part $K^{0}(\text{pt}) \simeq \bZ$  classifies the Romans mass of the ten dimensional gravity.}

More generally, brane charges are not necessarily described by cohomology groups. Any topologically nontrivial configuration on the manifold $W_n$ at infinity gives a topological charge.\footnote{
In holography, not only charges, but in fact everything is determined by the behavior at infinity.} 
One class of charges is as follows.
Let us consider the case in which the theory contains gauge fields with gauge group $G$. The gauge bundle can be nontrivial on $W_n=S^n$.
The topologies of gauge bundles on $S^n$ are classified by the homotopy group $\pi_{n-1}(G)$---this is because any gauge bundle can be constructed from configurations on the \emph{northern hemisphere} and \emph{southern hemisphere} by gluing them along the \emph{equator} $S^{n-1}$ by a gauge transformation 
\beq
g: S^{n-1} \to G~.
\eeq
Topologies of such gauge transformations are classified by $\pi_{n-1}(G)$. 
For instance, in a $\U(1)$ gauge theory, we have $\pi_1(\U(1)) \simeq \bZ$, which leads to the usual classification of the magnetic charges of magnetic monopoles, whose codimension are $n+1=3$. 

\subsection{Some brane charges in heterotic superstring theories}\label{sec:heteroticcharge}
In this paper, we mainly discuss the case of ten-dimensional heterotic superstring theories where the gauge group $G$ is either $\hete$ or $\hets$. We consider brane charges associated to $\pi_{n-1}(G)$. 
Let us first recall the global structures of these groups.

In $\hete$, the $\bZ_2$ is the symmetry that exchanges the two $E_8$ gauge groups. Therefore $\bZ_2$ acts nontrivially on $E_8 \times E_8$ as a group automorphism. This is the reason for the appearance of the semidirect product $\rtimes$.

In $\hets$, the group $\Spin(32)$ is the universal cover of $\SO(32)$, and the $\bZ_2 \subset \Spin(32)$ is defined as follows. Let $v$ be the vector (defining) representation of $\so(32)$, and let $s$ and $c$ be the two spinor representations of $\so(32)$.\footnote{Throughout this work, we will denote the Lie algebra of a Lie group $G$ by $\mathfrak{g}$, like $\e_8$ for $E_8$ or $\so(32)$ for $\SO(32)$.} Then the $\bZ_2 \subset \Spin(32)$ appearing in the quotient is the subgroup of the center which acts on $v, s$, and $c$ as
\beq
v: (-1)~, \qquad s : (+1)~, \qquad c: (-1)~. 
\eeq
The action of $\bZ_2$ on $v$ actually follows from the action on $s$ and $c$, since $v$ is contained in the tensor product $s \otimes c$. In a theory with gauge group $\hets$, particles that transform nontrivially  under the $\bZ_2$ (such as particles in the representations $v$ and $c$) are forbidden.

The nontrivial homotopy groups of the above groups are given by
\beq
&\pi_0(\hete) \simeq \bZ_2~, \quad \pi_3(\hete) \simeq \bZ \times \bZ~, \nonumber \\
& \pi_1(\hets) \simeq \bZ_2~, \quad \pi_3(\hets) \simeq \bZ~, \quad  \pi_7(\hets) \simeq \bZ~.  \label{eq:homotopy}
\eeq
To obtain these homotopy groups, we have used the following mathematical facts:
\begin{itemize}
\item $\pi_0(G)$ is given by the number of disconnected components of the group $G$.
\item $\pi_n(G)$ for $n \geq 1$ is completely determined by the connected component $G_0 \subset G$ containing the identity. 
\item We can always represent $G_0$ as $G_0=\widetilde{G}_0/\Gamma$, where $\widetilde{G}_0$ is simply connected (i.e. $\pi_1(\widetilde{G}_0)=0$) and $\Gamma \subset \widetilde{G}_0$ is a discrete subgroup. We then have 
$\pi_1(\widetilde{G}_0/\Gamma)  \simeq   \pi_0(\Gamma) \simeq \Gamma$ and
$\pi_n(\widetilde{G}_0/\Gamma)  \simeq \pi_n(\widetilde{G}_0) $ for $n\geq 2$.
\item We have $\pi_n(G_1 \times G_2) \simeq \pi_n(G_1) \oplus \pi_n(G_2)$. 
\item $\pi_i(G)$ for simply-connected simple groups $G$ can be found in e.g.~ \cite[Appendix A]{Garcia-Etxebarria:2017crf}.
For us, we only need the following,\begin{itemize}
	\item $\pi_{1,2}(G)=0$ and $\pi_3(G) = \bZ$ for any simple compact Lie group $G$,
	\item $\pi_i(E_8)=0$ for $4\le i \le 7$,
	\item For $N \geq 9$, $\pi_{4,5,6}(\Spin(N))=0$ and $\pi_7(\Spin(N)) \simeq \bZ$.
\end{itemize}
\end{itemize}

Let us give slightly more detail on the cases involving $\pi_3$ and $\pi_7$ in (\ref{eq:homotopy}).
First, if we use $\pi_3(G) \simeq \bZ$ and $\pi_7(\Spin(N)) \simeq \bZ$ to construct gauge bundles on $S^4$ and $S^8$, the corresponding charges are measured by
\beq
\sQ_{4}=\int_{S^4} \frac{1}{2} \tr' \left( \frac{\i F}{2\pi} \right)^2~, \qquad \sQ_{8}= \int_{S^8} \frac{1}{4!} \tr_v \left( \frac{\i F}{2\pi} \right)^4~,
\eeq
where $F$ is the 2-form field strength, and the meanings of $\tr'$ and $\tr_v$ are as follows. The trace $\tr_v$ is taken in the vector representation $v$ of $\so(32)$. The fact that this representation is not allowed in $\hets$ is not a problem, since $F$ takes values in the Lie algebra $\so(32)$ rather than in the group. The symbol $\tr'$ for $\sQ_4$ is defined for $\e_8$ as $\frac{1}{60}$ times the trace in the adjoint representation, and for $\so(N)$ as $\frac{1}{2}$ times the trace in the vector representation.\footnote{
The algebra $\e_8$ has a subalgebra $\su(2) \times \e_7$ under which the 248-dimensional adjoint representation of $\e_8$ is decomposed as 
${\bf 248} = ( {\bf 3} \otimes {\bf 1} ) \oplus ( {\bf 1} \otimes {\bf 133} ) \oplus ({\bf 2} \otimes {\bf 56})$. 
By embedding an instanton in $\su(2) \subset \e_8$, one can check that 
$ \tr_{\bf 248} \left(  F \right)^2 =60   \tr_{\bf 2} \left( F_{\su(2)}  \right)^2$, where $F_{\su(2)}$ is the $\su(2)$ field strength.
This is the reason for including the factor $\frac{1}{60}$ for $\e_8$. In the same way, by using $\su(2) \times \su(2) = \so(4) \subset \so(N)$, we can see that $\sQ_4$ measures the instanton numbers also for $\so(N)$.
} 
In this normalization, $\sQ_{4}$ is the instanton number of a $G$ bundle on $S^4$ and hence $\sQ_{4} \in \bZ$. Likewise, $ \sQ_{8}$ is the integral of the Chern character of an $\so(N)$ bundle on $S^8$, and it can be shown using the Atiyah-Singer index theorem that $\sQ_{8} \in \bZ$. An explicit example with the minimum charges $\sQ_8=\pm 1$ can also be constructed and will be discussed below in Sec.~\ref{sec:n=8}.
This bundle underlies the Bott periodicity of KO theory.

\subsection{Explicit charge-carrying configurations} \label{sec:gaugeconfig}
Now let us discuss gauge bundles on $S^n$ constructed using the homotopy groups \eqref{eq:homotopy}
for the $\hets$ and $\hete$ heterotic superstring theories.
We start from $n=8$ and then move on to $n=4,2,1$.

\subsubsection{The case ${ n=8}$}\label{sec:n=8}
If we use $\pi_7(\hets) \simeq \bZ$, the gauge bundle is such that $\sQ_{8}$ defined above is nonzero.
For $\sQ_{8}=\pm 1$, we have the following geometric (i.e.~not only topological) construction of a $\hets$ gauge field configuration with the same rotational symmetry as the sphere. 

We consider the tangent bundle $TS^n$ of the sphere $S^8$, whose structure group is $\so(8)$. There are two associated spin bundles $S_\pm$  whose ranks are both 8.
We take one of them and embed it in $\so(32)$ by using $32=8+24$. Then the configuration has either $\sQ_{8}=1$ or $\sQ_{8}=-1$ depending on which spin bundle $S_\pm$ we use. This claim can be shown in the following way. 

Let $v$ be the vector representation of $\so(8)$, and let $s$ and $c$ be the two spinor representations. 
Now let us consider Chern roots of the $\so(8)$
field strength as follows. In the vector representation, we formally denote the eigenvalues of $ \si F_{\so(8)}|_v/2\pi$ by $\pm x_i~(i=1,2,3,4)$. In other words, in the standard block diagonal form of real antisymmetric matrices, we formally have
\beq
 \frac{F_{\so(8)}|_v}{2\pi} \to \bigoplus_{i=1}^{4} \begin{pmatrix}  0 & x_i \\  -x_i & 0 \end{pmatrix}~.
\eeq
These $x_i$ are the Chern roots. By using them, we define the $\ell$-th Pontryagin class $p_\ell~(\ell=0,1,2,\cdots)$ as the degree $2\ell$ part of the following polynomial,
\beq
\sum_{\ell \geq 0} p_\ell = \prod_i (1+x_i^2)~. \label{eq:Pontryagin}
\eeq
In particular, $p_0=1$ and
\beq
p_1 = \sum_{i} x_i^2~, \qquad p_2 = \sum_{i <j} x_i^2 x_j^2~.
\eeq
We also define the Euler density $e$ as
\beq
 \qquad e = \prod_{i} x_i~.
\eeq
Each of these characteristic classes can be given in terms of gauge invariant polynomials of $F_{\so(8)}$. 

In the spinor representations, the eigenvalues of $\si F_{\so(8)}|_s/2\pi$ and $\si F_{\so(8)}|_c/2\pi$ are given in terms of the Chern roots as
\beq
\frac{\pm x_1 \pm x_2 \pm x_3 \pm x_4}{2}~,
\eeq
where the product of the four signs is constrained to be $+$ for $s$ and $-$ for $c$. 
By a straightforward computation, one then finds that
\beq
\label{eq:Q8derivation1}
\frac{1}{4!} \tr_v \left( \frac{\i F_{\so(8)}}{2\pi} \right)^4 &= \frac{2}{4!} (p_1^2 - 2p_2) ~, \nonumber \\
\frac{1}{4!} \tr_s \left( \frac{\i F_{\so(8)}}{2\pi} \right)^4 &= \frac{1}{2 \cdot 4!} ( p_1^2+4p_2) + \frac{1}{2} e ~, \nonumber \\
\frac{1}{4!} \tr_c \left( \frac{\i F_{\so(8)}}{2\pi} \right)^4 &= \frac{1}{2 \cdot 4!} ( p_1^2+4p_2) - \frac{1}{2} e ~.
\eeq
On the sphere $S^8$, we have
\beq
\label{eq:Q8derivation2}
\int_{S^8} p_1^2 =0~, \qquad  \int_{S^8} p_2 = 0~, \qquad \int_{S^8} e =2~,
\eeq
where in the first two equations we have used the fact that the sphere is symmetric under an orientation-reversal diffeomorphism, and in the last equation we have used the fact that the Euler number of the sphere is 2.
Combining (\ref{eq:Q8derivation1}) and (\ref{eq:Q8derivation2}) then gives 
\beq
\int_{S^8}  \frac{1}{4!} \tr_v \left( \frac{\i F_{\so(8)}}{2\pi} \right)^4 =0~,  \quad
\int_{S^8}  \frac{1}{4!} \tr_s \left( \frac{\i F_{\so(8)}}{2\pi} \right)^4 = 1~,  \quad
\int_{S^8}  \frac{1}{4!} \tr_c \left( \frac{\i F_{\so(8)}}{2\pi} \right)^4 = -1~,
\eeq
and after embedding these bundles into $\so(32)$, the results for the spinor representations $s$ and $c$ give the desired bundles with $\sQ_8 = \pm 1$.

We remark that it is necessary to insert fundamental strings to make the above configuration consistent. Indeed, we have a supergravity equation (which follows from the supergravity action and in particular the Green-Schwarz coupling given in e.g. \cite{Polchinski:1998rr}) 
\beq
\d \star H = \frac{1}{4!} \tr \left( \frac{\i F}{2\pi} \right)^4  + \text{[gravitational]}+ J_\text{F1}~, \label{eq:gauss}
\eeq
where $H$ is the gauge invariant 3-form field strength of the $B$-field,\footnote{Our normalization of the $B$-field is different from standard textbooks such as \cite{Polchinski:1998rr} by a factor $(2\pi)^2 \alpha'$. For instance, when gauge and gravity fields are trivial, the Dirac quantization condition for $H$ is $\int H \in \bZ$ in our normalization. } 
the term [gravitational] represents gravitational and mixed gauge-gravitational corrections which are zero for the sphere, and $J_\text{F1}$ is the current for the charge of fundamental strings. By integrating this equation on $S^8$, we see that the insertion of $-\sQ_8$ fundamental strings is necessary. 

The presence of fundamental strings does not significantly modify supergravity configurations, since gauge and gravitational configurations have energy densities of order $g^{-2}_s$, while fundamental strings have energy densities of order $g_s^{0}$, and thus have negligible backreaction when the string coupling $g_s$ is taken to be small.

\subsubsection{The case ${ n=4}$}\label{sec:n=4}
We next consider the case of $n=4$. Corresponding to $ \pi_3(\hete) \simeq \bZ \times \bZ$, we have two instanton numbers $\sQ_{4,1}$ and $\sQ_{4,2}$ for the two $E_8$ groups. For $\hets$, we have just one instanton number $\sQ_4$.
In the present paper, unless otherwise stated, we consider the case $\sQ_{4,1} +\sQ_{4,2}=0$ for $\hete$ and $\sQ_4=0$ for $\hets$. The reason is as follows. There is a supergravity equation which for $\hete$ is given by (see e.g. \cite{Polchinski:1998rr})
\beq
\d H =  \frac{1}{2} \tr' \left( \frac{\i F_{1}}{2\pi} \right)^2+   \frac{1}{2} \tr' \left( \frac{\i F_{2}}{2\pi} \right)^2 + \text{[gravitational]}+\text{[NS5]}~, \label{eq:dH}
\eeq
where $H$ is the gauge invariant 3-form field strength of the $B$-field, $F_{1}$ and $F_{2}$ are the two field strengths for the two $E_8$ groups, and [gravitational] and [NS5] represent contributions from gravity and NS5-branes. On the sphere $S^4$, the gravitational contribution is zero because this term is parity odd, while the sphere is symmetric under orientation-reversing diffeomorphisms. Also, unless otherwise stated, we will consider the case that NS5-branes are absent. Therefore, the integration of the above equation over $S^4$ gives the condition $\sQ_{4,1} +\sQ_{4,2}=0$. The case of $\hets$ is similar. Notice that if we introduce NS5-branes, then their backreaction would be significant since their energy densities are of order $g^{-2}_s$.

For $\sQ_{4} = \pm 1$, we have the following geometric (i.e.~not only topological) construction of an $E_8$ gauge field configuration with the same rotational symmetry as the sphere. First we take $\su(2)  \subset \su(2) \times \e_7 \subset \e_8$. The sphere $S^4$ has two spin bundles associated to the tangent bundle, and we embed one of them into $\su(2)$. By an analogous computation as in the case of $n=8$ discussed above, we get $\sQ_{4} = \pm 1$ for each of these two spin bundles. Indeed, let $x_1,x_2$ be Chern roots of $\so(4)$,
\beq
 \frac{F_{\so(4)}|_v}{2\pi} \to \begin{pmatrix}  0 & x_1 \\  -x_1 & 0 \end{pmatrix} \oplus \begin{pmatrix}  0 & x_2 \\  -x_2 & 0 \end{pmatrix}~.
\eeq
In the spinor representations, the eigenvalues of $\si F_{\so(4)}|_s/2\pi$ and $\si F_{\so(4)}|_c/2\pi$ are
\beq
\frac{\si F_{\so(4)}|_s}{2\pi} \to \frac12 \begin{pmatrix} x_1 +x_2 & 0 \\ 0 &  -x_1-x_2     \end{pmatrix} , 
\qquad \frac{\si F_{\so(4)}|_c}{2\pi} \to \frac12 \begin{pmatrix} x_1 -x_2 & 0 \\ 0 &  -x_1+x_2     \end{pmatrix} .
\eeq
We thus have
\beq
\frac{1}{2} \tr_s \left( \frac{\i F_{\so(4)}}{2\pi} \right)^2 &=\frac14  (x_1+x_2)^2 = \frac14 p_1 + \frac12 e~, \nonumber \\
\frac{1}{2} \tr_c \left( \frac{\i F_{\so(4)}}{2\pi} \right)^2 &= \frac14 (x_1-x_2)^2 = \frac14 p_1 - \frac12 e~,
\eeq
where $p_1=x_1^2+x_2^2$ and $e=x_1x_2$ for the case of $\so(4)$.
By using the fact that the Euler number of $S^4$ is given by $\int_{S^4} e = 2$, we get the desired configuration with $\sQ_4 =\pm 1$. 

For $\sQ_{4,1} = - \sQ_{4,2} = 1$, a simple way to summarize this result is as follows. We take $\su(2) \times \su(2) \subset \e_8 \times \e_8$, and use the isomorphism $\su(2) \times \su(2) = \so(4)$.
By embedding the tangent bundle of the sphere into $\so(4)$, the two $\su(2)$ bundles are given by the two spin bundles discussed above, and hence we get $\sQ_{4,1} = 1$ and $ \sQ_{4,2} = -1$.

\subsubsection{The case ${ n=2}$}\label{sec:n=2}
We next consider the case of $n=2$. We can construct a gauge field configuration on $S^2$ associated to the elements of $ \pi_1(\hets) \simeq \bZ_2$ in the following way. 
We take the Cartan subalgebra $\so(2)^{16} \subset \so(32)$, and put a magnetic flux for each $\so(2) = \u(1)$ on $S^2$.
Then the $\so(32)$ gauge field in the vector representation is of the form
\beq
\frac{F}{2\pi} = \bigoplus_{i=1}^{16} \begin{pmatrix} 0 & \sq_i \\ -\sq_i & 0 \end{pmatrix} \frac{\epsilon}{4\pi}~, \label{eq:magneticflux}
\eeq
where $\sq_i$ capture the magnetic fluxes, and $\epsilon$ is the volume form on $S^2$ satisfying $\int_{S^2} \epsilon = 4\pi$. 

In fact, one can show that the solutions of the equations of motion for the two-dimensional Yang-Mills action $S \propto \int \tr (F_{\mu\nu}F^{\mu\nu})$ are always of the form \eqref{eq:magneticflux}, up to gauge transformations. This fact may be shown as follows. The Yang-Mills equation is given by $D^\mu F_{\mu\nu}=0$. In two dimensions, we have $F_{\mu\nu} = f \epsilon_{\mu\nu}$ where $f = \frac{1}{2}\epsilon^{\mu\nu} F_{\mu\nu}$. Thus the equation of motion becomes $D_\mu f=0$ and hence $f$ is covariantly constant.  After diagonalizing $f$ by gauge transformations into the Cartan subalgebra, a covariantly constant $f$ should be given by \eqref{eq:magneticflux} for constant parameters $\sq_i$.

If a particle in the vector representation $v$ is allowed, the $\u(1)$ charges of the particles contained in the vector representation are $\pm 1$. Then by the Dirac quantization condition, we would need to require that $\sq_i \in \bZ$. However, for the group $\hets$, there is no particle in the vector representation. We only have the adjoint representation, one of the spinor representations $s$, and representations that are constructed from them by tensor products and irreducible decompositions. In these representations, the $\u(1)^{16}$ charges are
\beq
\text{adjoint}~:~& ( 0, \cdots,0, \pm 1, 0,\cdots, 0,\pm 1, 0, \cdots, 0) \nonumber \\
\text{spinor }s~:~& ( \pm \frac{1}{2}, ~ \cdots ~, \pm \frac{1}{2})
\eeq
where for $s$ the signs are constrained by the condition that the product of all the signs is $+$. Then the Dirac quantization condition requires
\begin{equation}
\begin{aligned}
\pm \sq_i \pm \sq_j &\in \bZ \quad (i \neq j,~ i, j =1,\cdots,16)~,   \\
\frac{\pm \sq_1 \cdots \pm \sq_{16}}{2} &\in \bZ \quad (\text{the number of $-$ is even}).
\end{aligned}
\end{equation}
From these conditions it follows that
\beq
(\sq_1,\ldots,\sq_{16}) \equiv (0,\ldots,0) \quad \text{or} \quad \left( \frac12,\ldots,\frac12 \right) \mod \bZ^{16}~.
\eeq
The second case, namely $\sq_i \equiv \frac12 \mod \bZ$ for $i=1,\dots,16$, would be forbidden if the group were $\Spin(32)$ for which the vector representation would be allowed. Thus this is a configuration associated to the  nontrivial element of $\pi_1(\hets) \simeq \bZ_2$. On the other hand, the first case, namely $\sq_i \equiv 0 \mod \bZ$ for $i=1,\dots,16$, is associated to the trivial element of $\pi_1(\hets) $. 

If we want to minimize the energy of the configuration, we should take $|\sq_i|$ to be the minimal possible values.
For the topologically nontrivial case, such a configuration is given by $\sq_i = \frac12$ for all $i=1,\cdots,16$, up to gauge transformations (by the Weyl group) that change the signs of even numbers of $\sq_i$. All other configurations in the same topological class can flow to this configuration.

\subsubsection{The case ${ n=1}$}\label{sec:n=1}
Finally, we consider the case of $n=1$. If we use the nontrivial element of $\pi_0(\hete)$ to construct a gauge bundle on $S^1$, there is a $\bZ_2$ holonomy around $S^1$ such that the two $E_8$ gauge fields (and gauginos) are exchanged. More explicitly, let $\theta$ be a coordinate of $S^1$ with the periodicity $\theta \sim \theta +2\pi$, and let $A_1$ and $A_2$ be the gauge fields of the two $E_8$ gauge groups. Then the boundary condition is that
\beq
A_1(\theta+2\pi) = A_2(\theta)~, \qquad A_2(\theta+2\pi) = A_1(\theta)~,
\eeq
where we have omitted coordinates other than that of $S^1$. For the gauginos $\lambda_1$ and $\lambda_2$, there is an additional minus sign,
\beq
\lambda_1(\theta+2\pi) = -\lambda_2(\theta)~, \qquad \lambda_2(\theta+2\pi) = -\lambda_1(\theta)~.
\eeq
The reason is that, in the present paper, we consider the case of anti-periodic (or NS) spin structure on $S^1$.

\section{Stability of branes}

Having introduced the topological charges carried by our branes, we now discuss their stability. The discussion of stability involves both topological and dynamical aspects.
By topological aspects, we mean the question of whether a configuration can be continuously deformed to a trivial configuration.
By dynamical aspects, we mean whether the spherically symmetric gauge field configurations discussed in the previous section are stable in a given topological class.

In Sec.~\ref{sec:bordism}, we mainly focus on the topological aspects, in particular discussing the role of the concept of bordism. 
In Sec.~\ref{sec:dynstability} we discuss dynamical aspects of stability.

\subsection{Bordism and stability of branes}\label{sec:bordism}

\subsubsection{Abelian groups $\sugra_n$ and $\sqft_{n-32}$}

Consider a possibly off-shell supergravity configuration. We denote an $n$-dimensional configuration by $W_n$, which includes the information of the spacetime manifold as well as all other (bosonic) fields such as gauge fields.
Suppose we are interested in a brane whose codimension is $n+1$. As mentioned before, the topological charge of the brane is specified by the topology of the configuration $W_n$ which is the \emph{boundary at spatial infinity}. For our particular applications, $W_n$ will be $S^n$ with a topologically nontrivial gauge field configuration on it, but for the moment we work more generally.

We now introduce the concept of a supergravity bordism group $\sugra_n$ as follows.\footnote{
It would be better to denote it as $\sugra_n(\text{pt})$, but we omit $(\text{pt})$ for notational simplicity. In the case of supergravity theories obtained from heterotic string theories, this bordism group is also called the bordism group for twisted string structure. }
Elements of this group are (possibly off-shell) $n$-dimensional supergravity configurations like $W_n$, with the following equivalence relation.
We denote the orientation reversal\footnote{In general cases such as in the presence of pin structures, we need a generalization of the concept of orientation reversal, though we do not discuss this here. See \cite{Freed:2016rqq,Yonekura:2018ufj}.}
of $W_n$ as $\overline{W_n}$. We also denote the disjoint union of two manifolds $W_n$ and $V_n$ (including all field configurations) by $W_n \sqcup V_n$. Now consider two configurations $W_n$ and $W'_n$. 
Suppose that $\overline{W_n} \sqcup W'_n$
is realized as the boundary of an $(n+1)$-dimensional configuration $Y_{n+1}$,
\beq
\partial Y_{n+1} = \overline{W_n} \sqcup W'_n~.
\eeq
Then $W'_n$ is considered to be equivalent to $W_n$, 
\beq
W'_n \sim W_n~.
\eeq
We denote the equivalence class of $W_n$ as $[W_n]$. As a set, $\sugra_n$ is defined by
\beq
\sugra_n = \left\{\, [W_n]~~|~~W_n \text{ : $n$-dimensional configuration} \right\}~.
\eeq
We may endow $\sugra_n$ with an Abelian group structure by defining the sum of  $[W_n]$ and $[V_n]$ to be
\beq
[W_n]+[V_n] = [W_n \sqcup V_n]~;
\eeq
it is straightforward to see that the equivalence class of the empty manifold $\varnothing$ gives the identity element of the group, and that $[\overline{W_n}]$ is the inverse of $[W_n]$,
\beq
[\varnothing]=0~, \qquad [\overline{W_n}] = - [W_n]~.
\eeq

Now let us restrict our attention to heterotic string theories.
Given a supergravity configuration $W_n$, we can use it to construct a worldsheet $\cN=(0,1)$ supersymmetric quantum field theory (SQFT) by regarding $W_n$ as the target space of a sigma model. For the present discussion, we neglect the issues of conformal invariance and UV completeness of the worldsheet theories. 

From the point of view of the worldsheet, the parameter $n$ specifying the spacetime dimension  
is related to the gravitational anomaly of the worldsheet theory. For each boson $\phi$ describing a target space coordinate, there is a right-moving Majorana-Weyl fermion $\tilde \psi$ which is the superpartner of $\phi$. This fermion  has a worldsheet gravitational anomaly. Recall that heterotic superstring theories also have the current algebra CFT for the group $G = \hete$ or $\hets$, which can be constructed in terms of 32 left-moving Majorana-Weyl fermions. Thus the total gravitational anomaly is proportional to $n - 32$ if we consider an $n$-dimensional target space. 

We may define the concept of a bordism group of $\cN=(0,1)$ SQFTs with such gravitational anomaly, which we denote as $\sqft_{n-32}$, in a way analogous to $\sugra_n$. Its elements consist of equivalence classes of SQFTs with the coefficient of the worldsheet gravitational anomaly given by $n-32$. The equivalence relation is defined as follows. 
First, an SQFT $\cY_{n-31}$ is said to have a boundary $\cW_{n-32}$ if $\cY_{n-31}$ has a noncompact direction\footnote{Here, \emph{noncompactness} is in the sense of energy spectrum. For instance, consider a harmonic oscillator with a potential $V(x) =\frac12 x^2$. This has a discrete energy spectrum and we regard it as compact, even though the target space is $\bR$ spanned by $x$. In a compact theory, the potential energy must grow in directions which are noncompact in the sense of target space manifolds. For more discussions, see e.g. \cite{Gaiotto:2019asa,Gaiotto:2019gef,Johnson-Freyd:2020itv,Yonekura:2022reu}. } 
and in that direction the theory looks like $\bR_{>0} \times \cW_{n-32}$, where by abuse of notation $\bR_{>0}$ denotes the ${\cal N}=(0,1)$ sigma model with target space $\bR_{>0}$. In particular, the theory $\cY_{n-31}$ quantized on $S^1$ has both discrete and continuous parts in the energy spectrum, and the continuous part of the energy spectrum matches that of $\bR_{>0} \times \cW_{n-32}$. In this situation, we denote $\cW_{n-32} = \partial \cY_{n-31}$. By using this concept, equivalence classes $[\cW_{n-32}]$ can be defined in the same way as for $\sugra_n$.

As mentioned above, we can construct an SQFT from a given supergravity configuration. This implies that there is a homomorphism
\beq
\sF:\, \sugra_n \to \sqft_{n-32}~.
\eeq
One can check that this map $\sF$ is well-defined, in the sense that the equivalence class of the SQFT obtained from $W_n$ depends only on the equivalence class $[W_n]$ and not on the representative $W_n$. Indeed, suppose that we have a configuration $Y_{n+1}$ with boundary $\partial Y_{n+1} = W_n$. We extend $ Y_{n+1}$ by attaching a cylinder of the form $\bR_{\geq 0} \times W_n$ so that the boundary is at spatial infinity. We can then use this manifold to construct an SQFT $\cY_{n-31}$ such that  $\partial \cY_{n-31} = \cW_{n-32}$, where $\cW_{n-32}$ is the SQFT constructed from $W_n$, from which we conclude that if $[W_n]=0$, then $[\cW_{n-32}]=0$, thereby implying the well-definedness of the map $\sF$.

\subsubsection{Implications of groups $\sugra_n$ and $\sqft_{n-32}$ on stability}
We now discuss the implications of the above rather abstract concepts to the stability of branes. For a given topological charge at spatial infinity specified by a supergravity configuration $W_n$, there are three different possibilities:
\begin{enumerate}
\item The equivalence class $[W_n] \in \sugra_n$ is trivial, i.e. $[W_n]=0$.
\item The equivalence class $[W_n] \in \sugra_n$ is nontrivial, but its image $\sF ([W_n] ) \in \sqft_{n-32}$ is trivial, i.e. $\sF ([W_n] )=0$.
\item The image $\sF ([W_n] ) \in \sqft_{n-32}$ is nontrivial.
\end{enumerate}

We now consider the implications of each of these cases for the stability of branes. Before doing that, however, let us comment that the configurations $W_n$ appearing above are not necessarily on-shell, and likewise the SQFT $\cW_{n-32}$ is not necessarily conformally invariant. However, even if they are off-shell or conformally non-invariant at a given time slice, they might still correspond to allowed initial conditions for time evolution. Recall that $W_n$ is interpreted as a configuration in a plane transverse to the brane, and in particular lives at a constant time slice. It may be possible to evolve the configuration in directions parallel to the brane so that they become on-shell. This point is particularly nontrivial for SQFTs, but in any case we neglect this issue in the following discussion. 

\paragraph{The first possibility:}
Let us first consider the case $[W_n]=0$. By definition of $\sugra_n$, there exists a smooth configuration $Y_{n+1}$ with $\partial Y_{n+1} = W_n$. In this case, the issue of stability depends on whether the energy density is localized or broadly distributed over spacetime. If the energy density is localized, it is appropriate to regard the configuration as a brane. On the other hand, if the energy density is broadly distributed, it means that the brane has dissipated into a continuous configuration. Let us discuss examples. 

One class of examples is D-branes in Type~I string theory. In this case, $W_n$ is $S^n$ with a topologically nontrivial RR field configuration, as mentioned in Sec.~\ref{sec:generalcharge}. We can replace a brane by a smooth $\O(32)$ gauge field configuration without changing its charge measured at spatial infinity.\footnote{Topologies of branes are determined KO-theoretically by bundles $\O(32+K) \times \O(K)$, where $K$ is the number of virtual anti-D9-branes, which is arbitrary. However, in spacetime dimensions less than $32$, we can always represent KO-theory elements by using only the case $K=0$. } 
We take $Y_{n+1}$ to be $\bR^{n+1}$ with an $\O(32)$ gauge field configuration associated to elements of $\pi_n(\O(32))$. 
We remark that the $\O(32)$ gauge field configuration is trivial at infinity; only RR fields are nontrivial on $W_n=S^n$. The homotopy group $\pi_n$ (rather than $\pi_{n-1}$) appears because the gauge field configuration is trivial at infinity and hence we can formally regard $\bR^{n+1}$ as $S^{n+1}$ as far as the $\O(32)$ gauge field (but not the RR field) is concerned. Let $A_\mu$ be the $\O(32)$ gauge field. Then the energy $\rho$ of the configuration per unit worldvolume is given by
\beq
\rho \propto -\int_{\bR^{n+1}} \d^{n+1} x\, \tr F_{\mu\nu}F^{\mu\nu} + \cO(\alpha')~,
\eeq
where $\cO(\alpha')$ represents $\alpha'$ corrections, and
the minus sign in the first term is just due to the fact that we take gauge fields to be anti-hermitian rather than hermitian.
Now, whether a configuration will be localized or not depends on the codimension and is determined by  Derrick's theorem \cite{derrick1964comments} as is standard. If we change the scale of the gauge field by using a constant $R$ as
\beq
A_\mu(x) \to A^R_\mu(x) : = \frac{1}{R}\, A_\mu\left( \frac{x}{R}\right),
\eeq
then the energy per unit worldvolume is rescaled as
\beq
\rho \to \rho_R = R^{n+1 - 4} \rho ~, \label{eq:Derrick}
\eeq
neglecting $\alpha'$ corrections. 
We thus conclude that: (i) when $n+1 >4$, the energy can be made smaller by taking $R \to 0$. This implies that a localized configuration is energetically preferred. The actual configuration depends on all the $\alpha'$ corrections; (ii) when $n+1=4$, the situation is marginal. In fact, both a localized brane and ``instanton'' gauge field configurations of any size saturate the BPS bound; (iii) when $n+1 < 4$, the energy can be made smaller by taking $R \to \infty$. Therefore, the brane completely dissipates into a broad gauge field configuration. For instance, the D7-brane of Type~I string theory has a nontrivial topological charge associated to $\pi_1(\O(32)) \simeq \bZ_2$~\cite{Witten:1998cd}, but this brane is unstable. In fact, it is known that a tachyon exists in the open strings stretching between the D7-brane and the 32 D9-branes \cite{Frau:1999qs,Loaiza-Brito:2001yer}.

A second example is the NS5-brane in heterotic string theories. In this case, $W_3$ is $S^3$ with a unit of flux for the 3-form field $H$. In this case, we can replace the NS5-brane by a gauge instanton for the gauge group $\hete$ or $\hets$. Both an NS5-brane and gauge instantons of any size saturate the BPS bound. 

We remark that both of the above examples have nonzero charges, and yet the corresponding elements of the bordism groups are zero. D-branes have charges of the RR-fields, and the NS5-branes have magnetic charges of the $B$-field. The elements of the bordism groups are zero because there exist smooth configurations in the interior region that interpolate a given configuration $W_n$ at spatial infinity.

\paragraph{The second possibility:}

Next let us consider the case that $[W_n] \neq 0$ and $\sF ([W_n] )=0$. 
In this case, we do not have a smooth low energy supergravity configuration for a given topological charge $W_n$. By definition of the bordism group $\sqft_{n-32}$, it may still be possible (up to the issue of conformal invariance) that the interior region is described by a nontrivial SQFT that is not realized by a low energy supergravity configuration. If this is the case, we expect that the region described by the SQFT has a very high energy density determined by inverse powers of $\alpha'$. This energy density is presumably positive in a supersymmetric theory in an asymptotically flat spacetime.\footnote{We do not have a complete justification of this claim, but let us make the following remarks. The problem can be asked even if there is no brane at all. A vacuum decay can happen in nonsupersymmetric theories. In supersymmetric theories, a decay may be forbidden due to the positivity of energy implied by the supersymmetry algebra of asymptotically flat spacetimes. For some of the branes, like the 6-brane of the present paper, we believe that this argument forbids a decay. However, for other branes of the present paper, asymptotic behavior is modified rather significantly as we discuss in the next subsection, and this argument is not immediately applicable. } Then the configuration should be localized rather than broadly distributed, since if it were not localized, we would have a large region with a very high positive energy density. Therefore, we expect to have a localized brane in this case.

\paragraph{The third possibility:}

Finally, let us consider the case that $\sF ([W_n] ) \neq 0$. In this case, even the existence of any configuration with a given topological charge $W_n$ is a nontrivial question. In quantum gravity, and in particular by the cobordism conjecture \cite{McNamara:2019rup}, the existence of such a configuration is expected. By a similar reason to the case of $[W_n] \neq 0$, we expect that the configuration should be a localized brane. 

We conclude that we expect the existence of a brane if $[W_n] \neq 0$ (regardless of $\sF([W_n])$), although we cannot prove it. We remark that there still remain some subtleties in this analysis, which we will discuss in the next subsection \ref{sec:dyn-stab}.

\subsubsection{How our branes fit in this picture}
\label{sec:topological-classes}

Let us study whether the condition $[W_n] \neq 0$ is satisfied or not for the $W_n$ discussed in Sec.~\ref{sec:heteroticcharge}. For this purpose, we use characteristic classes of gauge fields.\footnote{Usual characteristic classes are described as follows. Let $G$ be a gauge group, and let $BG$ be the classifying space of $G$-bundles. We consider a cohomology group $H^k(BG, \bA)$ with coefficient group $\bA$ (or more generally any generalized cohomology group). Let $v$ be any element of $H^k(BG, \bA)$.
Given a $G$-bundle on a manifold $W_n$, let $f : W_n \to BG$ be the classifying map. Then the pullback $f^* v \in H^k(W_n, \bA)$ is the characteristic class for that bundle associated to $v$. However, we remark that the characteristic class for the case $n=4$ discussed below is more subtle than $H^k(BG, \bA)$, since we need to take into account the effects of the $B$-field.} 
In particular,

\begin{itemize}
\item $\bZ_2$ holonomies of the group $\hete$ on a manifold $X$ are described by a cohomology element $v_1 \in H^1(X, \bZ_2)$.

\item For $\hets$, we can consider magnetic flux modulo 2 because, as discussed in Sec.~\ref{sec:heteroticcharge} in the case of $S^2$, all magnetic fluxes $\sq_i~(i=1,\cdots,16)$ are the same modulo integers, and are either $0$ or $\frac12$. More generally, on any manifold $X$, magnetic fluxes modulo 2 are given by a cohomology element $v_2 \in H^2(X, \bZ_2)$. This is a special case of the more general fact that for any connected group $G$, magnetic fluxes are classified by $H^2(X, \pi_1(G))$.

\item We will consider $v_4 \in H^4(X, \bZ_2)$, which is the characteristic class associated to the instanton number of one of the two $E_8$ groups modulo 2. The reason that we reduce by modulo 2, rather than using the instanton number itself, is roughly as follows.

 The two $E_8$ fields are exchanged by the action of the $\bZ_2$ symmetry, so each of the instanton numbers $\sQ_{4,1}$ and $\sQ_{4,2}$ is not gauge invariant, but rather behaves as $\sQ_{4,1} \leftrightarrow \sQ_{4,2}$ under the $\bZ_2$ gauge symmetry. On the other hand, as discussed around (\ref{eq:dH}), we have the constraint $\sQ_{4,1}=-\sQ_{4,2}$ in the absence of NS5-branes.\footnote{%
 The gravitational correction in \eqref{eq:dH} does not affect the following discussion since it turns out to be even when $\dim X< 8$. In more detail, the gravitational correction is given by a characteristic class of spin manifolds which is half the Pontryagin class $p_1/2$ at the rational level as determined by the Green-Schwarz mechanism (see e.g. \cite{Polchinski:1998rr}). The mod 2 reduction of this characteristic class is known to be the fourth Stiefel-Whitney class $w_4$. 
 Now, on spin manifolds with less than eight dimensions, we have $w_4=0$, 
 since on a spin manifold $w_4=\nu_4$ where $\nu_k$ is the $k$-th Wu class,
 and $v_k$ is zero when $k$ is more than half the dimension of the manifold.
 Therefore, the gravitational correction term $p_1/2$ is even.} 
 Together, we conclude that the gauge invariant information should be invariant under $\sQ_{4,1} \leftrightarrow -\sQ_{4,1}$. The appropriate invariant quantity is $\sQ_{4,1} \mod 2$. The quantity $v_4$ is a valid characteristic class as far as we neglect nonperturbative transitions between gauge field configurations and NS5-branes.

 A mathematically more precise discussion goes as follows. An $\hete$ bundle over $X$ can be considered as an $E_8$ bundle over the double cover $\pi: \tilde X\to X$, whose structure is specified by $v_1$ above. We have the characteristic class $\tilde v_4 \in H^4(\tilde X,\bZ)$ for the instanton number of the $E_8$ bundle on $\tilde X$. 
 
The $\tilde X$ can be regarded as a bundle over $X$ with fiber $S^0$. Let $L$ be the bundle $D^1 \to L \to X$ associated to $v_1$, where $D^1=[-1,1]$ is the interval whose boundary is $S^0$. In particular, the boundary of the total space $L$ is $\tilde X$. Now we consider the Gysin sequence which follows from the long exact sequence for $\tilde X \to L \to (L,\tilde X)$ and the Thom isomorphism $H^k(L,\tilde X,\bZ_2) \simeq H^{k-1}(X,\bZ_2)$. Here the bundle $L$ is not orientable if $v_1$ is nonzero, so we use the coefficient group $\bZ_2$. The Euler class of the bundle $L$ (i.e. the restriction of the Thom class to the zero section $X \subset L$) is given by $v_1$, and  hence we have the Gysin sequence 
\begin{equation}
\begin{aligned}
\cdots
\xrightarrow{\pi^*} &H^3 (\tilde X,\bZ_2)  
\xrightarrow{\pi_*} H^3 (X, \bZ_2)  \\
\xrightarrow{v_1\cup}  H^{4} (X,\bZ_2)  
\xrightarrow{\pi^*} &H^4 (\tilde X,\bZ_2)  
\xrightarrow{\pi_*} H^4 ( X,\bZ_2)  \\
\xrightarrow{v_1\cup}  H^{5} (X,\bZ_2)  
\xrightarrow{\pi^*} &H^5 (\tilde X,\bZ_2)  
\xrightarrow{\pi_*} \cdots 
\end{aligned} 
\end{equation}
where $\pi_*$ is given by the composition of the connecting homomorphism $H^k(\tilde X,\bZ_2) \to H^{k+1}(L,\tilde X,\bZ_2)$ and the Thom isomorphism $H^{k+1}(L,\tilde X,\bZ_2) \simeq H^{k}(X,\bZ_2)$, the $v_1\cup$ is the cup product with $v_1$, and the $\pi^*$ is the pullback along the double cover.

We consider the mod 2 reduction of the $\tilde v_4 \in H^4 (\tilde X,\bZ) $ which we denote by the same symbol by abuse of notation. The element $\pi_*(\tilde v_4) \in H^4(X,\bZ_2)$ is given by the sum of the characteristic classes for the instanton numbers of the two $E_8$ bundles reduced modulo 2.\footnote{This statement is explicitly checked from the definition of the connecting homomorphism and the Thom isomorphism when $v_1=0$ and hence $\tilde X = S^0 \times X$. On the other hand, when $v_1 \neq 0$, the characteristic class for each of the two $E_8$'s is not defined separately, and we may just define what we mean by ``the sum of the characteristic classes for the instanton numbers of the two $E_8$ bundles reduced modulo 2'' by $\pi_*(\tilde v_4)$.} Then the constraint \eqref{eq:dH} says that $\pi_*(\tilde v_4)$ is zero (when the gravitational correction $p_1/2$ is zero modulo 2). From the exact sequence, $\tilde v_4$ is an image by $\pi^*$ of an element $v_4\in H^4(X,\bZ_2)$,
which is well-defined modulo classes of the form $v_1\cup H^3(X,\bZ_2)$.\footnote{An intuitive explanation for the ambiguity $v_1\cup H^3(X,\bZ_2)$ is as follows. Let us demonstrate it when $X$ is a 4-manifold with nonzero $v_1$. First consider the case that there are no instantons on $X$. Intuitively, this is the situation in which we want to regard $v_4=0$. Now we create a pair of an instanton and an anti-instanton of one $E_8$, and move the instanton along a closed path on which $v_1$ is nonzero. After going around the closed path, the instanton of one $E_8$ becomes an instanton of the other $E_8$ because $v_1$ is the holonomy for the exchange of the two $E_8$'s. Thus, we have an anti-instanton of one $E_8$ and an instanton of the other $E_8$. Intuitively this is the situation in which we want to regard $v_4 \neq 0$. We discuss more details of this phenomenon in an example in Sec.~\ref{sec:7brane4brane}.
}

\item Finally, we define $v_8 \in H^8(X, \bR)$ by $v_8 =  \frac{1}{4!} \tr \left( \frac{\i F}{2\pi} \right)^4+\cdots$, where the ellipses represent gravitational corrections. Ideally, one would include gravitational corrections so that $v_8$ may be regarded as the image by the Chern character of a KO-theory element of $X$,  but this is not essential for the purposes of the present discussion. The quantity $v_8$ is a valid characteristic class as far as we neglect nonperturbative transitions between gauge field configurations and fundamental strings.
\end{itemize}

The connection between these characteristic classes and our previous discussion of stability is as follows. Suppose that we have a gauge field configuration on an $n$-manifold $W_n$ on which the integral of the characteristic class $v_n$ is nontrivial,
\beq
\int_{W_n} v_n \neq 0~.
\eeq
Then $W_n$ cannot be a boundary of some $Y_{n+1}$. Indeed, suppose on the contrary that $W_n = \partial Y_{n+1}$. The $v_n$ is also defined on $Y_{n+1}$. (For $n=4$, there is ambiguity of the form $v_1\cup H^3(Y_5,\bZ_2)$ and we just choose an arbitrary $v_4$ on $Y_5$. We assume that $v_1=0$ on the boundary $\partial Y_5=W_4$.) Then by Stokes' theorem (or its analog for cohomology with more general coefficients) we would have
\beq
\int_{W_n} v_n  = \int_{Y_{n+1}} \d v_n =0~,
\eeq
contradicting the original assumption. Since the configurations discussed in Sec.~\ref{sec:heteroticcharge} have nonvanishing characteristic classes, we conclude that $[W_n] \neq 0$ for these configurations, leading us to expect that there exist stable, localized branes sourcing them. (In Sec.~\ref{sec:GScoupling}, we will further discuss the stronger property that $\sF([W_n]) \neq 0$ as an element of $\sqft_{n-32}$.)

\subsection{Dynamical aspects}\label{sec:dynstability}
In the previous subsection, we have mainly focused on topological aspects of stability. In the current subsection, we move on to discuss the dynamical aspects of stability.

\subsubsection{Energy per unit worldvolume}
We start from a remark on the terminology ``brane'' used in the present paper. Some of the ``branes'' we discuss in the present paper have an infinite energy per unit worldvolume. The reason is as follows. Let us use a polar coordinate system
\beq
\d s_{n+1}^2 = \d r^2 + r^2 h_{ij}(\theta) \d \theta^i \d \theta^j~,
\eeq
where $\d s_{n+1}^2$ is the metric in a plane transverse to the brane, $r$ is the radial coordinate, $\theta^i~(i=1,\cdots,n)$ are coordinates of $S^n$, and 
\beq
\d \Omega_n^2:=h_{ij}(\theta) \d \theta^i \d \theta^j
\eeq
is the metric of the sphere with unit radius. We have neglected gravitational corrections to the flat metric, which is justified in the region $r \to \infty$. In this coordinate system, we consider a gauge field configuration 
\beq
A = A_i(\theta) \d \theta^i
\eeq
 that depends only on $\theta$. Then the Yang-Mills kinetic term is proportional to
\beq
S \propto -\int \d r   \d^n \theta\,\, r^{n} \sqrt{h} \left( r^{-4}   h^{ij} h^{k\ell} \tr F_{ik} F_{j \ell} \right), \label{eq:divergence}
\eeq
where we have neglected the $r$ dependence of the dilaton, which is justified in the region $r \to \infty$. Notice that $h^{ij} h^{k\ell} \tr F_{ik} F_{j \ell}$ is a function of $\theta$ and does not depend on $r$. As a result, the integral \eqref{eq:divergence} has an IR divergence when $n \geq 3$, coming from the region $r \to \infty$. (In the region $r \to 0$ the expression is not reliable and this argument does not apply.) 

This infrared divergence is not a fundamental problem. Indeed, the energy density in ten-dimensional spacetime is proportional to $ \left( r^{-4}  h^{ij} h^{k\ell} \tr F_{ik} F_{j \ell} \right) \propto r^{-4}$ and goes to zero as $r \to \infty$. We may compare this situation with that of a cosmological constant, which is clearly physically sensible and has an energy density which behaves as $r^{0}$. The asymptotic structure of spacetime is modified in the presence of a cosmological constant, such as to asymptotic AdS in the case of a negative cosmological constant. In our case of the gauge field $A = A_i(\theta) \d \theta^i$, the asymptotic behavior is modified by the energy density $ \left( r^{-4}  h^{ij} h^{k\ell} \tr F_{ik} F_{j \ell} \right) \propto r^{-4}$, and this modification is much milder than a cosmological constant. Corrections to the metric from the local energy density of order $r^{-4}$ are of order $r^{-2}$, because Einstein's equations relate second derivatives of the metric to the energy density.

Terms of the order $r^{-2}$ are the leading terms of the perturbation of the metric when $n \geq 3$; 
a completely neutral black brane with a constant tension per unit worldvolume gives a contribution of order $r^{1-n}$ to the metric, since it is a solution to the Laplace equation in $n+1$ dimensions.
When $n=2$, this becomes of order $r^{-1}$.
When $n=1$, more severe terms of order $\log r$ are also possible. 

Although the IR divergence for $n \geq 3$ is not a fundamental problem, it may be necessary to think that the Hilbert space of the theory is different from the Hilbert space without the brane
 when a brane with IR divergent energy per worldvolume is present. For instance, let us again consider the case of a cosmological constant, or even more concretely the case of Type IIB string theory on $AdS_5 \times S^5$. Different fluxes on $S^5$ give different cosmological constants and different asymptotic behavior at infinity, and they have different corresponding Hilbert spaces, i.e. the Hilbert spaces of ${\cal N}=4$ $\su(N)$ super Yang-Mills for different $N$. 

Another, quite different, example is that of an isolated very heavy quark in a confining gauge theory without a light quark. An isolated quark should be attached to a color flux tube which goes to infinity. A color flux tube is a dynamical string, and because of its tension, a semi-infinite string gives a  divergent energy. The Hilbert space with a semi-infinite string is different from the Hilbert space without such a string. On the other hand, a configuration containing both a quark and an anti-quark is in the same Hilbert space as the one without a semi-infinite string. In the same way, when both a brane and an anti-brane are present, the configuration is described by the same Hilbert space as the one without a brane because of the cancellation of gauge fluxes at infinity. 

Despite the infinite energy per unit worldvolume for $n \geq 3$, we still refer to our configurations as ``branes''. This is analogous to calling an isolated heavy quark a ``particle'' in a confining gauge theory.

We emphasize that the situation for our branes for $n \geq 3$ is not exactly the same as in the cases of a cosmological constant or a heavy quark in a confining gauge theory. The message is simply that such configurations are physically sensible even if they have infinite energy per unit worldvolume.

Let us close by recalling that in string theory, O-planes are well-known objects which change the asymptotic behavior significantly---in fact, the manifold at infinity is $W_n = S^n/\bZ_2$. O-planes play important roles in various ways. In the same way, the branes studied in this paper may play important roles in string theory. Also, we remark that the 6-brane (i.e. $n=2$) has no problem about infinite energy per unit worldvolume, and is just like a usual brane.

\subsubsection{Dynamical stability}
\label{sec:dyn-stab}

We now return to the issue of the stability of brane configurations. In Sec.~\ref{sec:heteroticcharge}, we have discussed geometric configurations of the gauge field $ A_i(\theta) \d \theta^i$ on $S^n$ which realize a given topology and also have spherical symmetry. We can ask whether these configurations are energetically stable or not. This question is somewhat subtle, and the answer depends on the precise meaning of ``stability'', and also on whether we only consider the sphere $S^n$ or also take into account the radial direction $r$. 

First, let us consider the stability of the gauge field configurations when they are put on a rigid sphere $S^n$, without taking into account the radial direction. 
\begin{itemize}

\item The configuration for $n=1$ discussed in Sec.~\ref{sec:n=1} is just a $\bZ_2$ holonomy around $S^1$, and hence it has no energy cost (at the classical level). We can also include holonomies of the diagonal subgroup $E_8 \subset E_8 \times E_8$ without energy cost, so there are moduli.

\item The configuration for $n=2$ discussed in Sec.~\ref{sec:n=2}  is given by the magnetic flux \eqref{eq:magneticflux}. It satisfies the equations of motion, and the lowest energy solution in a given topological class is stable. In particular, the solution with $\sq_i= \frac12$ for all $i=1,\cdots,16$ is stable.

\item The configuration for $n=4$ discussed in Sec.~\ref{sec:n=4} is given by an instanton for one $\su(2) \subset \e_8$ and an anti-instanton for the other $\su(2) \subset \e_8$. As is well-known, an instanton solution has the lowest energy in a given topological class, and it has moduli.

\item The configuration for $n=8$ discussed in Sec.~\ref{sec:n=8} is constructed by using one of the spin bundles on $S^8$ associated to the tangent bundle. This is not stable for the following reason. As discussed around \eqref{eq:Derrick}, Derrick's theorem states that the energy of a gauge field configuration with codimension $8$ is lowered if we localize the configuration into a small region. This means that the spherically symmetric configuration does \textit{not} have the minimum energy. In fact, one can show explicitly that the spherically symmetric configuration has a tachyonic mode on $S^8$, see appendix~\ref{sec:instability}. 

\end{itemize}
At first, the instability for the case $n=8$ may seem problematic. A completely localized gauge configuration is indistinguishable from a fundamental string piercing the $S^8$. There is also a fundamental string required by the Gauss law \eqref{eq:gauss} which has the opposite charge to the gauge field configuration. In the lowest energy configuration, these strings seem to annihilate with each other and disappear completely. 

We argue that the spherically symmetric configuration for $n=8$ is actually stable in the sense discussed below if we take into account the radial direction. In the presence of the radial direction, we formulate the question in the following way. We impose the spherically symmetric gauge field configuration as a boundary condition at infinity, $r \to \infty$. This makes sense because the configuration satisfies the equations of motion. Subject to this boundary condition, we ask whether there is any tachyonic mode that is normalizable (in particular in the $r$ direction). If there is no tachyonic mode, we regard the configuration as stable. 

Even if a gauge field configuration has tachyonic modes on $S^n$, it is still possible that tachyonic modes are absent when the radial direction is included. As an analogy, let us recall the case of AdS space with a scalar field. If the mass term of the scalar field is negative, it seems at first that there is a tachyonic instability. However, it is well-known that a scalar field with a negative mass term is stable (in the sense of normalizable modes)  in AdS space as long as the mass satisfies the Breitenlohner-Friedmann (BF) bound \cite{Breitenlohner:1982bm,Breitenlohner:1982jf}. An analogous thing can happen in our situation. Even if the mode expansion on $S^n$ has a negative eigenvalue mode, that mode may be stabilized by the effect of the radial direction. 

It is technically hard to study tachyonic modes for the entire $r$, so we separately study the limits $r \lsim \alpha'^{1/2}$ and $r \gg \alpha'^{1/2}$. 
\begin{itemize}
\item For the region $r \gg \alpha'^{1/2}$, we study the issue in detail in appendix~\ref{sec:instability} and find that there is no tachyonic mode for the configuration discussed in Sec.~\ref{sec:n=8}, which uses a spin bundle to construct a gauge configuration of $ \so(32)$ on $S^8$. In the appendix we also study another example in which we use the tangent bundle (rather than the spin bundle) to construct a gauge configuration of $ \so(32)$ on $S^8$, and in that case we \textit{do} find a tachyonic instability. 

\item For the region $r \lsim \alpha'^{1/2}$, stringy $\alpha'$ corrections become important. We will obtain exact worldsheet descriptions of this region and find no tachyonic modes from the worldsheet analysis, as discussed in Sec.~\ref{sec:throatstability}. 
\end{itemize}
The moduli mentioned for $n=4$ and $n=1$ are also stabilized by the effects of the radial direction. 

Finally, we remark that if we also introduce an anti-brane so that the total charge is zero, then the instability on $S^n$ may reappear. However, this is just a problem of the IR region $r \to \infty$ and hence the core region $r \lsim \alpha'^{1/2}$ still makes sense. This core region is the one which is  expected to contain interesting new physics.

\section{Six-brane solutions}\label{sec:6brane}

Having introduced the charges carried by the proposed branes and discussed their stability, we now provide some more details on the specific case of the 6-brane. 
In fact, supergravity solutions for the 6-brane have been known for general $\sq_i$ since the late 1980s and the early 1990s \cite{Gibbons:1987ps,Garfinkle:1990qj,Horowitz:1991cd},
and are similar to the well-known black NS5-brane solutions.
After reviewing these supergravity solutions,  we perform a worldsheet analysis for general $\sq_i$, and compare it to the results of supergravity.
Such a worldsheet analysis was also done in the past in \cite{Giveon:1993hm,Giddings:1993wn}, but our choice of the monopole charge vector seems new and leads to a somewhat different conclusion from those old results.
Our comparison between supergravity and worldsheet predictions provides a nice consistency check for the discussion of worldsheet dynamics for the other branes, which will appear in later sections.  

\subsection{Supergravity solutions}

In this subsection, we review the black brane solutions for the 6-brane. Solutions for the 4- and 0-branes will be discussed in \cite{KFWY}.

The relevant part of the supergravity action is
\begin{multline}
S = \frac{2\pi}{(2\pi \alpha'^{1/2})^{8}} \int \d^{10} x\, \sqrt{-G} e^{-2\Phi} \Bigl( R+4\partial_\mu \Phi \partial^\mu \Phi \\
 - \frac{1}{2}(2\pi \alpha'^{1/2})^2\tr' \left(\frac{\si F}{2\pi} \wedge \star \frac{\si F}{2\pi} \right)+\cdots \Bigr)~,
\end{multline}
where $G_{\mu\nu}$ is the metric, $R$ is the Ricci scalar, $\Phi$ is the dilaton, $F$ is the field strength for the gauge group $G$ which is taken to be anti-hermitian (and hence $\i F$ is hermitian), and $\star$ is the Hodge star. As discussed in Sec.~\ref{sec:heteroticcharge}, the $\tr'$ is defined to be
$\frac{1}{60}$ times the trace in the adjoint representation for $\e_8$, and $\frac12$ times the trace in the vector representation for $\so(32)$. The overall factor of the action depends on the conventions for $\Phi$ and is not essential in the following discussion.\footnote{For instance, the tension of the solitonic 5-brane constructed from a gauge instanton configuration $F = \star_4 F$ (where $\star_4$ is the four dimensional Hodge star in a plane $\bR^4$ transverse to the brane) is
\beq
 \frac{2\pi}{(2\pi \alpha'^{1/2})^{6} e^{2\Phi}} \int_{\bR^4} \frac{1}{2}\tr' \left(\frac{\si F}{2\pi} \wedge \star_4 \frac{\si F}{2\pi} \right) =  \frac{2\pi}{(2\pi \alpha'^{1/2})^{6}e^{2\Phi}} |\sQ_4|
\eeq
where $\sQ_4 \in \bZ$ is the instanton number. This reproduce the tension of the NS5-brane with charge $\sQ_4$.}
On the other hand, the relative coefficients of terms in the action are important. 

\begin{figure}[t]
    \centering
            \includegraphics[width=0.48\hsize]{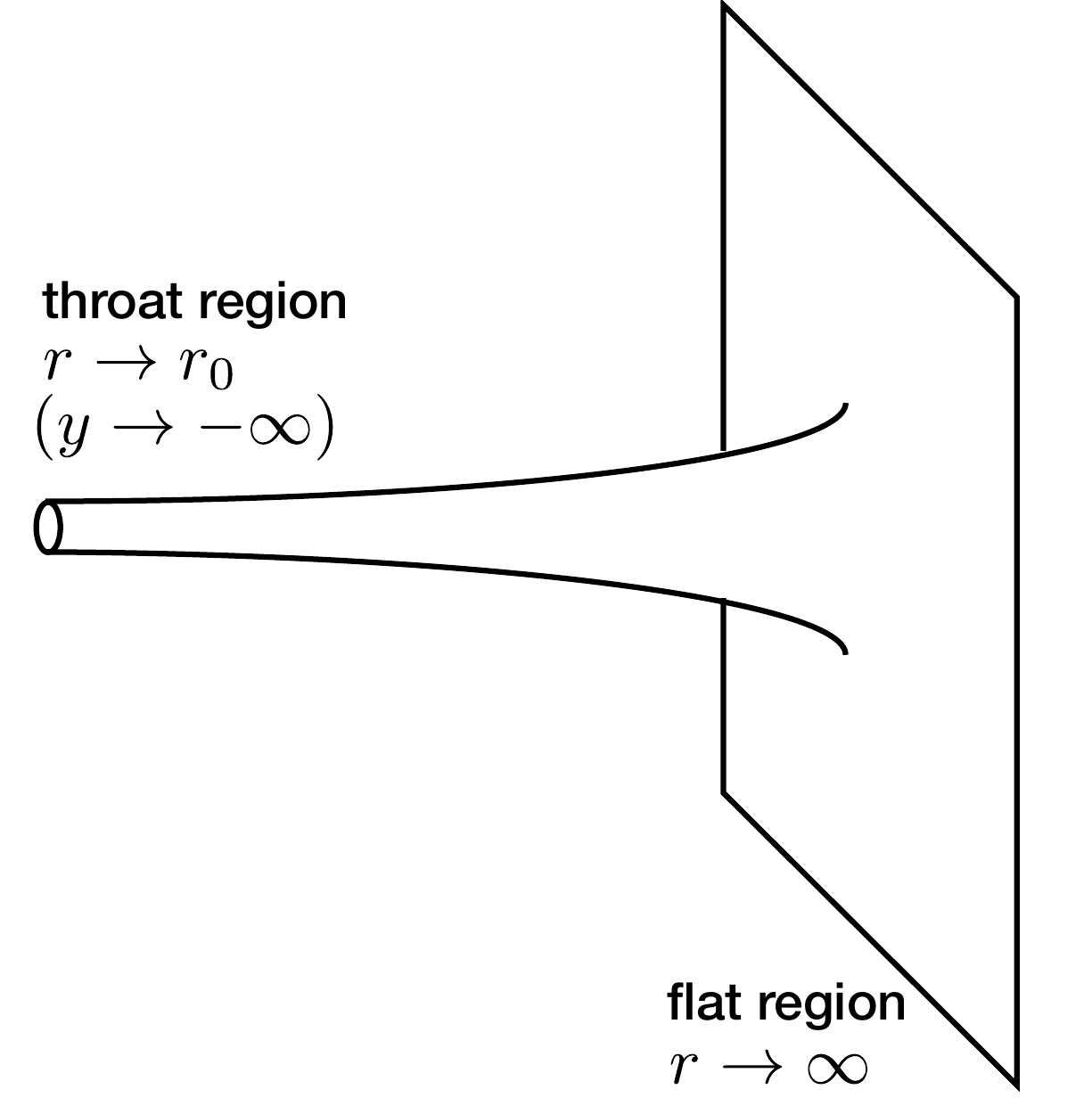}
    \caption{
A schematic picture of the extremal 6-brane solution. The directions $x^\mu$ parallel to the brane are suppressed, and the two-dimensional angular sphere $S^2$ is simplified to be one-dimensional. }
    \label{fig:6-brane}
\end{figure}

Now we write down supergravity solutions for the 6-brane. We first recall that the $\hets$ gauge field configuration is taken to be \eqref{eq:magneticflux}, which we write as
\beq
F= \bigoplus_{i=1}^{16} \begin{pmatrix} 0 & \sq_i \\ -\sq_i & 0 \end{pmatrix} f~, \qquad f= \frac{\epsilon}{2}~. \label{eq:magneticflux2}
\eeq
The lowest energy, stable case is given by $\sq_i=\frac12~(i=1,\cdots,16)$, but a solution is known for general $\sq_i$. The $f$ in \eqref{eq:magneticflux2} is regarded as the field strength for a $\u(1)$ gauge field (normalized to be hermitian), in terms of which the above action is given by
\beq
S  \propto \int \d^{10} x\, \sqrt{-G} e^{-2\Phi} \left( R+4\partial_\mu \Phi \partial^\mu \Phi - \frac{1}{2}  \alpha' \left( \sum_{i=1}^{16}\sq_i^2 \right) \left( f \wedge \star f \right)+\cdots \right)~.
\eeq
Thus the fields are reduced to the set $(G_{\mu\nu}, \Phi, f)$. For this set, black brane solutions are known \cite{Horowitz:1991cd}, and take the form
\beq
\d s^2 &= - \frac{(1- \frac{r_-}{r})}{(1- \frac{r_+}{r})} \d t^2+ \d x^i \d x^i + \frac{\d r^2}{(1- \frac{r_-}{r})(1- \frac{r_+}{r})} + r^2 \d \Omega^2_2~, \nonumber \\ 
e^{-2\Phi} &= g_s^{-2} \left( 1- \frac{r_-}{r} \right)~, \quad\qquad f= \frac{\epsilon}{2}~,
\eeq
where $t$ is a time coordinate, $x^i$ are coordinates parallel to the brane, $r$ is the radial coordinate perpendicular to the brane, $\d \Omega^2_2$ is the metric for the $S^2$ surrounding the brane, $g_s$ is the string coupling constant at infinity, and $r_\pm$ are parameters of the solution related to the $\sq_i$ by
\beq
r_+r_- = \frac{1}{8} \alpha' \sum_{i=1}^{16} \sq_i^2~, \qquad r_+ \geq r_-~.
\eeq
The extremal limit is given by taking $r_+=r_- =: r_0$,
\beq
\d s^2 =  \d x^\mu \d x_\mu + \frac{\d r^2}{(1- \frac{r_0}{r})^2} + r^2 \d \Omega^2_2~, \quad
e^{-2\Phi} = g_s^{-2} \left( 1- \frac{r_0}{r} \right)~, \quad r_0^2 =  \frac{1}{8} \alpha' \sum_{i=1}^{16} \sq_i^2~, \label{eq:6extremal}
\eeq
where $(x^\mu) = (t, x^i)$. This extremal solution is schematically shown in Figure~\ref{fig:6-brane}. This behavior is similar to that of the extremal NS5-brane solution.

We can take the near horizon limit of the extremal solution. For this purpose, we introduce a new coordinate $y$ defined by
\beq
y = r_0 \log \left(  \frac{r-r_0}{r_0}\right) ,
\eeq
and take the limit $r \to r_0$ or in other words $y \to -\infty$ to obtain
\beq\label{eq:6throat}
\d s^2 =  \d x^\mu \d x^\mu + \d y^2 + r_0^2 \d \Omega^2_2~, \qquad
\Phi =  -  \frac{y}{2r_0}  + \log g_s~.
\eeq
This is the throat region in Figure~\ref{fig:6-brane}.
We see that the dilaton has a linear dependence on $y$. 

As discussed in the previous section, these solutions have tachyonic instabilities unless $\sq_i=\frac12~(i=1,\cdots,16)$.\footnote{%
The analysis of the stability of monopole configurations in non-Abelian gauge theory goes back to \cite{Brandt:1979kk}.
There, it was shown that a tachyonic instability already exists in the non-gravitational limit when
there is a root such that its inner product with the monopole charge vector is $>1$.
} However, they are still valid solutions of the equations of motion. Moreover, if $r_+ \gg \alpha'^{1/2}$, the solutions are reliable in the sense that $\alpha'$ corrections are negligible. In the extremal limit, this condition is satisfied if $\sum_i \sq_i^2 \gg 1$. Let us analyze this case in a bit more detail.

The worldsheet action for the coordinate $y$ (neglecting its fermionic superpartner) is given by
\beq
S=\frac{1}{4\pi \alpha'} \int \d^2 \sigma \sqrt{h} h^{\alpha\beta} \partial_\alpha y \partial_\beta y + \frac{1}{4\pi}\int \d^2 \sigma \sqrt{h} \Phi(y) R(h)~, \label{eq:dilatonCFT}
\eeq
where $\sigma^\alpha$ is a worldsheet coordinate system, $h_{\alpha\beta}$ is the worldsheet metric, $R(h)$ is the worldsheet Ricci scalar, and the dilaton $\Phi$ is given by \eqref{eq:6throat} as a function of $y$. This is a linear dilaton CFT. The contribution to the central charge from the nontrivial dilaton profile $\Phi$ is given by the formula
\beq
\delta c  =  6\alpha' (\partial_y \Phi)^2 = \frac{3\alpha'}{2 r_0^2} 
 =  \frac{12}{\sum_{i=1}^{16} \sq_i^2}~, \label{eq:centralcharge}
\eeq
where we have used the value of $r_0$ given by \eqref{eq:6extremal}. This result is expected to be valid for large $\sum_{i=1}^{16} \sq_i^2$. In the next subsection, we will rederive this formula from a worldsheet analysis without using supergravity solutions.

\subsection{Worldsheet analysis in the throat region}\label{sec:6worldsheet}
We now turn to the worldsheet analysis of the near horizon limit of the extremal 6-brane solution.
The current algebra CFT for $\hets$ can be constructed from 32 left-moving Majorana-Weyl fermions $\lambda=(\lambda_i)$ where $i=1,\cdots,32$. They are gauged by a $\bZ_2$ symmetry $\lambda \to -\lambda$. 
On the other hand, the worldsheet sigma model fields for the directions transverse to the 6-brane will be denoted by $(X^\mu, \tilde \psi^\mu)~(\mu=1,2,3)$, where $\tilde \psi^\mu$ are right-moving superpartners of $X^\mu$. Before coupling them to the current algebra CFT, they are ${\cal N}=(0,1)$ sigma model fields for the target space $\bR^3$.

After coupling the sigma model fields to the current algebra theory, the action is roughly given by
\beq
S \sim \int \d^2 \sigma\, G_{\mu\nu}(X) (\partial_\alpha X^\mu \partial^\alpha X^\nu +\tilde \psi^\mu \partial  \tilde\psi^\nu) + \lambda \bar D \lambda + \cdots.\label{eq:worldsheetS}
\eeq
Here the ellipses represent some 4-fermi interaction terms, and $\bar D$ is given by
\beq
\bar D = \bar \partial + A_\mu (X) {\bar \partial}X^\mu~,
\eeq
where $A_\mu(X) \d X^\mu$ is the $\hets$ gauge field whose field strength is given by \eqref{eq:magneticflux2}. Without the $\bZ_2$ gauging of $\lambda$, this gauge field would be inconsistent if $(\sq_1,\ldots,\sq_{16}) \equiv (\frac12,\ldots,\frac12) \mod \bZ^{16}$.

We now study the behavior of the above sigma model. For this purpose, we make the following assumption. The target space $\bR^3$ can be decomposed into the radial direction $\bR_\text{radial}$ and the angular directions $S^2$, and we assume that the dynamics is essentially determined by the coupling between the $S^2$ part of the sigma model and $\lambda$. The quantum field theory for this system of $S^2$ and $\lambda$ has a nontrivial renormalization group flow. However, the full worldsheet theory must have conformal invariance. We assume that the radial direction $\bR_\text{radial}$ plays the role of the ``dilaton'', in the sense of a worldsheet scalar field used to recover conformal invariance of a theory with a mass scale.

This recovery of conformal invariance may be described very roughly as follows. Let $y$ be a worldsheet scalar field with action of the form \eqref{eq:dilatonCFT}, with a nontrivial profile for $\Phi(y)$. Then, if $\partial_y \Phi$ were constant, the operator $\exp(- 2\kappa y)$ would have left and right scaling dimensions $(h_L, h_R)$
\beq
h_L=h_R= - \alpha'  (\partial_y \Phi) \kappa-  \alpha'\kappa^2 ~.
\eeq
If $|\partial_y \Phi |$ is sufficiently large, we may take a small $\kappa$ so that $\exp( - 2\kappa y)$ has scaling dimensions $h_L=h_R=1$. Now, when there is another theory (like the coupled $S^2$-$\lambda$ system) which depends on a mass scale $M$, we may replace 
\beq
M^2 \to M^2\exp( -2\kappa y)
\eeq
to recover conformal invariance. This procedure would be rigorously valid if $|\partial_y \Phi| \to \infty$ and hence $\kappa \to 0$ with $(\partial_y \Phi) \kappa$ fixed. Our assumption is that conformal invariance is recovered in a way qualitatively similar to this mechanism, even though $|\partial_y \Phi |$ is not very large and not constant.

If this mechanism works, the radial direction roughly corresponds to the renormalization group flow of the coupled $S^2$-$\lambda$ system. We do not have a complete justification of this assumption, since the radial direction has nontrivial interactions with the $S^2$ via the term involving $G_{\mu\nu}(X)$. However, one partial justification of this assumption will come from the fact that we will correctly reproduce the result \eqref{eq:centralcharge} obtained by the supergravity analysis. 

To get a solution, we also use the following special property of the 6-brane which is not available for other branes. The $S^2$ can be regarded as the complex projective space $\mathbb{CP}^1$. The fermions $\lambda_i$ can also be regarded as 16 complex fermions via
\beq
\Lambda_i = \lambda_{2i-1} + \si \lambda_{2i}~, \qquad (i=1,\cdots,16)~.
\eeq
We then see that the matter content is that of ${\cal N}=(0,2)$ supersymmetry. The gauge field configuration $A_\mu(X) \d X^\mu$ is also consistent with the complex structure because it is the direct sum of $\u(1)$ gauge fields acting on each complex fermion $\Lambda_i$.

For ${\cal N}=(0,2)$ supersymmetry, we can consider R-symmetries under which the complex supercharge has R-charge $-1$. For the sigma model part, the bosonic field of the $\mathbb{CP}^1$ sigma model has R-charge 0, and hence the fermionic superpartner has R-charge $-1$. For the current algebra theory, we assign R-charges $\sr_i$ to $\Lambda_i$. These charges are constrained by a sigma model anomaly as follows. First, $\mathbb{CP}^1$ has a $\U(1)$ bundle whose field strength is given by $f$ in \eqref{eq:magneticflux2}.
If a fermion is coupled to the sigma model via flux $\sq f$ for some $\sq$, then a $\u(1)_R$ symmetry which assigns charge $\sr$ to that fermion suffers from an axial anomaly given by 
\beq
\pm  \sq\sr \frac{ f}{2\pi} ~,
\eeq
where the sign depends on whether the fermion is left- or right-moving.
Recall that the superpartner of the $\mathbb{CP}^1$ sigma model is coupled to the tangent bundle of $\mathbb{CP}^1$, whose field strength is $2f$. Also notice that
each left-moving $\Lambda_i$ is coupled to $\sq_i f$. Then the anomaly cancellation condition for the R-symmetry is given by
\beq
2+ \sum_{i=1}^{16} \sq_i \sr_i=0~. \label{eq:anomalyfree}
\eeq

We assume that the theory flows to a superconformal fixed point in the IR. The R-symmetry $\u(1)_R$ that appears in the superconformal algebra is determined via $c$-extremization \cite{Benini:2012cz},
which we quickly review below. 
First we consider the $\u(1)_R$--$\u(1)_R$ anomaly. Its coefficient $\cA_R$ is given by
\beq
\cA_R = 1 - \sum_{i=1}^{16} \sr_i^2~. 
\eeq
In two-dimensional CFTs, this coefficient is determined by the two-point function $\vev{j_R j_R}$ of the R-symmetry current $j_R$. By supersymmetry, this two-point function is related to the two-point function of the right-moving energy-momentum tensor $\widetilde T(\bar z)$, whose coefficient is determined by the right-moving central charge $c_R$. Therefore, for the superconformal R-symmetry, $c_R$ should be proportional to $\cA_R$,
\beq
c_R = 3\cA_R~,
\eeq
where the coefficient $3$ is determined e.g. by considering a simple example (such as a free ${\cal N}=(0,2)$ multiplet with target space $T^2$). Now we regard $c_R$ as a function of $\sr_i~(i=1,\cdots,16)$ and extremize $c_R$ with respect to $\sr_i$. This condition follows from the fact that the two-point functions of the superconformal R-symmetry current and ordinary (non-R) currents are zero; any two R-symmetry currents $j_R$ and $j'_R$ are related by $j'_R = j_R + \sum_i \delta \sr_i j_i$, where $j_i$ are non-R currents and $\delta \sr_i$ are constants, and hence the derivative of $c_R$ with respect to $\sr_i$ is proportional to the coefficient of a two-point function $\vev{j_R j_i}$. In this extremization, we must take into account the condition \eqref{eq:anomalyfree}, which we impose by using a Lagrangian multiplier $\lambda$. 

In our case, we extremize the function
\beq
c_R(\{\sr_i\}, \lambda) = 3\left(1 - \sum_{i=1}^{16} \sr_i^2 \right) + \lambda \left( 2+ \sum_{i=1}^{16} \sq_i \sr_i \right)~.
\eeq
The result is
\beq
\sr_i = - \frac{2\sq_i}{\sum_{j=1}^{16} \sq_j^2} ~, \qquad
c_R = 3 -  \frac{12}{\sum_{j=1}^{16} \sq_j^2} ~.  \label{eq:c_R}
\eeq
On the other hand, the non-interacting right-moving central charge of the ${\cal N}=(0,2)$ multiplet for $\mathbb{CP}^1$ (i.e a single free ${\cal N}=(0,2)$ multiplet) is given by $c_R^\text{free} = 3$. To maintain the correct central charge, we assume that the radial direction $\bR_\text{radial}$ becomes a linear dilaton CFT which accounts for the difference
\beq
\delta c = c_R^\text{free} -c_R =  \frac{12}{\sum_{j=1}^{16} \sq_j^2}~.
\eeq 
This is exactly the result \eqref{eq:centralcharge} obtained by the supergravity analysis. We remark that these results are expected to coincide only  in the limit $\sum_{j=1}^{16} \sq_j^2 \gg 1$, so the complete agreement is accidental.

The success of the above computation is encouraging. We assumed that the worldsheet dynamics is essentially determined by the interaction between the $S^2$ sigma model and the current algebra theory $\lambda_i$, with the radial direction playing only the role of recovering conformal invariance including the correct total central charge. We will use the same assumption for other branes in later sections.

Before ending this section, let us observe that when $\sq_i=\frac12~(i=1,\cdots,16)$, the right-moving central charge $c_R$ is exactly zero, as can be seen by using the formula \eqref{eq:c_R}. Therefore, there are no right-moving degrees of freedom, and the IR theory is purely left-moving. We will see that this is the case for other branes as well.

\section{Exact CFTs for the throat region of the branes}
\label{sec:throats}
In Sec.~\ref{sec:heteroticcharge} we discussed gauge field configurations on $S^n$. We are interested in branes that have charges described by these gauge field configurations at infinity. These branes are supposed to be of the form shown in Figure~\ref{fig:6-brane}. (See \cite{KFWY} for the 4- and 0-branes.)

In this section, we study the worldsheet CFTs that describe the throat region of the branes. Our conclusion will be that in all cases the CFT is given by
\beq
\bR^{p,1} \times \bR_\text{linear~dilaton} \times H_k~, \label{eq:exactCFT}
\eeq
where $H_k$ is the current algebra theory based on a group $H$ at some level $k$,
$\bR^{p,1}$ is the ${\cal N}=(0,1)$ free sigma model with target space $\bR^{p,1}$ which describes the directions parallel to the $p$-brane (with $p=8-n$), and $\bR_\text{linear~dilaton}$ is the ${\cal N}=(0,1)$ linear dilaton CFT such that the total CFT has the correct central charge required by heterotic string theory.

\begin{table}
  \centering
  \begin{tabular}{c||c|c|c|c}
  $p$ & $7$ & $6$ & $4$ & $0$\\
  \hline
  $G$ & $\hete$ & $\hets$ & $\hete$ & $\hets$   \\ 
  \hline
  $H$ & $E_8$ & $\SU(16)/\bZ_4$ & $(E_7 \times E_7)/\bZ_2$ & $\Spin(24)/\bZ_2$ \\
  \hline
  $k$ & $2$ & $1$ & $1$ & $1$  \\
  \hline
  $n$ & $1$ & $2$ & $4$ & $8$  \\
  \hline
  $c$ & $31/2$ & $15$ & $14$ & $12$
  \end{tabular}
  \caption{The combinations of $(G, H, k, n)$ for which the equivalence \eqref{eq:equivalence} holds. 
  We also tabulate the central charge $c=16-n/2$ of $H_k$ for future reference.}
  \label{table:equivalence}
\end{table}

The part $\bR^{p,1}$ is not involved in the nontrivial worldsheet dynamics. 
The part $\bR_\text{linear~dilaton}$ follows from the  assumption mentioned previously in the case of 6-brane solutions in Sec.~\ref{sec:6brane}. 
Below we will discuss how the part $H_k$ is obtained.

For this purpose, 
we will need to use the following equivalence on purely left-moving worldsheet theories:
\begin{equation}
  G_1 = [H_k \times \SO(n)_1]/(-1)^{\sF_\sL}~. \label{eq:equivalence}
\end{equation}
The list of $(G,H,k,n)$ is given in Table \ref{table:equivalence}.
Here, $H_k$ and $\SO(n)_1$ are current algebra theories which turn out to be spin-CFTs (i.e., CFTs that depend on spin structures of worldsheets). 
In particular, $\SO(n)_1$ is realized by $n$ Majorana-Weyl fermions which make sense even for $n=1$.
In general, spin-CFTs (or fermionic CFTs) have fermion parity symmetry $(-1)^{\sF_\sL}$ given by the ``$2\pi$-rotation''. The meaning of $/(-1)^{\sF_\sL}$ in \eqref{eq:equivalence} is that we gauge the $\bZ_2$ symmetry generated by $(-1)^{\sF_\sL}$. The right hand side of \eqref{eq:equivalence} is a bosonic CFT after gauging the $\bZ_2=\{1,(-1)^{\sF_\sL}\}$. The theory $G_1$ on the left-hand-side is also a bosonic CFT as is well-known in string theory.

We will  provide a review of basic facts about chiral CFTs and a derivation of \eqref{eq:equivalence} in Appendix~\ref{app:CFTsection}. (See also \cite{BoyleSmith:2023xkd} for a derivation from a slightly different perspective.)
Let us now proceed to the derivation of \eqref{eq:exactCFT}.

\subsection{Gauge field configurations}\label{sec:dualgaugeconfiguration}

Except for $n=1$, it turns out that our gauge field configurations on $S^n$ all have the property that the symmetry group $\SO(n)$ in the equivalence \eqref{eq:equivalence} is such that the $\SO(n)$ bundle on $S^n$ is given by the tangent bundle of $S^n$. Let us explain the reason in each case.

\subsubsection{The 0-brane}
For the case $n=8$, the gauge field configuration of Sec.~\ref{sec:n=8} means that we decompose $\so(32)$ as $\so(24) \times \so(8)'$ and then take the $\so(8)'$-bundle to be the rank 8 spin bundle of $S^n$. Here we have put a prime on $\so(8)'$ because we will need an automorphism of $\so(8)$
to relate it to the $\SO(n)$ that appears in \eqref{eq:equivalence}. This automorphism is necessary for the following reason.  

The equivalence \eqref{eq:equivalence} for the case $n=8$ is 
\beq
(\hets)_1 = [ (\Spin(24)/\bZ_2)_1 \times \SO(8)_1 ] /(-1)^{\sF_\sL}.~\label{eq:6.2}
\eeq
Recall that $(\Spin(8N)/\bZ_2)_1$ (for a positive integer $N$) has a realization in terms of $8N$ Majorana-Weyl fermions gauged by $\bZ_2$. On the other hand, $\SO(N)_1$ is realized by $N$ Majorana-Weyl fermions, without any gauging. We now use the duality\footnote{This is the duality that  appears in the relation between fermions in the Green-Schwarz and  NSR formalisms in  light-cone gauge. In both formalisms there are 8 fermions, and the $\bZ_2$ gauging is called the GSO projection in the context of the NSR formalism.}
\beq
(\Spin(8)/\bZ_2)_1 = \SO(8)_1~.
\eeq
This duality is related to the automorphism $\so(8) \xrightarrow{\sim} \so(8)$ under which the vector and one of the spinor representations are exchanged. In fact, the groups $\Spin(8)/\bZ_2$ and $\SO(8)$ are isomorphic under it. Thus the equivalence \eqref{eq:6.2} can also be written as
\beq
(\hets)_1 = [ (\Spin(24)/\bZ_2)_1 \times (\Spin(8)/\bZ_2)_1 ] /(-1)^{\sF_\sL}~. \label{eq:equivalence3}
\eeq

In general, $(\Spin(8N)/\bZ_2)_1 $ is a spin-CFT for odd $N$.
The fermion parity symmetry in $(\Spin(8N)/\bZ_2)_1$ for odd $N$ is realized as the $\widehat \bZ_2$ symmetry dual to the $\bZ_2$ used in the gauging. Here, the global $\widehat \bZ_K$ symmetry ``dual to'' a gauge $\bZ_K$ symmetry is defined in the standard way well-known for orbifolds; if a holonomy $a \in \bZ_K$ of the gauge $\bZ_K$ is put on a spatial $S^1$ on the worldsheet, then the states in that twisted sector carry dual global $\widehat \bZ_K$ charge $a$. More generally, the $\bZ_K$ gauge field $\sa \in H^1(\Sigma, \bZ_K)$ and a global $\widehat \bZ_K$ symmetry background field $\sb \in H^1(\Sigma,\widehat \bZ_K)$ are coupled on a Riemann surface $\Sigma$ by a term
\beq
\frac{2\pi \si}{K} \int_{\Sigma} \sa \wedge \sb~.\label{eq:dualcoupling}
\eeq

Now let us provide a derivation of the equivalence \eqref{eq:equivalence3} which is independent of the derivation in Sec.~\ref{sec:equivalenceintro}.
It is possible to show this in a greater generality in the following form: \begin{equation}
(\sX \times \sY)/\bZ_2 = [(\sX/\bZ_2) \times (\sY/\bZ_2) ]/\widehat\bZ_2 ~. \label{eq:moregeneral}
\end{equation}
Here $\sX$ and $\sY$ are two theories with $\bZ_2$ symmetry,
the quotient by $\bZ_2$ on the left hand side is by the diagonal subgroup,
and the quotient by $\widehat\bZ_2$ on the right hand side is by the diagonal subgroup
of the dual $\widehat \bZ_2$ symmetries of $\sX/\bZ_2$ and $\sY/\bZ_2$.
Our particular case \eqref{eq:equivalence3} is recovered 
by taking $\sX=\SO(24)_1$ and $\sY=\SO(8)_1$
and noticing that the dual $\bZ_2$ symmetries in this case are $(-1)^{\sF_\sL}$.

Starting from the right hand side of \eqref{eq:moregeneral},
let $\sa_1$ and $\sa_2$ be the $\bZ_2$ gauge fields used in the construction of the theories $\sX/\bZ_2 $ and $\sY/\bZ_2$, respectively, 
and let $\sb$ be the $\widehat \bZ_2$ gauge field used in gauging the diagonal $\widehat \bZ_2$. 
They are coupled by a term
\beq
\frac{2\pi \si}{2} \int_{\Sigma} (\sa_1 + \sa_2) \wedge \sb~.
\eeq
Integrating out $\sb$ plays the role of a Lagrange multiplier setting $\sa_1=-\sa_2$, and $-\sa_2$ is equal to $\sa_2$ since it is a $\bZ_2$ field. After integrating out $\sb$, we denote $\sa:=\sa_1 = \sa_2$. Then we have the theory $\sX\times \sY$  gauged by the diagonal $\bZ_2$ symmetry whose gauge field is $\sa$. This is exactly the left hand side of \eqref{eq:moregeneral}  and hence we have confirmed the equivalence \eqref{eq:equivalence3}. 

Coming back to our original question,  
we see from \eqref{eq:6.2} and \eqref{eq:equivalence3} that the rank 8 spin bundle embedded into $\Spin(8)/\bZ_2$ is the same thing as the tangent bundle embedded into $\SO(8)$. This is what we wanted to show. 

\subsubsection{The 4-brane}
For the case $n=4$, the gauge field configuration of Sec.~\ref{sec:n=4} is as follows. We take $(\su(2) \times \e_7) \times (\su(2) \times \e_7) \subset \e_8 \times \e_8$, and then insert an instanton of one $\su(2)$ and an anti-instanton of the other $\su(2)$.
On the other hand, the equivalence of CFTs in (\ref{eq:equivalence}) is
\beq
(\hete)_1 = [ ( (E_7 \times E_7)/\bZ_2)_1 \times \SO(4)_1 ] /(-1)^{\sF_\sL}~.
\eeq
Recalling that $\so(4) = \su(2) \times \su(2)$, together with the fact that the $\so(4)$ tangent bundle is exactly an instanton of one $\su(2)$ and an anti-instanton of the other $\su(2)$ as discussed in Sec.~\ref{sec:n=4}, we again see that the $\SO(4)$ bundle is identified with the tangent bundle of $S^4$. 

\subsubsection{The 6-brane}\label{eq:6u(1)}
For the case $n=2$, the equivalence \eqref{eq:equivalence} is given by 
\beq
(\hets)_1 = [ (\SU(16)/\bZ_4)_1 \times \SO(2)_1 ] /(-1)^{\sF_\sL}~.\label{eq:equivn=2}
\eeq
The origin of the part $\SO(2)_1$ is as follows. We take a subalgebra $\u(1) \subset \u(16) \subset \so(32)$ with the charge normalization such that
the 16 complex Weyl fermions from which we construct $(\hets)_1$ have $\u(1)$ charge $1/4$. 
In this normalization, the level of the $\u(1)$ current subalgebra is given by
\beq
k=16 \times \left( \frac{1}{4} \right)^2=1~.
\eeq
Such a level one $\u(1)$ current algebra is realized by a single complex Weyl fermion or equivalently two Majorana-Weyl fermions. This is the current algebra theory $\SO(2)_1$ appearing in \eqref{eq:equivn=2}.

The $\hets$ gauge field configuration \eqref{eq:magneticflux} of Sec.~\ref{sec:n=2} can be written in terms of the subalgebra $\u(16) \subset \so(32)$ as
\beq
\frac{\si F_{\u(16)}}{2\pi} =\begin{pmatrix} \sq_1 && \\ &\ddots& \\ && \sq_{16} \end{pmatrix} \frac{\epsilon}{4\pi}~. \label{eq:magneticflux3}
\eeq
For a stable, minimal energy configuration, we take $\sq_i=\frac12$ for all $i=1,\cdots,16$. Because 
\beq
\begin{pmatrix} \frac12 && \\ &\ddots& \\ && \frac12 \end{pmatrix} \frac{\epsilon}{4\pi} = \begin{pmatrix} \frac14 && \\ &\ddots& \\ && \frac14 \end{pmatrix} \cdot 2 \cdot \frac{\epsilon}{4\pi},
\eeq
this corresponds to magnetic flux $2$ for the above $\u(1)$ subalgebra. The magnetic flux 2 is exactly that of the complex tangent bundle of $S^2 \simeq \mathbb{CP}^1$. 

\subsubsection{The 7-brane}\label{sec:dual7gauge}

For the case $n=1$, the tangent bundle of $S^1$ is trivial and it should not be the answer for the gauge field configuration. Instead, the gauge field configuration of Sec.~\ref{sec:n=1} is given as follows. 

\paragraph{An equivalence we need:}
From the equivalence in (\ref{eq:equivalence}), we have\footnote{%
This equivalence was discussed at length in \cite{BoyleSmith:2023xkd,Saxena:2024eil},
including the necessity to include the effect of the discrete theta angle, or equivalently the Arf theory.}
\beq
(\hete)_1 = [(E_8)_2 \times \SO(1)_1]/(-1)^{\sF_\sL}~, \label{eq:equivn=1}
\eeq
where $\SO(1)_1$ means a single Majorana-Weyl fermion.
The original gauge field configuration on the left-hand side includes the $\bZ_2$ holonomy of $\hete$ around the target space $S^1$. We claim that this $\bZ_2$ symmetry is realized on the right-hand side of \eqref{eq:equivn=1} by the global $\bZ_2$ symmetry dual to the gauge $\bZ_2=\{1,(-1)^{\sF_\sL}\}$, where the concept of the dual symmetry has been briefly recalled around \eqref{eq:dualcoupling}. This dual $\bZ_2$ is basically the only (at least visible) non-anomalous $\bZ_2$ on the right-hand side.\footnote{The symmetry that uses the fermion parity of only one of $(E_8)_2$ or $\SO(1)_1$ has a mixed anomaly with the total $\bZ_2=\{1,(-1)^{\sF_\sL}\}$ of \eqref{eq:equivn=1} and hence is not a legitimate candidate.} 


Let us denote the $\bZ_2$ symmetry generated by $(-1)^{\sF_\sL}$ by $\bZ_2^{(0)}$, and the dual $\bZ_2$ by $\bZ_2^{(1)}$. Let $\sa$ and $\sb$ be the gauge fields for $\bZ_2^{(0)}$ and $\bZ_2^{(1)}$, respectively. They are coupled as in \eqref{eq:dualcoupling},
\beq
\frac{2\pi \si}{2} \int \sa \wedge \sb~.\label{eq:dualcoupling2}
\eeq
The symmetry $\bZ_2^{(1)}$ is a global symmetry in \eqref{eq:equivn=1}.
If we further gauge the $\bZ_2^{(1)}$ symmetry by integrating out the field $\sb$, then it plays the role of a Lagrange multiplier setting $\sa=0$, which implies that the gauging by $\bZ_2^{(0)}$ is undone. Therefore, on the right-hand side of \eqref{eq:equivn=1}, we get the theory $(E_8)_2 \times \SO(1)_1$ after gauging the $\bZ_2^{(1)}$. 

By doing the same gauging on the left-hand side, we expect to get the theory $(E_8)_2 \times \SO(1)_1$ if the identification $\bZ_2=\bZ_2^{(1)}$ is correct. However, there is a subtlety. The theory $(\hete)_1$ is a bosonic theory and its gauging by $\bZ_2$ should give a bosonic theory rather than a spin CFT. On the other hand, $(E_8)_2 \times \SO(1)_1$ is a spin CFT.

To resolve this problem, it is necessary to include a subtle discrete theta term in the equivalence \eqref{eq:equivn=1}. What we need is a  discrete theta term related to mod 2 indices,
which is called the Arf theory in the recent literature.

\paragraph{Arf theory, or the discrete theta term:}
In general, consider a chiral Dirac operator 
\beq
\cD = {\bar \partial}+\sb_{\bar z}~,
\eeq
possibly coupled to a real vector bundle $V$ with connection $\sb$. Such a chiral Dirac operator $\cD$ can be regarded as an ``antisymmetric matrix'' in the vector space of sections $\Gamma(S_+ \otimes V)$ of the chiral spin bundle $S_+$ on $\Sigma$ tensored with $V$. Now, the number of zero modes of an antisymmetric matrix modulo 2 is invariant under continuous deformations, as is evident if we transform it to a standard block diagonal form,
\beq
\cD =  
\begin{pmatrix}
0
\end{pmatrix} 
\oplus \cdots \oplus
\begin{pmatrix}
0
\end{pmatrix} 
\oplus
\begin{pmatrix}
0 & \lambda_1\\
-\lambda_1 & 0
\end{pmatrix}
\oplus
\begin{pmatrix}
0 & \lambda_2\\
-\lambda_2 & 0
\end{pmatrix}
\oplus\cdots
\eeq
Thus we can define the mod 2 index of $\cD$ as the number of zero modes of $\cD$ modulo 2. We denote this quantity by $\text{Index}_2(\cD) \in \bZ_2$.

We can now define the following discrete theta term. We consider two chiral Dirac operators $\cD$ and $\cD'$. The first, $\cD ={\bar \partial}$, is coupled to the trivial bundle $\bR$. The second, $\cD'= {\bar \partial}+\sb_{\bar z}$, is coupled to the real one-dimensional  bundle $L$ associated to a $\bZ_2$ bundle. Let 
\beq
I(\sb) := \text{Index}_2(\cD') - \text{Index}_2(\cD)
\eeq
 be the difference of the two indices, where $\sb$ is a $\bZ_2$ gauge field. It is a function of $\sb$. We can now add to the action a term
\beq
\pi \si I(\sb) \mod 2\pi \si~.
\eeq
This is the discrete theta term we mentioned. It is defined by using Dirac operators, and hence it depends on spin structures of Riemann surfaces. 
(The quantity $e^{\pi\si I(\sb) } $ is denoted by $(-1)^{q(\sb)}$ in some of the recent literature.)

\paragraph{A more precise version of the equivalence:}
Now the precise version of our claim is \begin{equation}
[(\hete)_1\times \Arf]/\bZ_2 =  (E_8)_2 \times \SO(1)_1 ~, \label{eq:equivn=1precise}
\end{equation}
where $\Arf$ represents the discrete theta term $e^{\pi\si I(\sb)}$,\footnote{This $\Arf$ is regarded as a theory in the expression \eqref{eq:equivn=1precise} because any discrete theta angle can be interpreted as a special class of theories known as invertible field theory~\cite{Freed:2004yc}. See \cite{Kapustin:2014tfa,Kapustin:2014dxa,Freed:2016rqq,Yonekura:2018ufj,Yamashita:2021cao} for more discussions.}
and $/\bZ_2$ means the $\bZ_2$ gauging.
When the $\bZ_2 =\bZ_2^{(1)}$ is just a global symmetry, this theta term does not play much of a role 
and hence it is not visible when we discussed \eqref{eq:equivn=1} in Appendix~\ref{sec:equivalenceintro}. 
However, once we gauge $ \bZ_2$, it has important consequences. 

In particular, consider physical states of the theory on $S^1$ with NS spin structure and $\bZ_2$ holonomy around $S^1$. Then there is an additional contribution to the $\bZ_2$ charge from the discrete theta term, as can be checked from the standard relation between the path integral formalism on $T^2$ and the Hilbert space formalism; if we consider $T^2=S^1 \times S^1$ with NS spin structures and nontrivial $\bZ_2$ holonomies on both $S^1$'s , then $I(\sb) $ has a nontrivial value, which is interpreted in the Hilbert space as a nontrivial $\bZ_2$ charge for states in the NS sector with $\bZ_2$ holonomy. Therefore, the condition of gauge invariance of states (i.e. neutrality under $\bZ_2$) is changed by the presence of the discrete theta term.

Let us study what continuous symmetries exist after gauging the $\bZ_2$ symmetry which exchanges the two $E_8$'s.
Suppose we have a symmetry current operator $J$ in a chiral CFT. From the state-operator correspondence, we should be able to find a physical state corresponding to the operator $J$. The energy of the state on an $S^1$ with circumference $2\pi$ is given by $1-\frac{c_L}{24}$, where $c_L(=16)$ is the central charge of the chiral CFT and $1$ is the scaling dimension of $J$. Thus from the study of physical states, we can determine the Lie algebra generators of the symmetries of the CFT. By explicitly studying physical states of $(\hete)_1$ gauged by $\bZ_2$ by using its free fermion construction, one find the following results after some computations:
\begin{itemize}
\item If we do not include the discrete theta term, there exist states corresponding to current operators for the symmetry $\e_8 \times \e_8$, implying that $(\hete)_1$ is self-dual under the gauging.  Half of the generators come from the diagonal part of the original $\e_8 \times \e_8$ which existed before gauging, and the other half comes from the twisted sector---that is, the sector with nontrivial $\bZ_2$ holonomy around $S^1$.\footnote{We use the realization of $(\hete)_1$ by $32=16+16$ free Majorana-Weyl fermions. In the twisted sector, the quantization of the fermions is done as if there are 16 fermions on an $S^1$ with circumference $4\pi=2\cdot 2\pi$. Then the Casimir energy in the antiperiodic boundary condition of the fermions is given by $-\frac12 \cdot \frac{8}{24}$, where the factor $\frac12$ is due to the fact that the effective circumference is twice the standard circumference $2\pi$. The momentum of the fermions on $S^1$ is given by $\frac{1}{2} (k+\frac12)~(k=0,1,2,\cdots)$ where the factor $\frac12$ is again due to the effective circumference. If we excite two fermions with momentum $\frac14$, the energy is given by $2 \cdot \frac14 -\frac12 \cdot \frac{8}{24} = \frac{1}{3}$, which is the correct energy $1-\frac{16}{24}$ for current operators. One can analyze the periodic boundary condition similarly, and see that an $E_8$ multiplet appears from the twisted sector. }

\item If we include the discrete theta term, there exist states corresponding to current operators for the symmetry $\e_8 $, consistent with the fact that $(E_8)_2 \times \SO(1)_1$ has just a single $\e_8$. The states corresponding to the other $\e_8$ current from the twisted sector mentioned above disappear because they are now odd under the $\bZ_2$ gauge symmetry due to the effect of the discrete theta angle, and hence are projected out.
\end{itemize}
This shows the consistency of identifying the $\bZ_2$ symmetry of the left-hand side of \eqref{eq:equivn=1} with the $\bZ_2^{(1)}$ on the right-hand side.

\subsection{Worldsheet dynamics for the 0, 4, and 6-branes}
\label{sec:worldsheetdynamics046}
We have seen that for $n=2,4,8$, the gauge field configuration is such that we take the $\SO(n)$ bundle to be the tangent bundle of $S^n$. We now explain how to obtain the worldsheet CFT given in \eqref{eq:exactCFT}. Our basic assumption is the same as in Sec.~\ref{sec:6worldsheet}; the worldsheet dynamics are essentially determined by the coupling between the sigma model with target space $S^n$ and the current algebra theory $G_1$. This coupled system has a nontrivial renormalization group flow. The radial direction $\bR_\text{radial}$ is assumed to play the role of the ``dilaton,'' in the sense that it recovers worldsheet conformal invariance, including the correct total central charge. Therefore, we will neglect the radial direction in our analysis, assuming that it provides a linear dilaton CFT $\bR_\text{linear~dilaton}$ at the end.

The $S^n$ sigma model and $G_1$ are coupled via the nontrivial gauge configurations described above. The action is schematically shown in \eqref{eq:worldsheetS}. First let us neglect the gauging $/(-1)^{\sF_\sL}$ by fermion parity. Then in the equivalence
$
G_1 = [H_k \times \SO(n)_1]/(-1)^{\sF_\sL},
$
only the part $\SO(n)_1$ is involved in the interaction, and the part $H_k$ is completely decoupled. The theory $\SO(n)_1$ is realized by $n$ left-moving Majorana-Weyl fermions. On $S^n$, we have the aforementioned $\SO(n)$ bundle that is isomorphic to the tangent bundle, and hence the $n$ fermions are sections of the tangent bundle. Recall that the ${\cal N}=(0,1)$ sigma model also has superpartner right-moving Majorana-Weyl fermions, which are once again sections of the tangent bundle. Thus by combining the $\SO(n)_1$ and ${\cal N}=(0,1)$ sigma models, we get the matter content of an ${\cal N}=(1,1)$ sigma model. 
This matter content is 
\beq
\{\phi^i, \psi^i, \tilde \psi^i \}~, \label{eq:matter}
\eeq 
where $\phi^i~(i=1,\cdots,n)$ are bosonic coordinates of $S^n$, and $\psi^i$ and $\tilde \psi^i$ are left-moving and right-moving Majorana Weyl fermions. The ${\cal N}=(0,1)$ supersymmetric sigma model Lagrangian for this ${\cal N}=(1,1)$ matter content also has ${\cal N}=(1,1)$ supersymmetry.\footnote{A related fact is often utilized in the context of compactification of heterotic strings on Calabi-Yau manifolds with gauge field configurations determined by the Calabi-Yau tangent bundle. In that context, ${\cal N}=(0,2)$ supersymmetry is enhanced to ${\cal N}=(2,2)$ supersymmetry as far as the sigma model part is concerned. 
This construction is usually called the \emph{standard embedding} of the spacetime connection into the gauge connection. See e.g. \cite{Green:1987mn,Polchinski:1998rr} for textbook accounts.}

The ${\cal N}=(1,1)$ sigma model with the target space $S^n$ is believed to have two gapped vacua which are exchanged by $(-1)^{\sF_\sL}$. The mass gap is generated by a nontrivial renormalization group flow, and may be seen via a large $n$ analysis if $n$ is sufficiently large. Even without solving the system, the fact that there are two vacua exchanged by $(-1)^{\sF_\sL}$ can be understood by an analysis of the Witten index (at least for even $n$) \cite{Witten:1982df,Witten:1982im}, as we now recall. 

Let us consider an ${\cal N}=(1,1)$ sigma model with an arbitrary compact target space $X$ with dimension $\dim X = n$, and quantize it on $S^1$ with R spin structure. By taking $S^1$ to be very small, we can focus on zero modes and hence the dependence of $\{\phi^i, \psi^i, \tilde \psi^i \}$ on the coordinate of $S^1$ is neglected. Thus we obtain a supersymmetric quantum mechanics. The commutation relations of fermions in this theory are
\beq
\{\psi^i, \psi^j\} = \{\tilde\psi^i, \tilde\psi^j\}  = 2G^{ij}~, \qquad \{\psi^i, \tilde \psi^j\} =0~,
\eeq
where $G_{ij}$ is the metric on the target space $X$ with respect to which indices are raised and lowered, and the factor of $2$ is just our normalization convention for the fermions. By combining the fermions as
\beq
a^i = \frac{ \psi^i - \si \tilde \psi^i}{2}~, \hspace{0.5 in} a_i^\dagger = \frac{ \psi_i + \si  \tilde\psi_i}{2}~, 
\eeq
we get
\beq
\{a^i , a_j^\dagger\} =\delta^i_j~, \qquad \{a^i, a^j\}=0~, \qquad \{a_i^\dagger, a_j^\dagger\}=0~.
\eeq
For a fixed $\phi$, we can construct the Hilbert space as a Fock space. Take a state $\ket{\phi}$ such that 
\beq
a_j^\dagger \ket{\phi}=0~, \qquad \hat{\phi}^i\ket{\phi} = \phi^i\ket{\phi}~, \qquad \bra{\phi }\phi'\rangle = \sqrt{G}^{-1}\delta^n(\phi-\phi')~,
\eeq
where $\hat{\phi}^i$ is the quantum mechanical operator corresponding to $\phi^i$. Also, to make the geometric meaning somewhat clearer, we denote 
\beq
\d \phi^i := a^i~. 
\eeq
These quantities anticommute with one another. The Fock space states are then given by
\beq
\d\phi^{i_1} \cdots  \d\phi^{i_p} \ket{\phi}~, \qquad (p=0,1,\cdots,n)~,
\eeq
with normalizable physical states being of the form
\beq
\int \sqrt{G}  \d^n \phi\, \frac{1}{p!} F_{i_i \cdots i_p}(\phi) \d\phi^{i_1} \cdots  \d\phi^{i_p} \ket{\phi}~,
\eeq
where $F_{i_i \cdots i_p}(\phi) $ is an antisymmetric tensor on $X$.
In other words, wavefunctions of this quantum mechanical system are described by $p$-forms 
\beq
F=\frac{1}{p!} F_{i_i \cdots i_p}(\phi) \d\phi^{i_1} \cdots  \d\phi^{i_p} ~.
\eeq

The ${\cal N}=(0,1)$ supercharge is given up to a constant factor by $Q = \dot{\phi}_i \tilde\psi^i$, where the dot represents a time derivative. As is usual in quantum mechanics, $\dot{\phi}_i$ becomes the derivative operator $-\si \frac{\partial}{\partial \phi^i}$. We also have $\tilde \psi^i = \si\,  \d \phi^i +\rm{h.c.}$. The combination
\beq
\d = \d \phi^i \frac{\partial}{\partial \phi^i}
\eeq
plays the role of the exterior derivative, and the supercharge is given by $Q = \d + \d^\dagger$. States annihilated by $Q$ form the space of de~Rham cohomology group $H^p_\text{dR}(X)$ represented by harmonic forms. 

The Witten index is defined as $\tr_{\cal H} (-1)^{\sF}$, where $(-1)^{\sF}$ is the total fermion parity operator which changes the signs of both $\psi$ and $\tilde \psi$, and the trace is taken over the Hilbert space ${\cal H}$. 
In the sigma model currently under consideration, 
$(-1)^{\sF}$ is given in terms of the form degree $p$ by $(-1)^{p}$. Then the Witten index is given by the Euler number,
\beq
\tr_{\cal H} (-1)^{\sF} =  \sum_{p=0}^n (-1)^p \dim H^p_\text{dR}(X)~.
\eeq

The operator $(-1)^{\sF_\sL}$ acts only on $\psi^i$ and not on $\tilde \psi^i$, because it is defined as a symmetry of the part $H_k \times \SO(n)_1$. We then see that it exchanges $\d\phi^i=(\psi^i - \si \tilde \psi^i)/2$ and $-(\d\phi^i)^\dagger = (-\psi^i - \si \tilde \psi^i)/2$.
Geometrically, $(\d\phi^i)^\dagger$ is a contraction operator of differential forms.
Such an operator exchanging $\d\phi^i$ and $-(\d\phi^i)^\dagger$ should be proportional to the Hodge star operator $\star$ acting on differential forms. Notice that $(-1)^{\sF_\sL}$ commutes with $Q=\dot{\phi}_i \tilde\psi^i$ which is consistent with the geometric fact that $\star$ (with an appropriate modification by a sign factor) commutes with $\d +\d^\dagger$.\footnote{
More precisely, let $\sP$ be an operator that takes value $p$ when it is acted on a $p$-form. Then, one can check that the operator $\star (-1)^{\frac12 \sP(\sP+1)}$ commutes with $\d + \d^\dagger$. Its square is given by $[\star (-1)^{\frac12 \sP(\sP+1)}]^2 = (-1)^{\frac12 n(n+1)}$, so a consistent identification may be $(-1)^{\sF_\sL}=\i^{\frac12 n(n+1)} \star (-1)^{\frac12 \sP(\sP+1)}$.\label{footnote:harmonic}
}

Let us restrict our attention to the case that the target space $X$ is given by $S^n$. In this case the only nonzero de~Rham cohomology groups are
\beq
H^0_\text{dR}(S^n) = H^n_\text{dR}(S^n) = \bR~.
\eeq
This means that there are two supersymmetric vacuum states annihilated by $Q$---one from $H^0_\text{dR}(S^n) $ and the other from $H^n_\text{dR}(S^n) $. These two are exchanged by the Hodge star. Therefore, we have shown that there are two supersymmetric vacuum states which are exchanged by $(-1)^{\sF_\sL}$, at least when the size of the spatial $S^1$ of the worldsheet is very small. The Witten index (or its refinement by $(-1)^{\sF_\sL}$) is preserved when the size of $S^1$ is changed because the Witten index is a deformation invariant. The reasonable assumption is then that this vacuum structure is preserved in the limit of infinite size of $S^1$ when $n \geq 2$, since this is the case in the large $n$ limit and also the theory is asymptotically free for $n\geq 2$. The case $n=1$ will be discussed later. 

We can now finally describe the worldsheet dynamics. After coupling $G_1$ and the ${\cal N}=(0,1)$ sigma model with the target space $S^n$, the theory is given by
\beq
\left[ H_k \times S^n_{{\cal N}=(1,1)} \right]/(-1)^{\sF_\sL}~,
\eeq
where $S^n_{{\cal N}=(1,1)}$ is the ${\cal N}=(1,1)$ sigma model explained above. By flowing to the IR of this theory, the part $S^n_{{\cal N}=(1,1)}$ gives rise to two gapped vacua exchanged by $(-1)^{\sF_\sL}$. However, after gauging $(-1)^{\sF_\sL}$, these two vacua are identified and there is just a single vacuum, leaving only $H_k$. Including the directions parallel to the brane as well as the radial direction, we obtain the worldsheet theory claimed in \eqref{eq:exactCFT}.

\subsection{Worldsheet dynamics for the 7-brane}
\label{sec:worldsheetdynamics7}
As we now discuss, the case of $n=1$ requires special treatment. To distinguish the target space $S^1$ from the worldsheet spatial $S^1$, we denote the target space as $S^1_T$ and the worldsheet spatial $S^1$ as $S^1_W$.\footnote{%
This issue was also investigated from a more computational perspective in \cite{Nakajima:2023zsh,Saxena:2024eil}.}

The starting point is the equivalence
\beq
(\hete)_1 = [(E_8)_2 \times \psi]/(-1)^{\sF_\sL}~, \label{eq:equivn=1-2}
\eeq
where we have denoted the single left-moving Majorana-Weyl fermion as $\psi$;
this was denoted by $\SO(1)_1$ until now.

 The ${\cal N}=(0,1)$ sigma model with target space $S^1_T$ has matter content $\{\phi,\tilde\psi\}$, where $\phi$ is a periodic scalar with periodicity $\phi \sim \phi+2\pi$ and $\tilde \psi$ is its superpartner. By adding $\psi$, we get the matter content $\{\phi,\psi,\tilde\psi\}$. The action is schematically
\beq
S \sim \frac{1}{4\pi} \int \d^2 \sigma \left( R^2 (\partial_\alpha \phi) (\partial^\alpha \phi) + \tilde \psi \partial \tilde \psi + \psi \bar{\partial} \psi \right)~,\label{eq:roughS1action}
\eeq
where $R = r/\alpha'^{1/2}$ is the radius $r$ of $S^1_T$ divided by $\alpha'^{1/2}$.

The main difference between the case $n=1$ and the cases $n\geq 2$ is that the $S^1_T$ sigma model itself is conformally invariant without any IR flow. However, we will now argue that tachyon condensation triggers a renormalization group flow, leaving two gapped vacua exchanged by $(-1)^{\sF_\sL}$. Having shown that, the argument for the exact worldsheet CFT (\ref{eq:exactCFT}) is completely the same as in the cases $n\geq 2$. 

\subsubsection{Tachyons} 

We would like to understand the tachyons in our setup.
We provide two analyses. The first is a down-to-earth argument, 
and the second is a more formal approach.

\paragraph{A down-to-earth approach:}
We need to include a $\bZ_2$ holonomy for $\hete$ around $S^1_T$. The $\bZ_2$ symmetry of the left-hand side of  \eqref{eq:equivn=1-2} corresponds to the symmetry $\bZ_2^{(1)}$ dual to $\bZ_2^{(0)}=\{ 1, (-1)^{\sF_\sL}\}$ on the right-hand side, as was discussed in Sec.~\ref{sec:dual7gauge}.

Including the holonomy of the $\bZ_2^{(1)}$ symmetry around the target space $S^1_T$ (rather than the worldsheet) has the following effect. Let $j$ be the current for the winding symmetry of the $S^1_T$ target space,
\beq
j = \frac{\d \phi}{2\pi}~.
\eeq
It is defined such that the winding number $w$ of a state on $S^1_W$ is measured by the integral
\beq
w= \int_{S^1_W} j~.
\eeq
When the holonomy of the $\bZ_2^{(1)}$ symmetry around $S^1_T$ is included in the target space, its consequence on the worldsheet is that the $\bZ_2^{(1)}$ gauge field $\sb$ is given by
\beq
\sb = j \mod 2~. \label{eq:bj-id}
\eeq
To understand this equation, consider for example the case that we wrap the worldsheet $S^1_W$ once around the target space $S^1_T$. Then the $\bZ_2^{(1)}$ holonomy around $S^1_T$ becomes a $\bZ_2^{(1)}$ holonomy around $S^1_W$. More generally, for a worldsheet $\Sigma$, the field $\sb$ is the pullback of the target space $\bZ_2^{(1)}$  gauge field by the sigma model map $\phi: \Sigma \to S^1_T$.

The coupling \eqref{eq:dualcoupling2} now becomes
\beq
\frac{2\pi \si }{2} \int \sa \wedge j~,\label{eq:dualcoupling3}
\eeq
where $\sa$ is the gauge field for $\bZ_2^{(0)}=\{ 1, (-1)^{\sF_\sL}\}$.
An implication of this coupling is as follows. If we consider a state on $S^1_W$ with an odd winding number, then there is an additional contribution to $(-1)^{\sF_\sL}$ charge from this coupling. 

Let $\widehat \phi$ be the scalar field that is T-dual to $\phi$. The operator $\exp(\i w \widehat \phi)$ has winding charge $w$ and hence transforms under $(-1)^{\sF_\sL}$ as $(-1)^w$. Then, for odd $w$, the operator $\psi \exp(\i w \widehat \phi)$ is invariant under $(-1)^{\sF_\sL}$,
and therefore is present in our theory.
This operator is the bottom component of the supermultiplet $\Psi \exp(\i w \widehat \Phi )$, where 
\begin{equation}
\begin{aligned}
  \widehat \Phi(\sigma,\theta) &\sim \widehat \phi(\sigma) + \si \theta \tilde\psi (\sigma)~, \nonumber \\
  \Psi(\sigma,\theta) &\sim  \psi(\sigma) + \theta f(\sigma)~.   
\end{aligned}
\end{equation}
Here $\theta$ is the Grassmann supercoordinate of the ${\cal N}=(0,1)$ supersymmetry, and $f$ is an auxiliary field. We can consider a term in the action of the form
\beq
\int \d^2 \sigma \d \theta\,\, \Psi \exp(\i w \widehat \Phi) \sim \int \d^2 \sigma \left( w \psi \tilde \psi \exp(\i w \widehat \phi) + f  \exp(\i w \widehat \phi) \right)~.\label{eq:tacyoncondense}
\eeq

The operator $ \psi \tilde \psi \exp(\i w \widehat \phi)$ for odd $w$ is invariant under the gauge symmetries $(-1)^{\sF_\sL} $,
since both $\psi$ and $\exp(\i w \widehat\phi)$ are odd under it.
Therefore this survives the $(-1)^{\sF_\sL}$ quotient in \eqref{eq:equivn=1-2}.

Let us now consider if this operator survives the GSO projection.
The answer depends on the spin structure on $S^1_T$. 
%
%
To explain this point, let us first consider compactification of the heterotic $\hete$ superstring theory on $S^1_T\times \bR^{8,1}$ with the $\bZ_2=\bZ_2^{(1)}$ holonomy around $S^1_T$ and with the periodic spin structure. In this case, it is well-known that the target space supersymmetry is preserved~\cite{Chaudhuri:1995fk,Chaudhuri:1995bf} and there is no tachyon. From the worldsheet point of view, the absence of tachyon implies that the operator \eqref{eq:tacyoncondense} should be forbidden since it would lead to a tachyon when $R$ is small. In fact, the operator \eqref{eq:tacyoncondense} is not invariant under the right-moving (rather than left-moving) fermion parity $(-1)^{\sF_\sR}$ because it contains $\tilde \psi$ and hence it is projected out. 

In the above supersymmetric case with the periodic spin structure on $S^1_T$, the facts that the operator \eqref{eq:tacyoncondense} is invariant under $(-1)^{\sF_\sL}$ but is not invariant under $(-1)^{\sF_\sR}$ imply that it is not invariant under the total fermion parity $(-1)^{\sF}$. This point may look strange since naively the operator $ \exp(\i w \widehat \phi)$ seems to be bosonic. The answer to this puzzle comes from the discrete theta term discussed in Sec.~\ref{sec:dual7gauge}. We can rewrite \eqref{eq:equivn=1precise} as
\beq
(\hete)_1   = \Arf \times  [ (E_8)_2 \times \SO(1)_1 ]/(-1)^{\sF_\sL}    ~.
\eeq
The $\Arf$ is given by the discrete theta term ${\pi \i I(\sb) }$. Because of the identification \eqref{eq:bj-id}, it gives a theta term for the $S^1$ sigma model $\phi$,
\beq
\pi \i I(\d \phi/2\pi) ~. \label{eq:spintheta}
\eeq
It has the effect that in the sectors with odd winding number $w=\int_{S_W} j=$ odd, it gives an additional contribution to the value of $(-1)^{\sF}$ by the relation between the path integral formalism and the Hilbert space formalism (see the paragraph below the one containing \eqref{eq:equivn=1precise}). The same theta term also gives the effect that in the sector with winding number $w$, the states get a minus sign $(-1)^w$ under the shift symmetry $\phi \to \phi+2\pi R$. This means that these states have momentum $P$ along $S^1_T$ that is quantized as $RP \in \frac12 w +\bZ$. Thus, it is not appropriate to consider an operator like $ \exp(\i w \widehat \phi) $ for odd $w$ without taking into account the value of the momentum which cannot be zero.

Next let us consider the case that the spin structure on $S^1_T$ is anti-periodic. The target space spin structure can be accounted for by adding the same theta term $\pi \i I(\d \phi/2\pi)$ as \eqref{eq:spintheta}.\footnote{It is discussed e.g.~in Sec.~3.2.2 of \cite{Gaiotto:2019gef}, and is explained in more general situations in \cite{Yonekura:2022reu}. }
The added theta term cancels the one that already existed for the periodic spin structure. After the cancellation, the term \eqref{eq:tacyoncondense} is allowed.
Below we give a more or less elementary explanation.

We are going to argue that
the operator $ \psi \tilde \psi \exp(\i w \widehat \phi)$ for odd $w$ is invariant under the right-moving fermion parity $(-1)^{\sF_\sR}$ when the spin structure on $S^1_T$ is anti-periodic. 
Let us consider a worldsheet torus $T^2_W=S^1_{W,1} \times S^1_{W,2}$, with R spin structure in the $S^1_{W,1}$ direction and NS spin structure in the direction $S^1_{W,2}$. 

First let us regard $S^1_{W,1}$ as a space direction and $S^1_{W,2}$ as a (Euclidean) time direction. In this case, physical states on $S^1_{W,1}$ are target space fermions. We consider a sigma model map $\phi : S^1_{W,2} \to S^1_T$ that has an odd winding number $w$. Then the target space fermions must be anti-periodic in the direction $S^1_{W,2}$ because they go an odd number of times (i.e. $w$) around $S^1_T$. In the quantum mechanical picture of a massive particle, the contribution of a massive fermion to the path integral may be represented by a Wilson loop $\tr \text{P} \exp (-\int \omega)$ of the spin connection $\omega$ along its worldline. This Wilson loop has different signs for periodic and anti-periodic spin structures. To account for this sign difference, the worldsheet path integral must have an additional sign factor $(-1)^w$ when the spin structure on $S^1_T$ is anti-periodic compared to the case of periodic spin structure. 

Now we switch the roles of $S^1_{W,1} $ and $ S^1_{W,2}$, by regarding $S^1_{W,1} $ as a time direction and $S^1_{W,2} $ as a space direction. Then we have the winding number $w$ in the space $S^1_{W,2} $. The R spin structure in the time direction $S^1_{W,1}$ means that we include a factor $(-1)^{\sF_\sR}$ when we represent the path integral as a trace over the Hilbert space. Then the sign factor $(-1)^w$ discussed above is interpreted as the fact that there is an additional contribution to the charge of $(-1)^{\sF_\sR}$ when the winding number $w$ of a state is odd. We conclude that the operator $\exp(\i w \widehat \phi)$ transforms as $(-1)^w$ under $(-1)^{\sF_\sR}$. 
As $\tilde \psi$ is odd under $(-1)^{\sF_\sR}$, the combination $\psi \tilde \psi \exp(\i w\widehat \phi)$ is even when $w$ is odd. This is what we wanted to show.

We summarize the results above in the following table, which shows the charges of the operators $\psi, \tilde\psi$, and $\exp(\i w \widehat \phi)$ under $(-1)^{\sF_\sL}$ and $(-1)^{\sF_\sR}$ when the spin structure on the target space $S^1_T$ is anti-periodic:
\beq
\begin{array}{c|c|c|c}
& \psi & \tilde \psi & \exp(\i w \widehat \phi) \\ \hline
(-1)^{\sF_\sL} &-1&+1&(-1)^w \\ \hline
(-1)^{\sF_\sR}  &+1&-1&(-1)^w
\end{array}
\eeq
The charge $(-1)^w$ of $ \exp(\i w \widehat \phi)$ under $(-1)^{\sF_\sL} $ comes from the $\bZ_2^{(1)}$ holonomy around $S^1_T$, while the charge $(-1)^w$ of $ \exp(\i w \widehat \phi)$ under $(-1)^{\sF_\sR} $ comes from the anti-periodic spin structure on $S^1_T$.

The scaling dimension of the operator $\psi \tilde \psi \exp(\i w \widehat \phi)$ is 
\beq
h_L=h_R= \frac12 + \frac14 R^2 w^2. \label{eq:sclng}
\eeq
When $R$ is small, some of the operators with odd $w$ are relevant and represent target space tachyons.

As mentioned above, there is no tachyon when the spin structure on $S^1_T$ is periodic. When $S^1_T$ has the anti-periodic spin structure and there is no $\bZ_2^{(1)}$ holonomy, we get a tachyon by omitting $\psi$ from the above operator~\cite{Atick:1988si}.

\paragraph{A more formal argument:}
Let us now rederive what we found above in a more formal way. To make the radius $R$ explicit, we denote the target space circle with radius $R$ as $S^1_{T,R}$.
Our setup is to have $(\hete)_1$ fibered over $S^1_{T,R}$ with the anti-periodic spin structure, such that 
we perform the exchange of two $E_8$ factors when we go around $S^1_{T,R}$.
Such a configuration can be obtained by 
starting from a trivial product of $(\hete)_1$ and $S^1_{T,2R}$ with a periodic spin structure,
and then performing a $\bZ_2$ gauging combining 
\begin{enumerate}
  \item the half-shift of $S^1_{T,2R}$ (i.e. the $2\pi R$ shift in the circle $S^1_{T,2R}$ with circumference $4\pi R$),
  \item the exchange of the two $E_8$ factors (i.e. the $\bZ_2$ of $\hete$), and
  \item the sign change of the target space fermion.
\end{enumerate}
The point (3) can be achieved as follows. Let $\sb$ be the gauge field for the $\bZ_2$ gauging. We include the discrete theta term $\pi \i I(\sb)$ or in other words the $\Arf$ theory. Let $\phi$ and $\tilde \phi$ be the bosons for $S^1_{T,R}$ and $S^1_{T,2R}$. Both of them are taken to have the periodicity $2\pi$, and their kinetic terms in the Lagrangians are proportional to $R^2$ and $(2R)^2$, respectively (see \eqref{eq:roughS1action}). The  currents $\d \phi/2\pi$ and $\d \tilde \phi/2\pi$ for the winding number are schematically related by 
\beq
\frac{\d  \phi}{2\pi} \sim \sb+ 2 \cdot \frac{\d \tilde \phi}{2\pi}
\eeq
Thus the $\pi \i I(\sb)$ is equivalent to the theta term \eqref{eq:spintheta}. Therefore, the effect of the anti-periodic spin structure can be accounted for by including the $\Arf$ theory $\pi \i I(\sb)$ in the $\bZ_2$ gauging.

From the above points, we see that the internal worldsheet theory is given by \begin{equation}
  \frac{(\hete)_1\times\Arf\times S^1_{T,2R}}{\bZ_2} \label{eq:7brane-internal}
\end{equation}
We now use the T-duality $S^1_{T,2R} = \widehat S^1_{T,  1/(2R)}$. 
Denoting $\widehat R = 1/R$, this implies \begin{equation}
  S^1_{T,2R}= \widehat S^1_{T,\widehat R/2} = \widehat S^1_{T,\widehat R}/\bZ_2
\end{equation} where the final $\bZ_2$ quotient is by the half-shift.
Using also \eqref{eq:equivn=1precise}, we see that 
the theory \eqref{eq:7brane-internal}
is equivalent to 
\begin{equation}
   \frac{ [ (E_8)_2 \times \psi]/(-1)^{\sF_\sL} \times \widehat S^1_{T,\widehat R}/\bZ_2 } {\bZ_2},
\end{equation}
which is equivalent,
via the general truism 
$[\sX/\bZ_2\times \sY/\bZ_2]/\bZ_2 = [\sX\times \sY]/\bZ_2$ 
derived in \eqref{eq:moregeneral},
to \begin{equation}
  \frac{  (E_8)_2 \times \psi  \times \widehat S^1_{T,\widehat R} } {\bZ_2}.
\end{equation} 
Here the $\bZ_2$ quotient acts on the first two factors by $(-1)^{\sF_\sL}$ and on the last factor by the half-shift.
Therefore the operator \begin{equation}
  \psi \exp(\si w \widehat \phi)
\end{equation} where $\widehat\phi$ is the scalar T-dual to the original circle direction,
survives the $\bZ_2$ projection when $w$ is odd.

Promoting both $\psi$ and $\phi$ into \Nequals{(0,1)} superfields $\Psi$ and $\Phi$, we can include the term \begin{equation}
  \int \d^2 \sigma \d \theta\, \Psi \exp(\si w \widehat \Phi)
\end{equation} in the action. 
The scaling dimensions are given by \eqref{eq:sclng}.
When $\widehat R$ is large enough, or equivalently when $R$ is small enough,
these provide the tachyons we were looking for.

\subsubsection{The vacuum structure of the worldsheet theory}
\label{sec:vacuumstructure-7brane}

Let us now consider the 7-brane. In the region far from the brane, the radius $r  = \alpha'^{1/2}R$ of $S^1_T$ is large enough so that no tachyon mode exists. However, in the region $r \lsim \alpha'^{1/2}$, there are tachyonic modes. We assume that generic tachyons are activated, and hence the worldsheet theory has a term
\beq
\int \d^2\sigma \d \theta\, \Psi F(\widehat \Phi ) \sim \int \d^2\sigma \left( -\si \psi \tilde \psi F'(\widehat \phi) +f F(\widehat\phi) \right)~,
\eeq
where $F$ is a generic real function of the form
\beq
F(\widehat \Phi ) = \sum_{n \in \bZ} c_n \exp(  \si (2n+1) \widehat \Phi )~.
\eeq
Such a function is characterized by the property that it is odd under $\widehat \phi \to \widehat \phi +\pi$, i.e.
\beq
F(\widehat \phi  + \pi )  = - F(\widehat \phi  )~.
\eeq
We want to study the vacuum structure of the theory when such a term is added. 

Notice first that integrating out the auxiliary field $f$ gives a potential for $\widehat \phi$,
\beq
V(\widehat \phi) = F(\widehat \phi)^2~.
\eeq
As a simple example, let us consider $F(\widehat \phi  )  = \sin (\widehat \phi )$. Then $V(\widehat \phi) $ has classical vacua at $\widehat \phi = 0, \pi$. We can trust the classical analysis when the dual radius $\widehat R  = R^{-1}$ is large enough, and in this case we get two vacua. Moreover, these two vacua must be exchanged by $(-1)^{\sF_\sL} $ 
--- 
indeed, the operator $ \sin (\widehat \phi ) = (\exp (\si \widehat \phi ) - \exp (-\si \widehat \phi ) )/2\si$ is odd under $(-1)^{\sF_\sL}$ as we have discussed above. The symmetry $(-1)^{\sF_\sL} $ should act on $\widehat \phi$ as
\beq
(-1)^{\sF_\sL} :\,\, \widehat \phi \mapsto \widehat \phi + \pi~,
\eeq
and hence the two vacua are exchanged. 

As another example, consider the case $ F=\sin ((2n+1)\widehat \phi )$ for a positive integer $n$. Then we have $4n+2$ vacua at the classical level. 
As this example suggests, it may not always be true that the number of vacua is exactly two. 

The above analysis is just at the classical level, and we also need to consider the possibility that vacuum degeneracy is lifted by quantum effects. In fact, the mechanism discussed in \cite{Witten:1981nf} can work in our case.

We claim that the existence of exactly two vacua is \textit{generically} true when the function $F(\widehat \phi  )$ is general enough,
as can be argued using a mod-2 Witten index of this system, which will be described in detail in Appendix~\ref{app:mod2}.
There, regarding $(-1)^{\sF_{\sL,\sR}}$ as two global $\bZ_2$ symmetries, 
we will show two facts:
\begin{itemize}
  \item Vacua always come in pairs, exchanged by the action of $(-1)^{\sF_\sL}$. In particular, the number of vacua $N_\text{vacua}$ is even.
  \item $N_\text{vacua}/2$ modulo 2 is the mod-2 Witten index and hence it is invariant under a continuous deformation.
\end{itemize}
For example, as mentioned before, when $ F=\sin ((2n+1)\widehat \phi )$ the number of classical vacua is given by
\beq 
N_\text{vacua}^\text{classical}/2 = 2n+1 \equiv 1 \mod 2~.
\eeq
Thus $N_\text{vacua} \equiv 2 \mod 4$ is the robust prediction from the mod 2 Witten index. The actual number of vacua depends on the function $F(\widehat \Phi)$ and quantum effects, but generically we expect to have exactly two vacua unless there is some fine tuning of energies of states. These two vacua are exchanged by the operator $ (-1)^{\sF_\sL}$.

After gauging $ (-1)^{\sF_\sL}$, two vacua are identified and we get a single vacuum in which $(E_8)_2$ is left. Therefore we get the theory \eqref{eq:exactCFT} with $H_k = (E_8)_2$ for the throat region, as claimed before.

\subsection{Stability of the throat region} \label{sec:throatstability}

Finally, let us return to all of the branes and check for the absence of tachyons in the throat region. (For the 7-brane we are considering the theory after the above tachyon condensation.) The absence of tachyonic modes in the region far from the brane is discussed in Appendix~\ref{sec:instability}.

We can study the near-horizon spectrum by using the CFT \eqref{eq:exactCFT}, namely
\beq
\bR^{p,1} \times \bR_\text{linear~dilaton} \times H_k~ \label{eq:exactCFT2}
\eeq 
where $p=8-n$ and the list of $(G, H, k, n)$ is given in Table~\ref{table:equivalence}.
To identify bosonic modes, we consider the worldsheet theory on $S^1$ with NS spin structure. For simplicity we work in the light-cone gauge and replace 
\beq
\bR^{p,1} \to \bR^{p-1}~.
\eeq
Let $(c'_L, c'_R)$ be the left and right central charges of the theory in light-cone gauge, from which the effect of the running dilaton $6\alpha' (\partial_y \Phi)^2$ is subtracted. They are given by 
\beq
(c'_L, c'_R)=\left(12+\frac{3(8-n)}{2}, \frac{3(8-n)}{2} \right) ~.
\eeq
As usual, to satisfy the level matching condition, we need to act an operator with scaling dimension $(h_L, h_R) = (h_L, h_L-\frac12)$ for some $h_L \geq \frac12 $. Then the mass squared $m^2$ of the corresponding state is given by
\beq
\frac{\alpha'}{4} m^2 = h_L - \frac{16-n}{16}~. \label{eq:massformula}
\eeq
We want to check whether this is positive for bosonic states.

Notice first that for level matching we need $h_L \geq \frac12$, so it is necessary to consider a nontrivial primary state of $H_k$. 
In the current algebra theory $\g_k$ based on some simple Lie algebra $\g$ at level $k$, primary states are labelled by representations of $\g$. Let us consider a representation $\rho$, and let $Q_\rho$ be the quadratic Casimir invariant of $\rho$, normalized in such a way that for the adjoint representation it is equal to the dual Coxeter number $h^\vee$. The left scaling dimension $h_L$ is given by
\beq
h_\rho = \frac{Q_\rho}{h^\vee +k}~.
\eeq
Alternatively, in the Narain construction of chiral CFTs for simply laced $\g$ at level $k=1$, $h_\rho$ is given by half the length of a weight vector $v_\rho$ of $\rho$, 
\beq
h_\rho = \frac12 v^2_\rho~.
\eeq
For instance, in the adjoint representation the length of a root is 2 and hence $h_\rho$ is 1, as it should be for a conserved current operator. We give some examples relevant for our purposes in Table~\ref{table:scaling}. (These examples are reviewed in Appendix~\ref{sec:someCFT}).

\begin{table}
\centering
\begin{tabular}{c||c|c|c|c|c|c}
$\g_k$ & $\so(24)_1$ & $\so(24)_1$ & $(\e_7)_1$ & $\su(16)_1$ & $\su(16)_1$  & $(\e_8)_2$ \\
\hline
$\rho$ & $ v$ & $ s$ & ${\bf 56}$ & $\wedge^2$ & $\wedge^4$  & ${\bf 248}$ \\
\hline
$h_\rho$ & $ \frac12$ & $\frac32$ &   $\frac34$ & $\frac78$ & $\frac32$  & $\frac{15}{16} $
\end{tabular}
\caption{Scaling dimensions of some primary states of current algebras. For $\so(N)$, $v$ is the vector representation and $s,c$ are spinor representations. For $\su(N)$, $\wedge^\ell$ is the $\ell$-th antisymmetric representation. For $\e_7$ and $\e_8$, we use the dimensions  in boldface to denote the representations. }
\label{table:scaling}

\end{table}

Table~\ref{table:scaling} contains states relevant for both NS and R sectors. We will postpone the discussion of the R sector to Sec.~\ref{sec:boundary}.
States of the theory $H_k$ in the NS sector must be a genuine representation of the group $H$ (rather than just the Lie algebra). For instance, the representation $v$ of $\so(24)$ cannot be a state in the NS sector for $H=\Spin(24)/\bZ_2$. 

Some states that are constructed from the ones in Table~\ref{table:scaling} are as follows;
\beq
\text{NS sector:} \qquad 
\begin{array}{c|c|c|c}
H_k & (\Spin(24)/\bZ_2)_1 &  ((E_7 \times E_7)/\bZ_2)_1 & (\SU(16)/\bZ_4)_1  \\ \hline
\rho & s & {\bf 56} \otimes {\bf 56} & \wedge^4  \\ \hline 
h_\rho & \frac32 & \frac32 & \frac32
\end{array}
\eeq
All of these states have positive mass terms. States not listed here are either states which have larger values of $h_\rho$, or states created by current operators and hence satisfying $h_\rho=1$. Therefore, all states have positive masses $m^2>0$ and there are no tachyons in the throat region. 

We remark that the states created by current operators (which have $h_\rho=1$) are gauge bosons. From \eqref{eq:massformula} we see that these states are massive. In the target space, this is caused by the effect of the linear dilaton background, as one can check by canonically normalizing the kinetic terms in the action.

\section{Branes as a boundary condition for anomalous fermions}\label{sec:boundary}

In the previous section, we studied the spectrum of bosonic fields living in the throat regions of the branes, and verified the absence of tachyons. In the current section, we will discuss the spectrum of massless fermionic fields living in the throat region, and in particular study the anomalies coming from them. 

\subsection{Massless fermion spectrum}
\label{sec:fermion-spectrum}
We continue our study of the spectrum of the theories \eqref{eq:exactCFT} describing the throat regions of the branes,
\beq
\bR^{p,1} \times \bR_\text{linear~dilaton} \times H_k~.
\eeq
In Table~\ref{table:scaling}, we have listed some primary states of these theories. From these states, we can construct the following states in the R sector of the current algebra theory $H_k$:
\beq
\text{R sector:}\qquad
\begin{array}{c||c|c|c|c}
H_k & (\Spin(24)/\bZ_2)_1 &  ((E_7 \times E_7)/\bZ_2)_1 & (\SU(16)/\bZ_4)_1 & (E_8)_2  \\ \hline
\rho & v & {\bf 56} \otimes {\bf 1} \oplus {\bf 1} \otimes {\bf 56} & \wedge^2 & {\bf 248}   \\ \hline 
h_\rho & \frac12 & \frac34 & \frac78 & \frac{15}{16}
\end{array} \label{eq:Rsectorstate}
\eeq
Here $v$ is the vector representation of $\so(24)$, $\wedge^2$ is the second antisymmetric representation of $\su(16)$, and ${\bf 56}$ and ${\bf 248}$ are the representations of $\e_7$ and $\e_8$ of dimensions $56$ and $248$, respectively.

That the R-sector of the theories $H_k$ has these states was already found in e.g.~\cite{BoyleSmith:2023xkd}.
Here we will give a derivation from a slightly different perspective.

The equivalence $G_1 = [H_k \times \SO(n)_1]/(-1)^{\sF_\sL}$ implies that the states of the theory $G_1$ are recovered by considering both the NS and R sectors of $H_k \times \SO(n)_1$. In the R sector, states of $\SO(n)_1$ are always spinor-type representations for which a ``$2\pi$-rotation'' in $\SO(n)$ gives a minus sign. This is due to a mixed anomaly between worldsheet Lorentz symmetry (in particular $(-1)^{\sF_\sL}$) and the $\SO(n)$ (in particular $\pi_1(\SO(n))$). By decomposing representations of the current algebra $G_1$ in terms of projective representations of $H$ and $\SO(n)$, we see that the states in \eqref{eq:Rsectorstate} are in the R sector since they are tensored with spinor-type representations of $\SO(n)$. In particular, one also finds that there is a mixed anomaly between $(-1)^\sF$ and $\pi_1(H)$, except for the case $H=E_8$.

By using the formula \eqref{eq:massformula} for the mass, we see that all the states listed in \eqref{eq:Rsectorstate} are massless. Since they are in the R sector, they are massless fermions. In fact, it is possible to understand them in terms of supergravity. By regarding the angular sphere $S^n$ as an internal manifold, we can think of this as a compactification of supergravity on $S^n$ with the particular gauge field configurations discussed in Sec.~\ref{sec:heteroticcharge}. Such a compactification has the following fermionic zero modes:
\begin{itemize}
\item For $n=1$, we include a $\bZ_2$ holonomy for $\hete$ around $S^1$. The gauginos $\lambda_1$ and $\lambda_2$ of the two $E_8$ gauge groups satisfy the boundary conditions 
\beq
\lambda_1(\theta + 2\pi) = - \lambda_2(\theta)~, \qquad \lambda_2(\theta +2\pi) = - \lambda_1(\theta)~,
\eeq
where $\theta$ is the coordinate on $S^1$, and the minus signs are due to the fact that $S^1$ is assumed to have antiperiodic spin structure. This gives a single zero mode from $\lambda_1 - \lambda_2$, which is in the $\mathbf{248}$ of $\e_8$. 

\item For $n=2$, we consider a flux of $\u(1) \subset \u(1) \times \su(16) =\u(16) \subset \so(32)$ through $S^2$. Under $\u(1) \times \su(16)$, the gaugino in the adjoint representation of $\so(32)$ is decomposed as 
\beq
\text{adj}(\so(32)) \to \wedge^2(\su(16) )_{1/2} + \overline{ \wedge^2(\su(16))_{1/2} }  + \text{adj}(\su(16)) +  \text{adj}(\u(1) )~,
\eeq
where the notation is that $\text{adj}(\g)$ is the adjoint representation of $\g$ and $\wedge^2(\su(16))_1$ is the second antisymmetric representation of $\su(16)$ with $\u(1)$ charge $1/2$. Here, the $\u(1)$ charge was normalized in the same way as in Sec.~\ref{eq:6u(1)}.
The $\u(1)$ magnetic flux is 2 as studied in Sec.~\ref{eq:6u(1)}. Because $(1/2) \cdot 2=1$, there is a single zero mode on $S^2$ in the representation $\wedge^2$ of $\su(16)$. 

\item For $n=4$, we insert an instanton of $\su(2) \subset \su(2) \times \e_7 \subset \e_8$ and an anti-instanton of the other $\su(2) \subset \su(2) \times \e_7 \subset \e_8$ on $S^4$. Under $ \su(2) \times \e_7 $, the gaugino of $\e_8$ is decomposed as
\beq
\text{adj}(\e_8) \to ({\bf 56} \otimes {\bf 2}) \oplus \text{adj}(\e_7) \oplus \text{adj}(\su(2))~. 
\eeq
An instanton of $\su(2)$ gives a single zero mode for the representation ${\bf 2}$ of $\su(2)$, and hence we get a zero mode in the ${\bf 56}$. From the other $\e_8$ we get an opposite chirality fermion.

\item For $n=8$, we consider $\so(8)' \times \su(24) \subset \so(32)$ and take the $\so(8)' $ bundle to be one of the spinor bundles associated to the tangent bundle of $S^8$. The gaugino is decomposed as
\beq
\text{adj}(\so(32))  \to {\bf 8}'_v \otimes v+  \text{adj}(\so(8)') +  \text{adj}(\so(24)) ~,
\eeq
where ${\bf 8}'_v$ is the vector representation of $\so(8)'$. From the spin bundle, we get a single zero mode which can be checked  by the Atiyah-Singer index theorem. Thus we get a zero mode in the $v$ of $\so(24)$. 

\end{itemize}

The fermions described above live in the spacetime after the compactification on $S^n$, that is 
\beq
\bR^{p,1} \times \bR_\text{radial}~,
\eeq
where $p=8-n$. The situation is somewhat analogous to the case of a magnetic monopole in four dimensions. In that case, there is a magnetic flux around the magnetic monopole, and if there are fermions charged under the $\u(1)$, we get zero modes in the spacetime $\bR_\text{time} \times \bR_\text{radial}$. The magnetic monopole is regarded as a boundary of the two dimensional spacetime. In the same way, our branes may be regarded as a boundary of the above fermions in a spacetime with dimension $p+2=10-n$. 

\subsection{Fermion anomalies }
\label{sec:fermion-anomalies}
The fermions discussed above have anomalies. For $n=2,4,8$, the anomalies are just perturbative ones. Let $\hat{A}(R)$ be the A-roof genus, 
\beq
\hat{A}(R) = \prod_i \frac{x_i/2}{\sinh (x_i/2)} = 1 - \frac{1}{24}p_1 + \frac{7p_1^2-4p_2}{5760} + \cdots,
\eeq
where $R$ is the Riemann curvature 2-form, $x_i$ are the Chern roots of $R$, and $p_\ell~(\ell=1,2,\cdots)$ is the $\ell$-th Pontryagin class of $R$ defined in \eqref{eq:Pontryagin}.  If a chiral fermion in $(10-n)$-dimensions is in a representation $\rho$ of a Lie algebra $\h$, then the anomaly polynomial is a $(12-n)$-form and is given by 
\beq
I_{12-n} = \pm \left[ \hat{A}(R) \tr_\rho \exp\left( \frac{\si F}{2\pi} \right)\right]_{12-n}~,
\eeq
where $F$ is the gauge field strength 2-form, the subscript ${12-n}$ means to take the $(12-n)$-form part, and the sign depends on the chirality of the fermion. If the fermion is Majorana, we should additionally divide by 2. For each of $n=2,4,8$, the anomaly polynomial is given more explicitly as follows:

\paragraph{The case $n=8$:} Let $c$ be defined by
\beq
c = - \frac14 \tr_{v} \left( \frac{\si F}{2\pi} \right)^2.
\eeq
Then the anomaly polynomial for $n=8$ is
\beq
I_4 = -\left( c+ \frac{1}{2} p_1 \right),
\eeq
where we have used the fact that anomalous fermions are Majorana-Weyl in two dimensions.

\paragraph{The case $n=4$:} Let $c_{(1)}$ and $c_{(2)}$ be 
\beq
c_{(a)} = - \frac{1}{24} \tr_{{\bf 56}}  \left( \frac{\si F_a}{2\pi} \right)^2~, \qquad (a=1,2)
\eeq
where $F_1 $ and $F_2$ are the field strengths of two $\e_7$'s, respectively. For $\e_7$, we have
\beq
\frac{1}{4!} \tr_{{\bf 56}}  \left( \frac{\si F_a}{2\pi} \right)^4 = c_{(a)}^2~,
\eeq
as follows from the fact that $\e_7$ does not have invariant polynomials at this degree other than $c_{(a)}^2$; the overall factor can be computed e.g.~by using the decomposition of ${\bf 56}$ under $\so(12) \times \su(2) \subset \e_7$ as ${\bf 56} \to ({\bf 2^5} \otimes {\bf 1}) \oplus ({\bf 12} \otimes {\bf 2})$.
Then the anomaly polynomial for $n=4$ is
\beq
I_8 &= \frac{1}{2} \left( c_{(1)}^2 - c_{(2)}^2\right) + \frac{1}{4} p_1 \left( c_{(1)} - c_{(2)}\right) \nonumber \\
& = \frac{1}{2} \left( c_{(1)} - c_{(2)} \right) \left( c_{(1)} + c_{(2)} + \frac{1}{2} p_1 \right) ~,
\label{4brane-anomaly}
\eeq
where we have used the fact that the fermions are Majorana-Weyl,\footnote{In six dimensions, if the gauge representation is a pseudo-real representation, we can consider a Majorana condition since the spinor representations of the Lorentz group $\Spin(5,1)$ are also pseudo-real.} and that the fermions for the two $\e_7$ factors have opposite chiralities. 

\paragraph{The case $n=2$:} Let $c_\ell$ and $\text{ch}_\ell$ be the $\ell$-th Chern class and Chern character of $\su(16)$, respectively. In particular, let $y_i$ be Chern roots of $\su(16)$,
\beq
\frac{\si F}{2\pi} \to \begin{pmatrix} y_1 && \\ & y_2 & \\ && \ddots \end{pmatrix}  ~,
\eeq
and define 
\beq
 \sum_{\ell \geq 0} \text{ch}_\ell &= \tr \exp\left( \frac{\si F}{2\pi} \right) = \sum_{\ell \geq 0} \frac{1}{\ell !} \sum_i y_i^\ell~, \nonumber \\
  \sum_{\ell \geq 0} c_\ell &=\det \left( 1+ \frac{\si F}{2\pi} \right) = \prod_{i} (1+ y_i)~,
\eeq
where $F$ is represented in the fundamental representation of $\su(16)$. Straightforward computations give
\beq
\tr_{\wedge^2}\exp\left( \frac{\si F}{2\pi} \right) = \frac{1}{2} \left( \sum_{i \geq 0}  \text{ch}_i \right)^2 - \frac{1}{2}\sum_{ i\geq0} 2^i  \text{ch}_i~.
\eeq
We also have (by taking the log of $\prod_i (1+y_i)$),
\beq
\log\left( \sum_{ \ell \geq 0} c_\ell \right) = \sum_{\ell \geq 1} (-1)^{\ell-1} (\ell-1)!\, \text{ch}_\ell~.
\eeq
By using these formulas, the anomaly polynomial for the case $n=2$ is computed to be
\beq
I_{10} = - \frac{1}{2} c_3 \left(c_2 + \frac{1}{2}p_1\right)~.
\eeq

\paragraph{The case $n=1$:} This case is subtler than the others. In this case, there is no perturbative anomaly. Instead, a Majorana fermion in the adjoint representation of $\e_8$, or more generally any semisimple Lie algebra $\g$, in nine (and in fact also eight) dimensions has a global anomaly. 
This can be seen as follows (the discussion parallels that given for eight dimensions in \cite{Lee:2022spd}).

Let $\mathbb{HP}^2$ be quaternionic projective space with quaternionic dimension 2 (i.e. real dimension $4 \times 2=8)$. Let $S^1_P$ be a circle with periodic spin structure. Now take $\su(2) \subset \g$ and take the $\su(2)$ bundle to be the canonical quaternionic line bundle on $\mathbb{HP}^2$. Then one can check using the Atiyah-Singer index theorem that a fermion in the adjoint representation has a single zero mode on $\mathbb{HP}^2 \times S^1_P$. 

In more detail, the canonical quaternionic line bundle is an $\su(2)$ bundle on $\mathbb{HP}^2$. Let $F$ be the field strength of this $\su(2)$ bundle, and let $c = -\frac12 \tr_{\bf 2} (\si F/2\pi)^2$ be its second Chern class. For $\su(2)$ representations ${\bf 2}$ and ${\bf 3}$, the numbers of zero modes of the corresponding Dirac operators on $\mathbb{HP}^2$ are given by the integral of
\beq
{\bf 2} : & \quad \left[ \hat{A}(R) \tr_{\bf 2} \exp\left( \frac{\si F}{2\pi} \right) \right]_8 = \frac{1}{12} (c^2 +\frac12 p_1 c) +2 \cdot \frac{7p_1^2-4p_2}{5760}~, \nonumber \\
{\bf 3} : & \quad  \left[  \hat{A}(R) \tr_{\bf 3} \exp\left( \frac{\si F}{2\pi} \right) \right]_8 = \frac{1}{12} (16c^2 +2p_1 c) + 3 \cdot \frac{7p_1^2-4p_2}{5760}~.
\eeq
As will be briefly explained below, we have 
\beq
p_1=-2c~, \qquad p_2=7c^2~, \qquad  \int_{\mathbb{HP}^2} c^2 = 1~. \label{eq:projectivep}
\eeq
Thus the representation ${\bf 2}$ (as well as the trivial representation ${\bf 1}$) gives no zero mode, while the representation ${\bf 3}$ gives one zero mode on $\mathbb{HP}^2$. We choose $\su(2) \subset \g$ such that the adjoint representation of $\g$ is decomposed as
\beq
\text{adj}(\g) \to {\bf 3} \oplus {\bf 2} \otimes (\cdots) \oplus {\bf 1} \otimes (\cdots)~.
\eeq
Therefore, there is a single zero mode in the adjoint representation.
This zero mode is preserved by the further compactification on $S^1_P$.

We have seen that there is a single zero mode on $\mathbb{HP}^2 \times S^1_P$. Then the path integral measure of the fermion changes sign under the fermion parity transformation. This is the global anomaly we mentioned for the case $n=1$.

To close this subsection, let us very briefly review one derivation of \eqref{eq:projectivep} (see \cite{MilnorStasheff} for a necessary background). Let $L$ be the canonical quaternionic line bundle. By definition of $\mathbb{HP}^n$, the bundle $L$ can be embedded in the trivial quaternionic bundle $\underline{\mathbb {H}}^{n+1}$, i.e. $L \subset \underline{\mathbb {H}}^{n+1}$. Let $L^\perp$ be the orthogonal bundle.  The tangent bundle of $\mathbb{HP}^n$ is given by
\beq
T\mathbb{HP}^n = \text{Hom}_\bH (L , L^\perp) = L^* \otimes_\bH L^\perp~,
\eeq
where $L^*$ is the dual bundle to $L$.
Since $L$ is a quaternionic line, we have $ L^* =L$. Also, by definition we have $L^\perp \oplus L = \underline{\mathbb {H}}^{n+1}$. From these two facts, we get $T\mathbb{HP}^n \oplus (L \otimes_\bH L) = (L \otimes_\bH \underline{\mathbb{H}} )^{n+1}$, where $L \otimes_\bH L$ and $L \otimes_\bH \underline{\mathbb{H}}$ are real bundles of rank 4. If we realize $L$ as a bundle of the representation ${\bf 2}$ of $\su(2)$, then $L \otimes_\bH L$ is realized by the representation ${\bf 2} \otimes {\bf 2} = {\bf 3} \oplus {\bf 1}$. From this information, one can compute the Pontryagin classes of $T\mathbb{HP}^n$. The total Pontryagin class is $p = {(1-c)^{2(n+1)}}/{(1-4c)}$, and in particular we get $p_1=-2(n-1)c$ and $p_2=(2n^2-5n+9)c^2$. It is known that $H^{\ell}(\mathbb{HP}^n, \bZ) \simeq \bZ$ for $\ell =0,4, 8, \cdots, 4n$, while the other cohomology groups are zero. This $c$ is the generator of the cohomology ring, and hence $\int_{\mathbb{HP}^n} c^n = 1$ (up to a sign depending on orientation).

\subsection{A higher dimensional Callan-Rubakov problem}
\label{sec:callan-rubakov}
In general, an anomalous theory cannot be put on a manifold with boundaries without breaking the anomalous symmetry. We have seen that the massless fermions in $\bR^{p,1} \times \bR_\text{radial}$ have anomalies, and our branes can be regarded as their boundaries. This would be inconsistent if there were no $B$-field which could cancel the anomalies via the Green-Schwarz mechanism. 

In reality we have the $B$-field, so there is no inconsistency. However, this consideration suggests that the boundary conditions at the branes must somehow mix the massless fermions and the $B$-field. In weakly coupled Lagrangian descriptions it is very hard (and presumably impossible) to write down such boundary conditions. Thus we expect that our branes will be very strongly coupled objects, which do not have a weakly coupled Lagrangian description.

Note that if we inject a fermion particle perpendicularly to a brane, it cannot simply be reflected back. To make the discussion concrete, let us consider the case of the 4-brane. Let $S_+$ and $S_-$ be the two spinor representations of the Lorentz group $\so(5,1)$ in six dimensions. Then there are two fermions whose representations under $\so(5,1) \times \e_7 \times \e_7$ are
\beq
S_+ \otimes {\bf 56} \otimes {\bf 1} ~, \qquad S_- \otimes {\bf 1} \otimes {\bf 56}~.
\eeq
The particle moves in the direction $\bR_\text{radial}$, and its spin is classified by the little group $\so(p)$ which is the rotation group of $\bR^p$. Let $s_+$ and $s_-$ be the two spinor representations of $\so(p)$. Depending on whether a particle is ingoing or outgoing, there are four cases;
\beq
\text{ingoing : }& \quad s_+ \otimes {\bf 56} \otimes {\bf 1}~, \qquad s_- \otimes {\bf 1} \otimes {\bf 56} \nonumber \\
\text{outgoing : }& \quad s_- \otimes {\bf 56} \otimes {\bf 1}~, \qquad s_+ \otimes {\bf 1} \otimes {\bf 56} \nonumber 
\eeq
Therefore, it is not possible to simply reflect an ingoing fermion particle to an outgoing particle, due to inconsistency of the quantum numbers.
(For the 0-brane, we will discuss a simple solution in Sec.~\ref{sec:0braneGS}.)

The above physical explanation shows that there is no elementary boundary condition for the fermions. Here, an elementary boundary condition for a fermion $\Psi$ is something like
\beq
P_\text{boundary} \Psi =0~, \label{eq:elementaryboundary}
\eeq
where $P_\text{boundary} $ is a matrix satisfying some conditions so that the boundary condition is physically sensible. For instance, for non-chiral fermions, one example of such a boundary condition is given by $P_\text{boundary} =\frac{1}{2}(1-\gamma_\text{boundary})$, where $\gamma_\text{boundary}$ is the gamma matrix in the direction perpendicular to the boundary.\footnote{This boundary condition plays an important role in a general discussion of anomaly inflow for fermions~\cite{Witten:2019bou}.} Any boundary condition of this type (assuming that it preserves symmetries) would reflect an ingoing fermion particle to an outgoing fermion particle, which we have seen to be impossible for our branes. 

The difficulty of writing down a boundary condition like \eqref{eq:elementaryboundary} exists for any chiral fermion, regardless of whether or not there is anomaly. For a magnetic monopole in a chiral gauge theory, this problem is known as the Callan-Rubakov effect~\cite{Rubakov:1982fp,Callan:1982au,Callan:1982ac}. In that case, a monopole is a boundary of two-dimensional fermions. Describing boundary conditions for two-dimensional chiral fermions required the use of bosonization and re-fermionization~\cite{Callan:1994ub,Maldacena:1995pq}. In our case, the situation is much more difficult for two reasons. First, the fermions are in higher dimensions (except for the 0-brane) and we cannot use techniques such as bosonization. Second, the fermions themselves are anomalous and the $B$-field must somehow be involved.


For the above reasons, we expect that the worldvolume theories on the branes and their interactions with the bulk fermions are very nontrivial. We can get an additional insight by compactifying the branes on $S^3$ with $k$ units of the $H$ flux~\cite{Yonekura:2024spl}. This means that the bulk theory is now given by
\beq
\bR^{p-3,1} \times  \bR_\text{linear~dilaton} \times H_k \times S^3_{H=k},
\eeq
where $S^3_{H=k}$ is the sphere $S^3$ with the 3-form flux $H$ given by $\int_{S^3} H=k$, which is realized as the $\cN=(0,1)$ Wess-Zumino-Witten (WZW) model for $\SU(2) \simeq S^3$ with the level $k$. We need $p \geq 3$ for the compactification to make sense.

For simplicity, we focus on perturbative anomalies, although the basic mechanism should also apply to global ones (see \cite{Tachikawa:2024ucm} for an analysis of global anomalies in an analogous situation). The fermion anomalies are of the form
\beq
-X_4 Y_{p}
\eeq
where $X_4=p_1/2+c$ (with $c$ the second Chern class) and $Y_p$ is given by
\beq
Y_{p} = \left\{ \begin{array}{ll} 
1 & p=0 \\
c_{(2)} - c_{(1)} & p=4 \\
 \frac{1}{2} c_3 & p=6
\end{array} \right.
\eeq
Then the Green-Schwarz coupling is schematically of the form
\beq
2\pi \i \int H \wedge \text{CS}_{p-1}~,
\eeq
where $ \text{CS}_{p-1}$ is the Chern-Simons form such that $\d  \text{CS}_{p-1}= Y_{p}$. Therefore, if we compactify our brane configuration on $S^3_{H=k}$, 
the bulk after the compactification contains the Chern-Simons term
\beq
2\pi \i k \int  \text{CS}_{p-1}~.
\eeq
This implies that the worldvolume theory on the $p$-brane after the compactification on $S^3_{H=k}$ has the anomaly determined by anomaly inflow from $ \text{CS}_{p-1}$. The bulk fermions cause no problem since they have no zero mode on $S^3$. 

Therefore, the worldvolume theory on the $p$-brane must have the property that its compactification on $S^3_{H=k}$ has an anomaly with the anomaly polynomial given by $Y_p= \d  \text{CS}_{p-1}$. This is a very nontrivial property.\footnote{If the $B$-field were an ordinary $2$-form field, an example of a $(p+1)$-dimensional theory having this property would be given by a $\U(1)$ gauge theory. Let $A$ be the $\U(1)$ gauge field. Then, we couple the $B$ field and the non-abelian field to $A$ as 
\beq
-\frac{1}{2g^2} \int  (\d A +B) \wedge (\d A+B) +2\pi \i \int \d A \wedge  \text{CS}_{p-1},
\eeq
where $g$ is the $\U(1)$ coupling.
Formally, the compactification of this theory on $S^3_{H=k}$ gives the anomaly determined by $ \text{CS}_{p-1}$. However, this cannot be the complete answer for two reasons. First, the pure $\U(1)$ field forbids compactifications with a nontrivial $H$ flux because the field strength $F=\d A+B$ of the $\U(1)$ gauge field satisfies $\d F=H$. In the heterotic string, the theory $S^3_{H=k}$ (which is just the WZW model) is not forbidden. Thus we need an additional dynamical object that is a magnetic source for $F$ so that the equation $\d F=H$ is modified. Second, the $B$-field is not an ordinary 2-form but a \emph{twisted string structure} such that $\d H=X_4$. We thank M.~Montero for helpful discussions on this model.
} 

Although the worldvolume theory on the $p$-brane itself is very difficult, its compactifications may be more tractable. (See \cite{Yonekura:2024spl} for more discussions on the case of the above compactification on $S^3_{H=k}$.) In this respect, the situation may be somewhat similar to that of M5-branes in M-theory; the worldvolume theory on M5-branes is the difficult six-dimensional $\cN=(2,0)$ theory and it may not have a Lagrangian description, but its compactifications are sometimes given by ordinary Lagrangian theories. For instance, the $T^2$ compactification of the $\cN=(2,0)$ theory is ordinary $\cN=4$ super-Yang-Mills in four dimensions, and compactifications on more nontrivial manifolds are also investigated extensively in the literature. It would also be interesting to investigate compactifications of our heterotic branes.

\section{Green-Schwarz couplings and invariants of SQFTs}
\label{sec:GScoupling}

In the previous section, various properties of our branes were studied by using the symmetry $H$. In this section, we will see that the branes have nontrivial properties even if one forgets this symmetry. Our aim in this section is to argue that the current algebra theories $H_k$ are nontrivial elements of the group $\sqft_{n-32}$ (where $n=8-p$), which was introduced in Sec.~\ref{sec:bordism}. They are detected by $\bZ_2$ (for the 7, 6, and 4-branes) or $\bZ$ (for the 0-brane) invariants. As discussed in Sec.~\ref{sec:bordism}, such invariants suggest the stability of the branes. 

Since this section is somewhat lengthy, let us begin with a brief summary of its contents. The main fact utilized in this section is that the Green-Schwarz coupling, to be defined later, is a bilinear pairing between the worldsheet theory $T$ and the spacetime string manifold $M_d$ (or rather their corresponding bordism classes), and hence by computing a non-trivial Green-Schwarz coupling, we will be able to prove that the worldsheet theory is non-trivial. For the case of the 0-brane the discussion is rather straightforward as we will see in Sec.~\ref{sec:0braneGS}; 
but for the cases of the 4-, 6-, and 7-branes the proper definition of the Green-Schwarz coupling turns out to be more involved. 

In more detail,
we first show in Sec.~\ref{sec:4braneprelims}
that the perturbative Green-Schwarz coupling for the 4-brane is trivial 
and therefore useless for our purpose.
Then, in Sec.~\ref{sec:generalGS}, 
we will give a general definition (valid for topologically non-trivial $B$-field configurations) of the global part of the Green-Schwarz coupling.
It is expressed as a bilinear pairing,
\beq
\mathrm{GS}: \, \mathbb{A}_{-d-22} \times \mathrm{Ker}\left( \Omega_d^\mathrm{string} \rightarrow \Omega_d^\mathrm{spin}\right) \, \rightarrow \U(1)~,
\eeq
where $\mathbb{A}_{-d-22} \subset \sqft_{-d-22}$ is the subgroup of elements $[T] \in \sqft_{-d-22}$ whose ordinary and mod 2 elliptic genera vanish, while $\mathrm{Ker}\left( \Omega_d^\mathrm{string} \rightarrow \Omega_d^\mathrm{spin}\right) $ is the set of bordism classes $[M_d] \in \Omega_d^\mathrm{string}$ such that $M_d$ can be realized as the boundary of a spin manifold $W_{d+1}$, i.e. $\partial W_{d+1} = M_d$,
where we need to assume $d\not \equiv 3$ modulo 4 from a technical restriction.

In order to apply this definition to our branes, we must show that the SQFT classes describing the angular parts of the 4-, 6-, and 7-branes are elements of the appropriate $\mathbb{A}_{-d-22}$, in particular that their elliptic genera vanish. 
This somewhat technical computation will be done in Appendix~\ref{app:elliptic}. 
With this condition confirmed,
our goal is then to identify an appropriate element of $\mathrm{Ker}\left( \Omega_d^\mathrm{string} \rightarrow \Omega_d^\mathrm{spin}\right) $ such that the Green-Schwarz pairing is non-zero. Let $S^3_{H=N}$ be a sphere with the flux of the field strength 3-form $H$ of the $B$-field given by $\int_{S^3} H =N$. Our strategy here is to first show that, for the angular part of the 4-brane, the class $[S^3_{H=1} \times S^3_{H=1}] \in \mathrm{Ker}\left( \Omega_6^\mathrm{string} \rightarrow \Omega_6^\mathrm{spin}\right)$ gives rise to a non-trivial Green-Schwarz coupling, thereby proving its non-triviality; this is carried out in Sec.~\ref{4-brane-GS-computation}. We then use a series of relationships between the angular parts of the 4-, 6-, and 7-branes, derived in Sec.~\ref{sec:bordism-relations}, to identify bordism classes giving non-trivial Green-Schwarz couplings for the 6- and 7-branes as well. This is performed in Sec.~\ref{sec:67braneGS}. Finally, we close in Sec.~\ref{sec:tmf} by commenting on the connection between these non-trivial classes and the theory of topological modular forms. 

Before entering into the discussion outlined above, let us briefly contrast the upcoming results with the analogous results for the NS5-brane. Neglecting the directions $\bR^{5,1}$ parallel to the branes, the worldsheet theory for $N$-coincident NS5-branes is given by~\cite{Callan:1991dj,Callan:1991at}
\beq
\bR_\text{linear~dilation} \times S^3_{H=N} \times G_1~, \label{eq:intNS5}
\eeq
where $S^3_{H=N}$ is realized by the Wess-Zumino-Witten model with level $N$.
It is straightforward to see that the theory $S^3_{H=N} \times G_1$ is a trivial element of $\sqft_{-29}$. The physical manifestation of this fact comes from the well-known fact that heterotic NS5-branes can be continuously deformed to instantons of $G$. 
Indeed, we can replace the above configuration, without changing the behavior at the positive infinity region of $\bR_\text{linear~dilation}$, with a configuration that is a ``fiber bundle,''  
\beq
G_1 \to \cX \to \bR^4_\text{cigar}~,
\eeq 
where $\bR^4_\text{cigar}$ is a cigar whose asymptotic behavior at infinity is the same as $\bR_\text{linear~dilation} \times S^3_{H=N}$, and the fiber bundle is determined by an $N$-instanton configuration of $G$ on $\bR^4_\text{cigar}$. In the notation of Sec.~\ref{sec:bordism}, we write
\beq
\partial \cX = S^3_{H=N} \times G_1~.
\eeq
Thus the theory $S^3_{H=N} \times G_1$ is trivial in $\sqft_{-29}$, which we write as $[S^3_{H=N} \times G_1]=0$.
In this case, the symmetry $G$ is partly broken in $ \cX$, i.e. the 2d SQFT $\cX$ does not have the full global symmetry $G$. Thus, if we do not try to preserve the symmetry $G$, we can replace the negative infinity region of $\bR_\text{linear~dilation}$ in \eqref{eq:intNS5} by a smooth configuration such that the negative infinity region is ``cutoff''.\footnote{For NS5-branes in Type~II string theories there also exist configurations in which the negative infinity region is cutoff---namely, the double scaling limit \cite{Giveon:1999px,Giveon:1999tq}.}

On the other hand, in the case of our non-supersymmetric branes, the worldsheet theories (neglecting the factor of $\bR^{p,1}$) are
\beq
\bR_\text{linear~dilation} \times H_k~. 
\eeq
The fact that $H_k$ is nontrivial in $\sqft_{n-32}$, which we write as $[H_k] \neq 0$, implies that we cannot replace $\bR_\text{linear~dilation} \times H_k$ by a configuration in which the negative infinity region of $\bR_\text{linear~dilation} $ is cutoff. This is not possible even if we do not try to preserve the symmetry $H$.

\subsection{Internal worldsheet theories}
\label{sec:names}
For convenience, we start by naming the $\SQFT$ classes corresponding to the angular directions of our branes.  
The topological classes of the gauge bundles were discussed in Sec.~\ref{sec:gaugeconfig}, \ref{sec:topological-classes} and \ref{sec:dualgaugeconfiguration}.
This leads to the following four cases:
\begin{itemize}
\item The worldsheet theory for the angular part of the 7-brane is given by 
the $\hete$ level-1 theory fibered over $S^1$ with antiperiodic spin structure,
such that two $E_8$ factors are exchanged when we go around $S^1$.
We denote the corresponding $\SQFT$ class by $[X_1]$.
\item The worldsheet theory  for the angular part of the 6-brane is given by 
the  $\hets$ level-1 theory  fibered over $S^2$ with nontrivial Stiefel-Whitney class $\int_{S^2} v_2=1\in \bZ_2$. 
More explicitly, we embedded the $\mathfrak{so}(2)$ spatial curvature of $S^2$ into $\mathfrak{u}(1)\subset \mathfrak{u}(16)\subset \mathfrak{so}(32)$ in a certain way, given in detail above.
We denote  the corresponding $\SQFT$ class  by $[X_2]$.
\item The worldsheet theory  for the angular part of the 4-brane is given by
the worldsheet theory $\hete$ fibered over $S^4$ such that the characteristic class $\int_{S^4}v_4=1\in \bZ_2$.
More explicitly,
the $\mathfrak{so}(4) = \su(2) \times \su(2)$ spatial curvature of $S^4$ is embedded into $\mathfrak{e}_8\times \mathfrak{e}_8$ so that the commutant is $\mathfrak{e}_7\times \mathfrak{e}_7$. 
We denote  the corresponding $\SQFT$ class  by $[X_4]$.
\item The worldsheet theory  for the angular part of the 0-brane is given by
the $\hets$ fibered over $S^8$ such that the characteristic class $\int_{S^8}v_8=1\in\bZ$.
More explicitly,
$\mathfrak{so}(8)$ spatial curvature of $S^8$ is embedded into the $\mathfrak{so}(8)$ part of $\mathfrak{so}(8)\times \mathfrak{so}(24)\subset \mathfrak{so}(32)$ via a triality of $\so(8)$ exchanging the vector and one of the spinor representations.
We denote  the corresponding $\SQFT$ class by $[X_8]$.
\end{itemize}

Recall that the {\it degree} of an SQFT is defined by the coefficient of the gravitational anomaly $\nu$ which, in the case of a CFT, is given by $\nu=-2(c_L-c_R)$.
The current algebra part $G_1$ always has the central charge $(c_L,c_R)=(16,0)$ and therefore has $\SQFT$ degree $-2(c_L - c_R)=-32$,
whereas the spatial part $S^n$ has $n$ left-moving bosons, $n$ right-moving bosons and $n$ right-moving Majorana-Weyl fermions, giving $\SQFT$ degree $n$.
Therefore the class $[X_n]$ has $\SQFT$ degree $-32+n$, i.e.~$[X_n] \in \SQFT_{-32+n}$.

The arguments in Sec.~\ref{sec:worldsheetdynamics046} for the 0, 4 and 6-branes and in Sec.~\ref{sec:worldsheetdynamics7} for the 7-brane give the following equivalences: \begin{align}
[X_1]  &= [(E_8)_2] &\in \SQFT_{-31}~,\label{X1}  \\
[X_2]  &= [(\SU(16)/\bZ_4)_1] &\in\SQFT_{-30}~,\label{X2} \\
[X_4]  &= [((E_7\times E_7)/\bZ_2)_1] & \in\SQFT_{-28}~,\label{X4} \\
[X_8]  &= [(\Spin(24)/\bZ_2)_1] & \in \SQFT_{-24}~.\label{X8}
\end{align}
Note that the current algebra theories on the right-hand side have central charges $c_L=31/2$, $15$, $14$, and $12$ according to Table~\ref{table:equivalence},
which match the degree $n-32$ of $\SQFT_{n-32}$ given above via the fact that the SQFT degree for a CFT is given by $-2(c_L-c_R)$.

\subsection{The 0-brane and the global \texorpdfstring{$B$}{B}-field anomaly}
\label{sec:0braneGS}

The Green-Schwarz coupling is most easily understood in the case of the 0-brane, which we focus on now. 
The near-horizon regime of the 0-brane is described by the particular worldsheet theory $(\Spin(24)/\bZ_2)_1$ recalled above,
but for the moment let us work in a slightly more general setting.

Let $T$ be any 2d \Nequals{(0,1)} superconformal field theory (SCFT) with left and right central charges $c_L$ and $c_R$ satisfying $c_L-c_R=12$, so that it can be combined with a two-dimensional non-compact part with a suitable dilaton gradient to give a legitimate string worldsheet theory. 
Suppose that the elliptic genus of $T$ is given by\footnote{%
The order of the poles is bounded by $\lfloor c_L/24\rfloor$, and can in principle be higher than $1$.
Such cases arise when the dilaton gradient is timelike.
} \begin{equation}
Z_\text{ell}(T;q)= \tr (-1)^\sF q^{L_0-c_L/24} \bar q^{\bar L_0-c_R/24} = a q^{-1} + b + \cdots ~,
\end{equation} 
where as usual, $q$ is related to the complex modulus $\tau$ of the torus $T^2$ in the path integral formalism as $q=e^{2\pi \i \tau}$, the $L_0$ and $\bar L_0$ are the obvious Virasoro generators, and $Z_\text{ell}(T;q)$ is independent of $\bar q$ due to the existence of supercharge $Q$ satisfying $Q^2 = \bar L_0-c_R/24$~\cite{Witten:1986bf}.

Each state contributing at $q^{-1}$ gives a target space gravitino and a dilatino,
and each state contributing at $q^0$ gives a target space massless fermion,
whose spacetime chirality is specified by the worldsheet fermion number $(-1)^\sF$.
From the standard result that the anomaly polynomial of a gravitino $\psi_\mu^\pm$
and a fermion $\psi^\pm$ are given by $\pm 23p_1/48$ and $\mp p_1/48$  respectively \cite{Alvarez-Gaume:1983ihn}, we see that the total anomaly polynomial is \begin{equation}
(-24 a + b) (-p_1/48)~.
\end{equation} 
As $\d H=p_1/2$, we see that the  Green-Schwarz coupling is $2\pi \i N\int B$ with \begin{equation}
N=-a + \frac{b}{24}~. \label{2dGS}
\end{equation}
The $N$ gives a $B$-field tadpole in two dimensions.
If $N$ is an integer, we can introduce $N$ space-filling fundamental strings to cancel the $B$-field tadpole.
But if $N$ is not an integer, this option is not available.
When $N$ is not an integer, we also have another issue that the coupling $e^{2\pi \i N\int B}$  has an anomaly under the large gauge transformation  shifting $\int B$ by $1$.

In our case of the 0-brane where $T = (\Spin(24)/\bZ_2)_1$, we have $a=0$, $b=24$ as is shown in Appendix~\ref{app:elliptic}, and therefore $N=1$.
This allows us to cancel the $B$-field tadpole by introducing a single fundamental string.
This necessity of inserting a fundamental string was already mentioned in Sec.~\ref{sec:n=8}
when we analyzed the charge of the 0-brane from a semi-classical perspective.
We also saw in Sec.~\ref{sec:fermion-spectrum} that in our case there are massless fermions transforming in $\mathbf{24}$ of $\mathfrak{so}(24)$, whose anomaly polynomial including the gauge part was computed in Sec.~\ref{sec:fermion-anomalies}.
Here we reproduced these two points from a slightly different angle (without taking into account the anomaly of the gauge part).

We also see that the higher-dimensional Callan-Rubakov problem we briefly mentioned in Sec.~\ref{sec:callan-rubakov} in the case of 0-brane has a straightforward solution.
We have $24$ spacetime massless fermions of one chirality on this two-dimensional spacetime. On the other hand,
the worldsheet of the introduced fundamental string has $24$ massless fermions of the opposite chirality.
A spacetime fermion flowing into the core of the 0-brane is simply reflected back as a fermion on the worldsheet, as first pointed out in \cite{Polchinski:2005bg}.

For a compactification down to the flat two-dimensional Minkowski space,
we need an internal SCFT $T$ of  $(c_L,c_R)=(24,12)$. 
Many such $T$ have been constructed and studied, see e.g.~ \cite{Vafa:1995fj,Sen:1996na,Bergman:2003ym,Font:2004et,Paquette:2016xoo,Apruzzi:2016iac,Florakis:2017zep,Melnikov:2017yvz}.
All of those examples satisfied the condition that $N$ is an integer,
or equivalently that $b$ is a multiple of $24$,
meaning that the global anomaly associated to the shift of $\int B$ by $1$ is always absent.

This property might not be too surprising if one believes that string theory should always be magically anomaly-free, but it is surprising when viewed purely from the point  of the worldsheet theory, at least to the authors.\footnote{In Type~II string theory, it is not guaranteed that an anomaly-free worldsheet theory gives a consistent target space background. 
For example, Type IIA string on a Calabi-Yau four-fold such that its Euler number $\chi$ is not divisible by 24 is inconsistent \cite{Vafa:1995fj,Sethi:1996es}. 
For a recent work investigating this issue, see also \cite{Yonekura:2024bvh}.}
An  explanation of this property was given in \cite{Tachikawa:2021mvw} using the relation to topological modular forms.
We will come back to this issue later in this section.

\subsection{The Green-Schwarz coupling for the 4-brane: preliminaries}
\label{sec:4braneprelims}
Let us now try to apply the same reasoning to the internal theory $((E_7\times E_7)/\bZ_2)_1$ of the 4-brane. 
As will be discussed later, 
in this case the presence of a topologically non-trivial $B$-field makes the proper definition of the 
Green-Schwarz coupling more subtle, but for the moment let us ignore this issue and proceed directly as before. 

We begin with a generalization of the 4-brane setup,  namely a generic internal worldsheet theory $T$ with $c_L-c_R=14$. 
Let us say that its elliptic genus is given by \begin{equation}
\label{eq:4branecaseeg}
Z_\text{ell}(T;q)= \tr (-1)^\sF q^{L_0-c_L/24}  \bar q^{\bar L_0-c_R/24}  = q^{1/6} (a q^{-1} + b + \cdots) ~.
\end{equation}
The worldsheet states contributing to $a$ give rise to spacetime gravitinos and dilatinos,
and those contributing to $b$ give spacetime fermions,
whose spacetime chirality are controlled by the worldsheet fermion parity $(-1)^\sF$ as before.
As the anomaly polynomials of six-dimensional chiral massless fermions and gravitinos are given by \begin{equation}
\frac{7p_1^2-4p_2}{5760}~, \qquad
\frac{275p_1^2-980 p_2}{5760}
\end{equation} respectively, we find that the total anomaly polynomial $I_{8}$ of the six-dimensional theory coming from the internal theory $T$ is \begin{equation}
I_8=\frac{(268 a+7b) p_1^2 -4(244a+b)p_2}{5760}~.
\end{equation}
Now, the Green-Schwarz mechanism can only cancel the term proportional to $p_1^2$.
What happens to the term proportional to $p_2$?

To study this issue, consider $\eta(q)^{-4}Z_\text{ell}(T;q)$, where 
\beq
\eta(q) = q^{1/24}\prod_{n= 1}^{\infty}(1-q^n)
\eeq
is the Dedekind $\eta$-function. The function $\eta(q)^{-4}Z_\text{ell}(T;q)$ is a weakly holomorphic integral modular form of degree $-2$ with poles at most of the form $q^{-1}$;
here the adjective ``weakly'' means that we allow poles at $q=0$.

In general, any weakly holomorphic modular form is given by polynomials of 
\beq
E_4(q)~, \quad E_6(q)~, \quad \Delta(q)~, \quad \Delta(q)^{-1}~, \label{eq:modularforms}
\eeq
where $E_{4}(q)$ and $E_{6}(q) $ are the Eisenstein series of weight $4$ and $6$,
\beq
E_{4} (q) = 1 + 240 \sum_{ n=1}^{\infty} \frac{n^3 q^n}{1-q^n}~, \qquad E_{6} (q) = 1 -504 \sum_{ n=1}^{\infty} \frac{n^5 q^n}{1-q^n}~,\label{eq:Eisenstein}
\eeq
and $\Delta(q) = \eta(q)^{24}$ is the modular discriminant with modular weight 12. They are related by $(E_{4} (q) )^3 - (E_{6} (q) )^2 = 1728\Delta(q)$. It is possible to use the inverse $\Delta(q)^{-1}$ of $\Delta(q)$, but we cannot use the inverse of $E_4(q)$ or $E_6(q)$ since these have zeros within the fundamental domain.

The only possibility for $\eta(q)^{-4}Z_\text{ell}(T;q)$ is that it is proportional to $E_4(q) E_6(q)/\Delta(q)$ because it has at most a simple pole $q^{-1}$.
Thus we have \begin{equation}
Z_\text{ell}(T;q) \propto \eta(q)^4E_4(q)E_6(q)/\Delta(q) \propto q^{1/6}(q^{-1}-244  + \cdots)~.
\end{equation}
This tells us that $b=-244a$, guaranteeing the absence of the term proportional to $p_2$, and giving
\begin{equation}
I_8 = \frac{(268 a+7b) p_1^2}{5760}=-a\frac{p_1^2}4.
\end{equation} 
Using $\d H=p_1/2$, we may now cancel the perturbative anomaly by introducing the Green-Schwarz coupling \begin{equation}
\exp\left( 2\pi \i a\int B \cdot \frac{p_1}2 \right) \label{6dGS-naive}
\end{equation} in the exponentiated Euclidean action.

In the particular case of the internal theory of the 4-brane, we have $a=b=0$, as will be discussed in more detail in Appendix~\ref{app:elliptic}.
This means that the anomaly polynomial and the perturbative Green-Schwarz coupling are simply zero
when we turn off the $(E_7\times E_7)/\bZ_2$ background gauge field.
Indeed, we already computed the spacetime fermion spectrum in Sec.~\ref{sec:fermion-spectrum} and found in Sec.~\ref{sec:fermion-anomalies} that the anomaly polynomial is \eqref{4brane-anomaly},
which  vanishes when $c_{(1,2)}$ are set to zero.

That said,  this does not mean that the exponentiated Green-Schwarz coupling is always trivial 
when $a=b=0$, because the form \eqref{6dGS-naive} is not applicable when the $H$ flux is nontrivial.
For example, as we will show below, we should really assign an exponentiated Green-Schwarz coupling of $-1$ 
to the six-dimensional manifold $S^3_{H=1}\times S^3_{H=1}$, which has $\int_{S^3} H=1$ for both factors of $S^3$.
Before explaining why this should be the case, we will need to give a more careful definition of the Green-Schwarz coupling,
applicable to the case when the $H$ flux is nontrivial.

\subsection{Green-Schwarz coupling: general theory}\label{sec:generalGS}

Consider a $d$-dimensional spacetime theory $\Psi$ with massless fermions,
whose anomaly is controlled by a $(d+1)$-dimensional invertible field theory $\cA$.\footnote{The $(d+1)$-dimensional theory $\cA$ is realized by massive fermions in the large mass limit. The Hilbert space $\cH_\cA(M_d)$ of the theory $\cA$ on a closed manifold $M_d$ is one-dimensional since it consists of only the ground state in the large mass limit. Massless anomalous fermions are realized as boundary modes of the massive fermion theory. See \cite{Witten:2019bou} for systematic discussions of the facts about $\cA$ used in this subsection.}

We denote the partition function of $\Psi$ on the spacetime $M_d$ as $Z_\Psi(M_d)$. When there is no anomaly, partition functions take values in complex numbers $\bC$. On the other hand, when there is an anomaly described by an invertible field theory $\cA$, 
the $Z_\Psi(M_d)$ takes values in the dual of the one-dimensional Hilbert space $\cH_\cA(M_d)$ of the invertible field theory $\cA$ on $M_d$
as \begin{equation}
Z_\Psi(M_d) \in \overline \cH_\cA(M_d)~,
\end{equation}
where $ \overline \cH_\cA(M_d)$ is the dual space to $  \cH_\cA(M_d)$. 

Let $N_{d+1}\colon M_d\to M'_d$ be a manifold with an incoming boundary $M_d$ and an outgoing boundary $M'_d$, i.e. $\partial N_{d+1} = \overline{M_d} \sqcup M'_d$, where the $\sqcup$ means disjoint union. Let
\beq
Z_\cA(N_{d+1}) \colon \cH_\cA(M_d)\to \cH_\cA(M'_d)
\eeq
be the evolution operator (obtained by the path integral on $N_d$ in the theory $\cA$) describing transition amplitudes of the theory $\cA$. It contains the nontrivial information of $\cA$ and hence the information of anomaly. In particular, when $\partial N_{d+1} =M_d$, the pairing $\langle Z_\Psi(M_d), Z_\cA(N_{d+1}) \rangle$ between $ \overline \cH_\cA(M_d)$ and $  \cH_\cA(M_d)$ takes values in $\bC$. This pairing depends on the choice of $N_{d+1}$ and hence it cannot be regarded as a partition function on $M_d$ unless $Z_\cA(N_{d+1}) $ is trivial.

We now assume that the invertible field theory trivializes when we introduce a $B$-field whose field strength $H$ satisfies (among other things) $\d H=X_4$, where $X_4$ is a certain degree-4 characteristic class. 
This in particular means the following: \begin{itemize}
\item The anomaly polynomial $I_{d+2}$ admits a factorization $I_{d+2}=-X_4 Y_{d-2}$ and hence it is a total derivative, $I_{d+2} = -\d ( H Y_{d-2})$.
\item There exists a family of distinguished norm-one states $v(M_d,B)\in \cH_\cA(M_d)$ continuously varying with $(M_d,B)$, which will be paired with $Z_\Psi(M_d)$ to give values in $\bC$.
\item The family of states satisfy the compatibility condition \begin{equation}
v(M'_d,B')=e^{2\pi \i \int_{N_{d+1}} H Y_{d-2}}Z_\cA(N_{d+1})v(M_d,B) \label{compatibility}
\end{equation}
where $N_{d+1}\colon M_d\to M'_d$ is a manifold with an incoming boundary $M_d$ and an outgoing boundary $M'_d$,
properly equipped with the $B$-field,
and $Z_\cA(N_{d+1}) $ is the evolution operator of the $\cA$ introduced above.
\end{itemize}
The necessary and sufficient condition for the existence of $v(M_d, B)$ is that for any closed manifold $N_{d+1}$ (i.e. $\partial N_{d+1}= \varnothing$) we have
\beq
e^{2\pi \i \int_{N_{d+1}} H Y_{d-2}}Z_\cA(N_{d+1}) =1 \quad \text{when $\partial N_{d+1}= \varnothing$}~,
\eeq
where $Z_\cA(N_{d+1}) \in \U(1)$ is the partition function of the theory $\cA$ on $N_{d+1}$. This is the anomaly cancellation condition. 

We can now form the following pairing between $Z_\Psi(M_d) \in \overline \cH_\cA(M_d)$ and $v(M_d,B)\in \cH_\cA(M_d)$, \begin{equation}
\langle Z_\Psi(M_d), v(M_d,B) \rangle  \in \bC \label{combination}
\end{equation} which is a complex number, as desired. 

This contains the standard Green-Schwarz coupling $2\pi \i \int_{M_d} B Y_{d-2}$ in the following sense.
Let us equip $M_d$ with $B$ and consider a small variation $B' = B+ \delta B$, where $\delta B$ is a globally well-defined 2-form on $M_d$.
Applying \eqref{compatibility} for $N_{d+1}=M_d \times [0,1]$ with a $B$-field configuration such that it restricts to $B$ on $M_d \times \{0\}$ and to $B'$ on $M_d \times \{1\}$, one finds from the compatibility condition \eqref{compatibility} that \begin{equation}
\langle Z_\Psi(M_d), v(M_d,B') \rangle
= e^{2\pi \i \int_{M_{d}} \delta B Y_{d-2}}\langle Z_\Psi(M_d), v(M_d,B) \rangle~,
\end{equation}
where we have used the fact that $Z_\cA(N_{d+1}) = 1$ if $N_{d+1}$ is just a simple product $M_d \times [0,1]$.
This shows the expected variation.

We can also use \eqref{combination} to define what we mean by the Green-Schwarz coupling
when the $B$-field is nontrivial, under the following conditions.
First we assume that  $M_d$ is a boundary of a spin manifold $W_{d+1}$, not necessarily equipped with the $B$-field.
In this case, we can rewrite \eqref{combination} as \begin{equation}
\langle Z_\Psi(M_d), v(M_d,B) \rangle
= 
\langle  Z_\Psi(M_d), Z_\cA(W_{d+1})\rangle  \langle  \overline{Z_\cA(W_{d+1})}, v(M_d,B)\rangle
\label{split-version}
\end{equation}
where  $Z_\cA(W_{d+1})$ is the norm-one state in $\cH_\cA(M_d)$ defined by the bordism $W_{d+1}: \varnothing \to M_d$.
The first factor of \eqref{split-version}
can now be thought of as a complex number which is the fermion partition function, while the second factor
can be thought of as a complex phase which is the exponentiated Euclidean Green-Schwarz coupling.
However, for general $W_{d+1}$ and general fermion content, this splitting depends on the choice of $W_{d+1}$. 

To avoid dependence on the choice of $W_{d+1}$, we further assume the following. Suppose that the above general construction has been done for general backgrounds with gauge field (of the group $H$ of the current algebra $H_k$ in our brane examples). Now we restrict $M_d$ and $W_{d+1}$ to be backgrounds where the gauge field is set to zero. Moreover, assume that the fermions are non-chiral and anomaly free when the gauge field is absent. Under these conditions, the factor $\langle  Z_\Psi(M_d), Z_\cA(W_{d+1})\rangle $ is the partition function of the non-anomalous fermions and hence it does not depend on the choice of $W_{d+1}$. Therefore, the Green-Schwarz coupling $ \langle  \overline{Z_\cA(W_{d+1})}, v(M_d,B)\rangle$ is also independent of $W_{d+1}$ and is determined only by $M_d$, including the $B$-field on it.  

More concretely, suppose that there is a manifold $N_{d+1}$ which is equipped with the $B$-field and in which the gauge field is possibly nonzero, such that $\partial N_{d+1} = M_{d}$. Then the compatibility condition \eqref{compatibility} in particular implies that
\beq
v(M_d,B)=e^{2\pi \i \int_{N_{d+1}} H Y_{d-2}}  Z_\cA(N_{d+1}) ~.
\eeq
Therefore, the Green-Schwarz coupling, which we denote as $\GS(M_d)$, is given by
\beq
\GS(M_d) &= 
\langle  \overline{Z_\cA(W_{d+1})}, v(M_d,B)\rangle \nonumber \\
&= e^{2\pi \i \int_{N_{d+1}} H Y_{d-2}} \langle \overline{Z_\cA(W_{d+1})}, Z_\cA(N_{d+1})  \rangle \nonumber \\
& = e^{2\pi \i \int_{N_{d+1}} H Y_{d-2}}  Z_\cA (\overline{W_{d+1}} \cup N_{d+1})~, \label{eq:formulaGS}
\eeq
where $\overline{W_{d+1}} $ is the orientation reversal of $W_{d+1}$, the manifold $\overline{W_{d+1}} \cup N_{d+1}$ is the closed manifold obtained by gluing $\overline{W_{d+1}} $ and $N_{d+1}$ along $M_d$, and $Z_\cA (\overline{W_{d+1}} \cup N_{d+1}) \in \U(1)$ is the partition function of the theory $\cA$ on the closed manifold $\overline{W_{d+1}} \cup N_{d+1}$. 

Let us recapitulate the conditions under which we obtained the formula \eqref{eq:formulaGS} for the Green-Schwarz coupling $\GS(M_d)$:
\begin{itemize}
\item $M_d$ is a manifold equipped with the $B$-field, on which the gauge field is set to zero.
\item $N_{d+1}$ is a manifold equipped with the $B$-field, on which the gauge field is possibly nonzero.
\item $W_{d+1}$ is a manifold on which the $B$-field is not necessarily defined, and the gauge field is set to zero. 
\item The massless fermions are non-anomalous in the absence of the gauge field.
\end{itemize}
The coupling $\GS(M_d)$ depends only on $M_d$, including the $B$-field on it.

The coupling $\GS(M_d)$ is actually a bordism invariant of manifolds equipped with a $B$-field on which the gauge field is set to zero. To see this, suppose $M_d =\partial N_{d+1}$ where $N_{d+1}$ is a manifold equipped with the $B$-field on which the gauge field is zero. In this case, we can take the $N_{d+1}$ in the definition of $\GS(M_d)$ to be this particular $N_{d+1}$. The absence of the fermion anomaly when the gauge field is turned off implies that both of $ Z_\cA (\overline{W_{d+1}} \cup N_{d+1})$ and $ e^{2\pi \i \int_{N_{d+1}} H Y_{d-2}} $ are trivial. Thus we get $\GS(M_d)=1$, establishing the bordism invariance. 

Let $\Omega^\text{string}_{d} = \Omega^\text{string}_{d}(\text{pt})$ be the bordism group of spin manifolds equipped with the $B$-field.
These manifolds are called string manifolds, and the structure determined by the $B$-field and spin structure is called string structure. Actually, we have defined $\GS(M_d)$ only in a subgroup of $\Omega^\text{string}_{d}$. In the construction, we have assumed that there exists a spin manifold $W_{d+1}$ such that $\partial W_{d+1} =M_d$ as a spin manifold. This means that the spin bordism class of $M_d$ is zero. Therefore, we have defined $\GS$ for elements of
\beq
\mathrm{Ker}( \Omega^\text{string}_d\to \Omega^\text{spin}_d) .
\eeq
Thus $\GS$ is actually a homomorphism
\beq
\GS:  \mathrm{Ker}( \Omega^\text{string}_d\to \Omega^\text{spin}_d) \to \U(1), \label{eq:bordisminvariance}
\eeq
where the homomorphism $\Omega^\text{string}_d\to \Omega^\text{spin}_d$ is given by forgetting the $B$-field, and $\text{Ker}$ means the kernel.

We remark that there are possible ambiguities when $d$ is of the form 
\beq
d \in 3 +4\bZ~.
\eeq
In this case, the massless fermion partition function has ambiguities because we can add gravitational Chern-Simons terms and also terms involving the field strength 3-form $H$. For instance, let us consider the case $d=3$. In this case, we can modify the massless fermion partition function by a term
\beq
\exp 2\pi \i\left( \alpha \eta(M_d) +   \beta  \int_{M_d} H   \right),
\eeq
where $\eta(M_d)$ is the Atiyah-Patodi-Singer (APS) $\eta$-invariant (which is basically a precise version of the gravitational Chern-Simons), and $\alpha \in \frac{1}{2}\bZ$ and $\beta \in \bR$ are parameters. We cannot determine the values of $\alpha$ and $\beta$ just from the perspective of low energy effective field theory. (The full string theory should determine these parameters, but we do not try to study it in the present paper.) This kind of ambiguities exists only in dimensions $d \in 3 +4\bZ$. Thus, in the present paper, we do not consider the Green-Schwarz coupling in these dimensions.

The above definition of the $\GS$ used the property that the internal worldsheet theory has a symmetry (which is the group $H$ of the current algebra theory $H_k$ in the case of our branes as we use more explicitly in later subsections). Then we used manifolds  $N_{d+1}$ equipped with the gauge field. It is not clear at this point if the Green-Schwarz coupling is a priori defined without using this additional information about the symmetry. Notice that even though we have used the gauge field on $N_{d+1}$, the term $\GS(M_d)$ only depends on $M_d$, on which the gauge field is turned off. Moreover, it is a bordism invariant as we showed above and hence we expect it to be robust under small deformations. Therefore, it makes sense to speculate that the Green-Schwarz coupling is determined without reference to symmetries at all, and is also invariant under deformations of the internal theory.

We claim that the Green-Schwarz coupling $\GS$ is actually determined by the bordism class $[T] $ of the internal theory $T$ in $ \sqft_{-d-22}$ when the dimension of $M_d$ is such that $d  \notin 3+4\bZ$. (Notice that $d=10-n$ for our branes.) In the definition of $\GS$, we have assumed that there is no gravitational anomaly of the massless fermions produced by $T$. A sufficient condition for this is to require that all elliptic genera, i.e.~both ordinary and mod-2 elliptic genus of $T$ vanish.
We define a subgroup $\bA_{-d-22} \subset \sqft_{-d-22}$ by
\beq
\bA_{-d-22} = \{ [T] \in \sqft_{-d-22}~|~ \text{ordinary and mod-2 elliptic genera of $[T]$ vanish} \}~. \label{eq:physicalA}
\eeq
Thus, $\GS$ is actually a bilinear form $\GS(T, M_d) \in  \U(1)$ depending only on bordism classes $[T] \in \bA_{-d-22}$ and $[M_d] \in  \mathrm{Ker}( \Omega^\text{string}_d\to \Omega^\text{spin}_d) $,
\beq
\GS:   \bA_{-d-22}  \times  \mathrm{Ker}( \Omega^\text{string}_d\to \Omega^\text{spin}_d) \to  \U(1)~. \label{eq:GSpairing}
\eeq
(A stronger claim is possible but we do not need it in the present paper.)
This is a nontrivial claim and it has its own interest in general studies of 2d ${\cal N}=(0,1)$ SQFTs, so it will be discussed elsewhere.

\subsection{The Green-Schwarz coupling for the 4-brane}
\label{4-brane-GS-computation}
Let us now apply the general theory developed in the last subsection to the particular case of our 4-brane.
Here, there is no net number of chiral fermions when we turn off the $E_7\times E_7$ gauge field, as we saw in Sec.~\ref{sec:4braneprelims},
so this case satisfies the condition under which $\GS(M_d)$ is defined.
This can also be verified by 
checking that the elliptic genera of the internal theory vanish, 
as we will do in Appendix~\ref{sec:4brane-elliptic}. 

The invertible field theory $\cA$ is nontrivial with the $E_7\times E_7$ gauge field turned on, with the anomaly polynomial given in \eqref{4brane-anomaly}.
In particular, we have\begin{equation}
X_4=\frac{p_1}2+ c_{(1)}+c_{(2)}~,\qquad
Y_4=-\frac12(c_{(1)}-c_{(2)})~.
\end{equation}

For explicitness, we consider the following $M_{d=6}$. Let $S^3_{H=N}$ be the sphere with a standard round metric and the $H$-flux $\int_{S^3} H=N$. (We will later make a brief comment on the case that the metric is not round.)
Then we take $M_6= S^3_{H=1} \times S^3_{H=1}$.

Recall that $S^3=\partial D^4$. To be definite, we take $D^4$ to be a hemisphere of $S^4$ with a round metric. Then we can just choose the $W_7$ in the formula \eqref{eq:formulaGS} to be 
\beq
W_7 = D^4 \times S^3~.
\eeq
We also choose the underlying manifold for $N_7$ to be $D^4 \times S^3$. For $N_7$, we cannot turn off the gauge field because we have
\beq
1 = \int_{S^3} H  = \int_{D^4} \d H = \int_{D^4} \left(\frac{p_1}2+ c_{(1)}+c_{(2)} \right) .
\eeq
For a round metric, we have $p_1=0$ because $p_1$ is a parity-odd quantity and the round sphere has a parity symmetry. Thus we need to turn on at least either $c_{(1)}$ or $c_{(2)}$. To be definite, we take
\beq
 \int_{D^4}  c_{(1)}  =1~.
\eeq
Let us denote the $D^4$ with an $E_7\times E_7$ fiber bundle $P$ having this instanton as $D^4_P$. Then we take
\beq
N_7 = D^4_P \times S^3_{H=1}~.
\eeq


Let $S^4_P$ be the sphere obtained by gluing $D^4_P$ and $\overline{D^4}$ along $S^3$.
Now the formula \eqref{eq:formulaGS} gives
\beq
\GS(S^3_{H=1} \times S^3_{H=1}) = e^{-\pi \i \int_{D^4_P \times S^3_1} H  (c_{(1)}-c_{(2)}) }  Z_\cA ( S^4_P \times S^3)~.
\eeq
The first factor on the right hand side is $-1$, as $\int_{D^4_P} c_{(1)}=1$ and $\int_{S^3} H=1$.
The second factor is determined by the properties of the fermions and it is independent of the $B$-field. The factor $S^3$ is parity symmetric and hence (from the explicit definition of $Z_\cA$ in terms of the APS $\eta$-invariant) we have $ Z_\cA ( S^4_P \times S^3)=1$. 
%
Combined, we found that 
$\GS(S^3_{H=1} \times S^3_{H=1}) = -1$,
showing that the exponentiated Green-Schwarz coupling
can be nontrivial even though the coefficient $a$ in \eqref{6dGS-naive} is zero. It is also clear by the same argument as above that 
\beq
\GS(S^3_{H=N} \times S^3_{H=M}) =(-1)^{MN} \qquad (N, M \in \bZ)~.
\eeq
for more general $H$ flux numbers $N$ and $M$.

Let us give a very brief comment on the case that the metric on $S^3$ is not a round one. In that case, the equation $\d H =  p_1/2 + \cdots $ implies that the integral $\int_{S^3} H$ is not an integer. (See \cite{Witten:1999eg,Yonekura:2022reu} for the precise condition which the $H$ must satisfy.) Thus the value of $\int_{S^3} H$ depends on the metric on $S^3$. The value of $Z_\cA ( S^4_P \times S^3)$ also depends on the metric as can be seen as follows. First, we take zero modes of the fermions on $S^4_P$ on the instanton background. From the representation ${\bf 56}$ of $E_7$, we get $24$ copies of Majorana-Weyl fermions in two dimensions. Then the value of $Z_\cA ( S^4_P \times S^3)$ is determined by the APS $\eta$-invariant in three dimensions corresponding to the 24 copies of Majorana-Weyl fermions in  two dimensions. The $\eta$-invariant depends on the metric, and its metric dependence precisely cancels the metric dependence of $ \int_{S^3} H$. Thus $\GS$ is a topological invariant. The topological invariance is of course an immediate corollary of the much stronger property that $\GS$ is a bordism invariant, as discussed around \eqref{eq:bordisminvariance}.

\subsection{Bordism relations among our heterotic branes}
\label{sec:bordism-relations}

We would now like to apply the general formalism of the global Green-Schwarz coupling
to the rest of our branes, i.e.~the 6- and 7-branes. 
For this, we need to show that the elliptic genera of the internal theories of these branes vanish.
As it is rather technical and outside of the main line of the development,
this will be done in Appendix~\ref{app:elliptic}.
The actual computation of the Green-Schwarz couplings for the 7- and 6-branes 
will be done in the next section.
The aim of this section is to prove the following two relations used in the next section,
namely the relation
\begin{equation}
[X_1]\times [S^1_\text{periodic}] = [X_2]
\label{31to30}
\end{equation}
and the relation \begin{equation}
[X_1]\times [S^3_{H=1}]=[X_4]
\label{31to28}
\end{equation}
as $\SQFT_\bullet$ elements.
Here $[X_n]$ are defined in Sec.~\ref{sec:names} and  \begin{itemize}
\item $[S^1_\text{periodic}]$ is the $\SQFT$ class for the \Nequals{(0,1)} supersymmetric sigma model on $S^1$ with periodic target-space spin structure, and
\item  $[S^3_{H=N}]$ is the $\SQFT$ class for the \Nequals{(0,1)} supersymmetric sigma model on $S^3$ with $N$ units of the $H$ flux.
\end{itemize}
This will be done by providing a continuous deformation from the left-hand side to the right-hand side
of these two equations.
Before proceeding, we note that the first relation \eqref{31to30} will use a T-duality, and therefore is a relation only in $\SQFT_\bullet$,
whereas the second relation \eqref{31to28} is a relation in $\sugra_\bullet$.

\subsubsection{The 7-brane and the 6-brane}
We start from the relation \eqref{31to30}. 
Clearly it suffices to show that \begin{equation}
[X_1]\times [S^1_\text{periodic}] + [(\hete)_1] \times [S^1_\text{periodic}] ^2  = [X_2]+ [(\hete)_1] \times [S^1_\text{periodic}] ^2~.
\label{31to30-intermediate}
\end{equation}
To do so, we start from the left-hand side of \eqref{31to30-intermediate}, \begin{align}
&[X_1]\times [S^1_\text{periodic}] + [(\hete)_1] \times [S^1_\text{periodic}] ^2 \nonumber \\
&= \left( [X_1]+ [(\hete)_1] \times [S^1_\text{periodic}] \right)\times [S^1_\text{periodic}] \nonumber  \\
&= [Y_1] \times [S^1_\text{periodic}]
\end{align} where $[Y_1]$ is now the $\hete$ level-1 theory 
fibered over $S^1$ with \emph{periodic} spin structure such that when we go around $S^1$ the two $E_8$ factors are interchanged. The equality
\beq
  [X_1]+ [(\hete)_1] \times [S^1_\text{periodic}]= [Y_1] 
\eeq
is essentially due to a bordism from the disjoint union of $S^1_\text{antiperiodic}$ and $S^1_\text{periodic}$ to $S^1_\text{periodic}$ via a sphere $S^2$ with three holes. More generally, it is known that for any manifolds $M_1$ and $M_2$, there is a bordism from the disjoint union of $M_1$ and $M_2$ to their connected sum $M_1 \# M_2$ (see e.g. \cite{Milnor}). We will use this fact a few times below.

When we use the worldsheet theory $[Y_1 \times S^1_\text{periodic}]$ for a heterotic compactification,
it will give us a fully-supersymmetric eight-dimensional theory with the gauge group rank $8+4$. 
Via the construction of  e.g.~\cite{Mikhailov:1998si},
it is known that any such heterotic 8d theories are continuously connected, and in particular we have
\begin{equation}
[Y_1] \times [S^1_\text{periodic}]
= [\text{$\hetss$ on $(S^1_\text{periodic})^2$ with flat gauge fields with nontrivial $v_2$} ]~,
\label{without-vec}
\end{equation} where the right hand side is the so-called \emph{compactification without vector structure} of \cite{Witten:1997bs},
which has e.g. $\mathfrak{u}(1)^4 \times \so(16)$ gauge group.\footnote{In the dual Type~I picture, we have three O$7^-$ planes and one O$7^+$ plane with eight D7-branes. Thus we can realize $\so$ and/or $\sp$ gauge groups depending on the positions of the D7-branes.} 
Let us see this equivalence a little more explicitly.

In the theory $Y_1 \times S^1_\text{periodic}$, we continuously turn on holonomies of the diagonal subgroup $E_8 \subset E_8 \times E_8$ around $S^1_\text{periodic}$. The diagonal subgroup is taken so that it is invariant under the exchange of the two $E_8$ factors and hence the holonomies are consistent with the structure of $Y_1$. By tuning the holonomy, we can realize a configuration such that each $\e_8$ is broken to $\so(16)$, giving $\so(16) \times \so(16)$. This is T-dual to the $\hets$ theory with $\so(32)$ holonomy, which is represented in terms of $\SO(32)$ matrices as
\beq
A=\begin{pmatrix} 1 & 0 \\ 0 & -1 \end{pmatrix} \otimes {\bf 1}_{16} \in \SO(32)~,
\eeq
where ${\bf 1}_{16}$ is the unit matrix in $\SO(16)$.
Moreover, along the $S^1$ direction in $Y_1$, there is a holonomy such that the two $\so(16)$ factors are exchanged. This is realized by a holonomy in $\so(32)$ given by
\beq
B=\begin{pmatrix} 0 & 1 \\ 1 & 0 \end{pmatrix} \otimes {\bf 1}_{16} \in \SO(32)~.
\eeq
The above holonomies $A$ and $B$ commute only up to a sign in $\SO(32)$,
\beq
AB = - BA \in \SO(32)~.
\eeq
The minus sign would be inconsistent if the gauge group were $\SO(32)$.
However, the minus sign is a trivial element of the actual group $\hets$, and hence these holonomies commute with each other, and are therefore consistent. Moreover, the minus sign implies that the gauge flux on the $T^2 = S^1 \times S^1$ has a nontrivial characteristic class $v_2 \in H^2(T^2; \pi_1(\hets))$, because this minus sign is the obstruction to uplifting the gauge configuration from $\hets$ to $\Spin(32)$, and in particular to $\SO(32)$. In this way we get \eqref{without-vec}. The remaining non-abelian gauge group in this case is $\so(16)$. 

Let us return to the proof of \eqref{31to30-intermediate}.
The class $[X_2]$ contains $\hets$ on $S^2$ with nontrivial $v_2$. By a bordism, this $S^2$ can be glued to $S^1_\text{periodic} \times S^1_\text{periodic}$ via a connected sum.
Thus we see that the right-hand side of \eqref{31to30-intermediate} is the same as the right-hand side of
\eqref{without-vec}, which was what we wanted to show.

\subsubsection{The 7-brane and the 4-brane}\label{sec:7brane4brane}

Let us now move on to the equality \eqref{31to28}, which relates the 7-brane and the 4-brane.
This can be shown within supergravity, since both sides of the equation use the $\hete$ theory.

We start from the right-hand side.
Since $[S^3_{H=0}]=0$, we have \begin{align}
[X_4] &= [X_4] + [X_1] \times [S^3_{H=0}]\\
&= \begin{tikzpicture}[baseline=(0),scale=.8]
\node (0) at (0,-.2) {};
\draw (-2,1)    --(2,1) ;
\draw (-2,-1)   --(2,-1) ;
\draw[dashed] (-2,1) --(-2,-1);
\draw[dashed] (+2,1) --(+2,-1);
\node[left,red] (r1) at (0,.5) {$+1$};
\node[right,ForestGreen] (g1) at (0,-.5) {$-1$};
\draw[red,fill=red] (0,.5) ellipse (.16 and .16);
\draw[ForestGreen,fill=ForestGreen] (0,-.5) ellipse (.16 and .16);
\end{tikzpicture}
\end{align}
where in the second line, the horizontal direction is the $S^1$ direction,
the vertical line is the $S^3$ direction,
and the red and green blobs represent the instanton density of two $E_8$ factors. Here we have used a bordism from the disjoint union of $S^4$ with instantons and $S^1 \times S^3$ without instantons to $S^1 \times S^3$ with instantons, via a connected sum.

We now move one of the instantons slightly to the right: \begin{equation}
=\,\, \begin{tikzpicture}[baseline=(0),scale=.8]
\path[fill=cyan!20!white] (0,1) rectangle (1,-1);
\node (0) at (0,-.2) {};
\draw (-2,1)    --(2,1) ;
\draw (-2,-1)   --(2,-1) ;
\draw[dashed] (-2,1) --(-2,-1);
\draw[dashed] (+2,1) --(+2,-1);
\node[left,red] (r1) at (0,.5) {$+1$};
\draw[red,fill=red] (0,.5) ellipse (.16 and .16);
\node[right,ForestGreen] (g1) at (1,-.5) {$-1$};
\draw[ForestGreen,fill=ForestGreen] (1,-.5) ellipse (.16 and .16);
\end{tikzpicture}
\end{equation} 
where now we have $\int_{S^3} H=1$ in the blue-shaded region due to the constraint \begin{equation}
\d H=n_1 + n_2~,
\end{equation} where $n_{1,2}$ are the instanton number densities for the two factors of $E_8$.
The unshaded region still has $\int_{S^3}H=0$.

We now move one of the instantons further to the right,
and bring it to the left using the periodic identification along the horizontal direction: \begin{equation}
=\,\, \begin{tikzpicture}[baseline=(0),scale=.8]
\path[fill=cyan!20!white] (0,1) rectangle (2,-1);
\path[fill=cyan!20!white] (-2,1) rectangle (-1.2,-1);
\node (0) at (0,-.2) {};
\draw (-2,1)    --(2,1) ;
\draw (-2,-1)   --(2,-1) ;
\draw[dashed] (-2,1) --(-2,-1);
\draw[dashed] (+2,1) --(+2,-1);
\node[left,red] (r1) at (0,.5) {$+1$};
\draw[red,fill=red] (0,.5) ellipse (.16 and .16);
\node[right,red,xshift=-4] (g1) at (-1,-.5) {$-1$};
\draw[red,fill=red] (-1.2,-.5) ellipse (.16 and .16);
\end{tikzpicture}
\end{equation}
where we now have instanton densities in the same $E_8$ for both points, 
since $[X_1]$ was such that two $E_8$ factors were exchanged when we go around $S^1$ once.
We now move it further to the right and we find \begin{equation}
=\,\, \begin{tikzpicture}[baseline=(0),scale=.8]
\path[fill=cyan!20!white] (-2,1) rectangle (2,-1);
\node (0) at (0,-.2) {};
\draw (-2,1)    --(2,1) ;
\draw (-2,-1)   --(2,-1) ;
\draw[dashed] (-2,1) --(-2,-1);
\draw[dashed] (+2,1) --(+2,-1);
\node[left,red] (r1) at (0,.5) {$+1$};
\node[right,red] (g1) at (0,-.5) {$-1$};
\draw[red,fill=red] (0,.5) ellipse (.16 and .16);
\draw[red,fill=red] (0,-.5) ellipse (.16 and .16);
\end{tikzpicture} \\
\,\,=\,\,\begin{tikzpicture}[baseline=(0),scale=.8]
\path[fill=cyan!20!white] (-2,1) rectangle (2,-1);
\node (0) at (0,-.2) {};
\draw (-2,1)    --(2,1) ;
\draw (-2,-1)   --(2,-1) ;
\draw[dashed] (-2,1) --(-2,-1);
\draw[dashed] (+2,1) --(+2,-1);
\end{tikzpicture}
\,\,=\,\, [X_1]\times [S^3_{H=1}]~.
\end{equation}
This was exactly what we wanted to establish.

\subsection{The Green-Schwarz coupling for the 6-brane and the 7-brane}
\label{sec:67braneGS}

We are finally in a position to compute the Green-Schwarz couplings associated to the angular parts of the 6- and 7-branes.
Around \eqref{eq:GSpairing}, we claimed that 
the Green-Schwarz coupling $\GS$ on the $d$-dimensional spacetime $M_d$ equipped with the $B$-field,
given by the worldsheet theory $T$ of degree $n-32=-d-22$, is well-defined for $d \notin 3+4\bZ$
if $[T] \in \bA_{-d-22}$ and $[M_d] \in  \mathrm{Ker}( \Omega^\text{string}_d\to \Omega^\text{spin}_d)$. 
The Green-Schwarz coupling is then a bilinear form
\beq
\GS:   \bA_{-d-22}  \times  \mathrm{Ker}( \Omega^\text{string}_d \to \Omega^\text{spin}_d) \ni ([T], [M_d]) \mapsto \GS([T],[M_d]) \in  \U(1)~.
\eeq

We show in Appendix~\ref{app:elliptic} that the elliptic genera of $[X_n]$ for $n=1,2,4$ indeed vanish, so that the above definition can be used. 
The Green-Schwarz coupling for the 4-brane case that we computed in subsection~\ref{4-brane-GS-computation} was an example of this,
where we use $M_d=S^3_{H=1} \times S^3_{H=1}$ and $[T]= [X_4]$,
\beq
 \GS([X_4],[S^3_{H=1}] \times [S^3_{H=1}]) = (-1) =  \exp\left( 2\pi \i \cdot \frac12 \right). \label{4-braneGS}
\eeq
We can now use the bordism relations found in Sec.~\ref{sec:bordism-relations} to compute the Green-Schwarz couplings in the case of the 6-brane and the 7-brane.

Before proceeding, let us determine what kind of $d$-dimensional manifolds $M_d$ will be of interest to us.
As mentioned in Sec.~\ref{sec:generalGS}, a spin manifold properly equipped with the $B$-field is mathematically known as a string manifold.
Their bordism groups were computed in \cite{Giambalvo1971}, which are tabulated together with the more familiar spin bordism groups, as well as the kernels of the map $\Omega^\text{string}_d \to \Omega^\text{spin}_d$, in Table~\ref{tab:bordism-groups}.

\begin{table}
\[
\begin{array}{c|ccccccccccccccccccccc}
d& 0 & 1 & 2 & 3 & 4 & 5 & 6 & 7 & 8 & 9 & 10 
\\
\hline
\Omega^\text{spin}_d &
\bZ &
\bZ_2 &
\bZ_2 &
0 &
\bZ &
0 &
0 & 
0 &
\bZ\oplus \bZ &
\bZ_2\oplus\bZ_2 &
(\bZ_2)^3\\
\Omega^\text{string}_d &
\bZ &
\bZ_2 &
\bZ_2 &
\bZ_{24} &
0 &
0 &
\bZ_2 & 
0 &
\bZ\oplus \bZ_2 &
(\bZ_2)^2 &
\bZ_6& \\
\mathrm{Ker}( \Omega^\text{string}_d\to \Omega^\text{spin}_d) &
0 & 0 & 0 & \bZ_{24} & 0 & 0 &  \bZ_2 & 0 & \bZ_2 & \bZ_2  & \bZ_3
\end{array}
\]
\caption{Spin and string bordism groups \label{tab:bordism-groups}}
\end{table}


Between $0\le d \le 10$,
we see that $\mathrm{Ker}( \Omega^\text{string}_d\to \Omega^\text{spin}_d)$ are as follows \cite{Hopkins2002}:
\begin{itemize}
\item $\bZ_{24}$ for $d=3$ is  generated by $[S^3_{H=1}]$,
\item $\bZ_2$ for $d=6$ by $[S^3_{H=1}]^2$,
\item $\bZ_2$ for $d=8$ by $[\SU(3)_{H=1}]$,
\item $\bZ_2$ for $d=9$ by $[S^3_{H=1}]^3 = [\SU(3)_{H=1}] \times [S^1_\text{periodic}]$, and
\item $\bZ_3$ for $d=10$ by $[\Sp(2)_{H=1}]$.
\end{itemize}
Here, the $\SU(3)_{H=1}$ is such that its underlying manifold is the manifold for the Lie group $\SU(3)$ which is 8-dimensional, and using the fiber bundle structure 
\begin{equation}
\SU(2) \simeq S^3 \to \SU(3) \to \SU(3)/\SU(2) \simeq S^5,
\end{equation}
we introduce a unit of $H$ flux on the fiber $S^3$. The meaning of $\Sp(2)_{H=1}$ is similar.

In view of these facts, we see that the Green-Schwarz coupling for the 4-brane \eqref{4-braneGS}
detects the non-trivial string bordism class at dimension 6. More importantly for us, this shows that the class $[X_4] \in \bA_{-28} \subset \sqft_{-28}$ is nontrivial. 

Let us now consider the 7-brane. 
We previously saw the bordism relation $[X_4]=[S^3_{H=1}] \times [X_1]$, which can be combined with \eqref{4-braneGS} to find 
\beq
\GS([X_1], [S^3_{H=1}]^3) &= \GS([X_1]  [S^3_{H=1}], [S^3_{H=1}]^2) =  \GS([X_4] , [S^3_{H=1}]^2) \nonumber \\
&=  \exp\left( 2\pi \i \cdot \frac12 \right),
\eeq
where in the first equality we regard one of the three factors of $S^3_{H=1}$ as an ``internal manifold,'' while the other two are regarded as ``spacetime''. 
Thus the Green-Schwarz coupling for the 7-brane detects the nontrivial $\bZ_2$ string bordism class in $d=9$. This also shows that the class $[X_1] \in \bA_{-31} \subset \sqft_{-31}$ is nontrivial. 

Next consider the case of the 6-brane.
Here we use the string bordism $[S^3_{H=1}]^3 = [\SU(3)_{H=1}] \times [S^1_\text{periodic}]$ known to mathematicians 
to do the following manipulations: 
\beq
\GS([X_2],  [\SU(3)_{H=1}]) & = \GS([X_1][S^1_\text{periodic}] , [\SU(3)_{H=1}]) = \GS([X_1] , [S^1_\text{periodic}][\SU(3)_{H=1}]) \nonumber \\
&=\GS([X_1] , [S^3_{H=1}]^3) = \exp\left( 2\pi \i \cdot \frac12 \right)~,
\eeq
where we have used \eqref{31to30}.
Thus we see that the Green-Schwarz coupling for the 6-brane detects the nontrivial $\bZ_2$ string bordism class in $d=8$. It shows that the class $[X_2] \in \bA_{-30} \subset \sqft_{-30}$ is nontrivial. 

We have shown in \eqref{eq:0branegenus} that the elliptic genus of $[X_8]$ is nontrivial. Therefore, we conclude that
\beq
[X_n] \neq 0 \quad \text{as an element of $\sqft_{n-32}$ }~~(n=1,2,4,8)~.
\eeq
The classes of the corresponding current algebra theories $[H_k]$ are also nonzero by \eqref{X1},  \eqref{X2},  \eqref{X4}, and  \eqref{X8}.

\subsection{Relation to the topological modular forms}
\label{sec:tmf}

So far, our discussions have been phrased in terms of the groups $\SQFT_{\bullet}$ of deformation classes of 2d \Nequals{(0,1)} SQFTs.
There is in fact a series of works by two mathematicians, Stolz and Teichner,
suggesting that $\SQFT_\bullet  (=\sqft_{\bullet}(\text{pt}) )$ is equivalent to the topological modular forms $\TMF_\bullet=\TMF_\bullet(\text{pt})$.
There are many recent works checking this conjectural equivalence, see e.g.~\cite{Gukov:2018iiq,Gaiotto:2018ypj,Gaiotto:2019asa,Gaiotto:2019gef,Johnson-Freyd:2020itv,Tachikawa:2021mvw,Lin:2021bcp,Yonekura:2022reu,Lin:2022wpx,Albert:2022gcs,Tachikawa:2023nne,Tachikawa:2024ucm,Saxena:2024eil}.


Taking the ordinary and mod-2 elliptic genera of SQFTs corresponds on the mathematical side to a morphism
\beq
\TMF_\bullet \to \KO_\bullet((q))~,
\eeq
where $ \KO_\bullet((q))$ is a formal Laurent series of $q$ with coefficients in $\KO_\bullet$. 
Interestingly, the kernel of $\TMF_{n-32} \to \KO_{n-32} ((q))$ 
in the range $1 \leq n \leq 10$ 
is known to be nonzero only at $n=1,2,4$, and is $\bZ_2$ in each case (see e.g.~Appendix E of \cite{Tachikawa:2023lwf}),
\beq
\bA_{n-32}:=\text{Ker}(\TMF_{n-32} \to \KO_{n-32}((q)))  \simeq \bZ_2  \quad \text{for $n=1,2,4$~.}
\eeq
These degrees are exactly where the angular parts $X_1$, $X_2$, and $X_4$ of our 7-, 6- and 4-branes appear,
see \eqref{X1}, \eqref{X2}, and \eqref{X4},
which we know to be in the physical definition of $\bA_{n-32}$ given in \eqref{eq:physicalA}.
We have seen above that the discrete Green-Schwarz coupling of $[X_1]$, $[X_2]$, and $[X_4]$ 
with respect to suitable string manifolds are nonzero,
meaning that these three classes in $\SQFT_{n-32}$ are actually nontrivial.
This strongly suggests that these angular parts
should correspond to the nontrivial elements of $\bZ_2 \simeq \mathrm{Ker}(\TMF_{n-32} \to\KO_{n-32} ((q)))$ at each degree.
For more on the relation to the theory of topological modular forms, the readers are referred to \cite{Tachikawa:2023lwf}.

\section*{Acknowledgements}

The authors would very much like to thank Kantaro Ohmori for helpful discussions during the entirety of the project.
YT thanks mathematical discussions with Mayuko Yamashita. KY thanks M.~Fukuda, S.~Kobayashi, J.~Maldacena, M.~Montero, and K.~Watanabe for discussions. JK thanks Jan Albert for discussions.
YT is supported in part  
by WPI Initiative, MEXT, Japan at Kavli IPMU, the University of Tokyo
and by JSPS KAKENHI Grant-in-Aid (Kiban-C), No.24K06883.
JK is supported in part by the Inamori Foundation through the
Institute for Advanced Study at Kyushu University. 
KY is supported in part by JST FOREST Program (Grant Number JPMJFR2030, Japan), 
MEXT-JSPS Grant-in-Aid for Transformative Research Areas (A) ”Extreme Universe” (No. 21H05188),
and JSPS KAKENHI (17K14265).

\appendix

\section{(In)stability of gauge field configurations at infinity}\label{sec:instability}

In this appendix, we discuss whether certain gauge field configurations $A_\mu$ on a sphere $S^n$ are stable near spatial infinity 
(rather than near the core regions of branes). 
We will only study tachyonic modes, and will not study non-perturbative instability.

In Sec.~\ref{sec:sphere} we first discuss the issue on $S^n$ without taking into account the radial direction. 
Then in Sec.~\ref{sec:radial} we include the radial direction and determine the condition for the absence of tachyonic modes.
Examples relevant to this paper are discussed in Sec.~\ref{sec:example}. 

\subsection{Gauge fields on a sphere}\label{sec:sphere}
In this subsection we recall some standard facts about perturbations of gauge fields around classical solutions on a closed manifold.
For definiteness we take the closed manifold to be $S^n$, although this assumption is not necessary in this subsection. 

Let us consider the Yang-Mills action on $S^n$,
\beq
S_\text{YM}(A) = \int_{S^n} \frac{1}{4} (F_{\mu\nu}, F^{\mu\nu})~,
\eeq
where 
$F_{\mu\nu} = \partial_\mu A_\nu - \partial_\nu A_\mu + [A_\mu, A_\nu]$ is the gauge field strength, 
and $(\bullet, \bullet)$ represents an invariant inner product on the Lie algebra, such as the negative of a trace $ - \tr$ in some representation.
We will not include higher derivative terms in the action because we are interested in spatial infinity, where curvatures are small and the leading derivative action is sufficient.
We remark that in the core regions of branes, we generally expect that such higher derivative terms are not negligible if the brane charge is of order 1.  

Suppose we are given a gauge field configuration $A_\mu$ on $S^n$ satisfying the equations of motion 
\beq
D_\mu F^{\mu\nu}=0~.
\eeq
We consider a perturbation $A'_\mu=A_\mu+\sa_\mu$ and compute the action at quadratic order in $\sa_\mu$. 
Covariant derivatives of $\sa_\mu$ will be denoted as $D_\mu \sa_\nu = \partial_\mu \sa_\nu  + [A_\mu , \sa_\nu]$. 
We also add a gauge fixing term  
\beq
S_\text{fix}(a) = \frac12 \int_{S^d} (D_\mu \sa^\mu, D_\nu \sa^\nu)~.
\eeq
This term kills zero modes coming from pure gauge modes. 

For later convenience, we use an orthonormal frame for spacetime indices $a, b, \cdots$ on $S^n$,
so we have e.g. $\sa_a = e_a^\mu \sa_\mu$ where $e^\mu_a$ are the basis of the orthonormal frame, 
\beq
g_{\mu\nu} e^\mu_a e^\nu_b = \delta_{ab}~.
\eeq
Upper and lower indices of $a,b,\cdots$ will not be distinguished. The gauge fixed action for $\sa_a$ after subtracting $S_\text{YM}(A)$ is given by
\beq
S(\sa) &= S_\text{YM}(A+\sa)  -S_\text{YM}(A) +S_\text{fix}(\sa)  \nonumber \\
&= \int_{S^d} \frac{1}{4} \left( F_{ab}+D_a\sa_b -D_b\sa_a +[\sa_a, \sa_b] , F_{ab} + D_a\sa_b -D_b\sa_a+ [\sa_a, \sa_b] \right) \nonumber \\
& \qquad    - \int_{S^d} \frac{1}{4} \left( F_{ab} , F_{ab}  \right)  +  \frac12 \int_{S^d} (D_a \sa_a, D_b \sa_b)     ~.
\eeq
This can be simplified by using
\beq
(D_a D_b - D_b D_a) \sa_c = R_{abcd} \sa_{d} + [F_{ab}, \sa_c]~,
\eeq
where $R_{abcd}$ is the Riemann curvature of $S^n$. We then have 
\beq
S(\sa) = \frac12 \int_{S^n} \Big[ (D_a \sa_b, D_a \sa_b) + R_{ab} (\sa_a, \sa_b) - 2(\sa_a , [ F_{ab}, \sa_b]) \Big] + \cO(\sa^3)~,
\eeq
where we have neglected terms of order $\sa^3$ and higher, and 
$R_{ab} = R_{acbc}$ is the Ricci tensor. Explicitly, on an $n$-sphere of unit radius, we have
\beq
R_{abcd} = \delta_{ac} \delta_{bd} - \delta_{ad}\delta_{bc}~, \qquad R_{ab} = (n-1) \delta_{ab}~.
\eeq

At each point on the sphere $p \in S^n$, let  $\text{Ad}_p $ be the fiber of the gauge bundle in the adjoint representation, and $T_p S^n$ be the tangent bundle. 
For later convenience, we define a linear operator $L $ on $\text{Ad}_p \otimes T_p S^n$ by
\beq
L:\,\, \sa_a \mapsto R_{ab}\sa_b -2[F_{ab}, \sa_b]~.  \label{eq:L-op}
\eeq
In terms of it, the action is
\beq
S(\sa) &=  \frac12 \int_{S^n} \Big[ (D_a \sa_b, D_a \sa_b) +  (\sa_a, L(\sa)_a) \Big] + \cO(\sa^3) \nonumber \\
&=  \frac12 \int_{S^n} \Big[ (\sa_a, \Delta  \sa_a)  \Big] + \cO(\sa^3)
\eeq
where
\beq
\Delta = - D_a D_a + L
\eeq
is a Laplacian.

As we will show below, eigenmodes of the Laplacian $\Delta$ can be decomposed into physical modes and gauge modes as $\sa_a = \sa'_a + D_a \sa''$,
where modes of the form $D_a \sa''$ are gauge modes, and modes $\sa'_a$ satisfying $D_a \sa_a'=0$ are physical modes. 
Let $ \widetilde \Delta = - D_a D_a$ be a Laplacian which acts on scalars (rather than vectors) in the adjoint representation of the gauge bundle.
By a straightforward computation, one can check that
\beq
\Delta (D_a \sa'') = D_a (\widetilde \Delta \sa'')~, \qquad D_a (\Delta \sa_a) = \widetilde \Delta (D_a \sa_a)~. \label{eq:twoLaplacerelation}
\eeq
The first equation implies that $D_a \sa''$ is an eigenmode of $\Delta$ if $\sa''$ is an eigenmode of $\widetilde \Delta$,
so gauge modes can be expanded in terms of eigenmodes of $\widetilde \Delta$. The second equation implies that if $\sa_a$ is an eigenmode of $\Delta$,
then $D_a \sa_a$ is an eigenmode of $\widetilde \Delta$ with the same eigenvalue. 

By using \eqref{eq:twoLaplacerelation}, we can see that any eigenmode $\sa_a$ of $\Delta$ can be decomposed into a physical mode and a gauge mode as follows.
Let $\Delta_0$ be the eigenvalue, $\Delta \sa_a = \Delta_0 \sa_a$. If $\Delta_0$ is nonzero,
then we can take 
\beq
\sa_a = \sa'_a+D_a \sa''~, \qquad \sa'_a=\sa_a + \Delta_0^{-1} D_a (D_b \sa_b)~, \qquad \sa''= -  \Delta_0^{-1} D_b \sa_b~.
\eeq
Then $D_a \sa'_a=0$ and hence $\sa'_a$ is a physical mode. If on the other hand $\Delta_0=0$, then $D_a \sa_a$ is annihilated by $\widetilde \Delta=-D_a D_a$
which implies that $D_a \sa_a$ is covariantly constant, i.e. $D_a (D_b\sa_b)=0$. 
Then the integral of $(D_a\sa_a, D_b\sa_b)$ is zero by integration by parts, and hence we get $D_a \sa_a=0$, which implies that $\sa_a$ is a physical mode.
Therefore, we can assume that each eigenmode of $\Delta$ is either a physical mode or a pure gauge mode.

In this appendix, we are interested in modes for which the eigenvalue of $\Delta$ is negative. For pure gauge modes, the eigenvalues of $\widetilde \Delta= - D_a D_a$ are nonnegative.
Thus, whenever we get a negative eigenvalue, the corresponding mode is always a physical mode.

\subsection{Implications for stability near infinity}\label{sec:radial}
One might at first think that if the Laplacian $\Delta$ of the previous subsection has a negative eigenvalue, then the gauge field configuration $A_a$ is not stable. However, the situation is not so simple if we take into account the radial direction $r$, as we now discuss.

The setup of the problem is as follows. We have a gauge field configuration $A_a$ on $S^n$ which solves the equations of motion. Then we consider perturbations $A_a +\delta A_a$. We impose a boundary condition on $A_a +\delta A_a$ that it goes to the $A_\mu$ at $r \to \infty$, or in other words $\delta A_a \to 0$ at $r \to \infty$. More precisely, we require that the fluctuation $\delta A_a$ is normalizable. Under this condition, the question is whether there is any tachyonic mode. As mentioned in Sec.~\ref{sec:dyn-stab}, the situation is somewhat analogous to the case of the BF bound for a scalar field in AdS space (see \cite{Klebanov:1999tb} for an explanation of the BF bound in the context of AdS/CFT correspondence). Even if the mass term of the scalar field is negative, the scalar field is still stable if we (i) impose an appropriate boundary condition at the AdS boundary, and then (ii) restrict our attention to normalizable modes. We will see a similar situation in our case.

The metric of the whole spacetime, near infinity, is given by
\beq
\d s^2 = \d x^\mu \d x_\mu + \d r^2 + r^2 \d \Omega_n^2~,
\eeq
where $x^\mu$ are the coordinates parallel to the brane, $r$ is the radial direction, and $\d \Omega_n^2$ is the metric of the sphere $S^n$. 
We also assume that the background dilaton is almost constant near infinity and hence we need not take it into account in this analysis.\footnote{The dilaton background becomes important in the throat region since its configuration is nontrivial there (namely, a linear dilaton background). }

Let us consider a physical eigenmode $\sa_a$ of the Laplacian $\Delta = - D_a D_a + L$ on $S^n$ with eigenvalue $\Delta_0$,
\beq
\Delta \sa_a = \Delta_0 \sa_a~, \qquad D_a \sa_a=0~.
\eeq
We now consider a fluctuation $\delta A_a = \phi(x, r) \sa_a$ of the gauge field around the background configuration. 
The action for $\phi(x, r)$ at the quadratic order is given by
\beq
S = - \frac12 \int  \d r\, r^n \left(  r^{-2} \partial_\mu \phi \partial^\mu \phi +  r^{-2} ( \partial_r \phi)^2 + r^{-4} \Delta_0 \phi^2 \right).
\eeq
where we have omitted $ \d^{p+1} x$.
We remark that the condition $D_a \sa_a =0$ can be used to show that $\phi$ does not mix with other components of the gauge field $A_r, A_\mu$. 
Also, possible mixings with metric and dilaton fluctuations are suppressed by $\alpha'^{1/2}$ after canonically normalizing the fields and hence we neglect them
since we are interested in the region $\alpha'^{1/2}/r \to 0$. 

We consider a Fourier mode in the directions $x^\mu$ with momentum $k^\mu$. By denoting $E= \partial_\mu \partial^\mu=- k^2$, the equation of motion is given by
\beq
-r^{-n+2} \partial_r (r^{n-2} \partial_r \phi) + r^{-2} \Delta_0 \phi = E \phi~.
\eeq
From the point of view of the direction $r$, this equation is an eigenvalue equation with eigenvalue $E$ for the differential operator appearing on the left-hand side. On the other hand,
from the point of view of the directions $x^\mu$, the quantity $E=-k^2$ is regarded as the ``mass squared''. 
Therefore, a negative eigenvalue $E<0$ would give a tachyonic instability. 

To get some intuition, it is convenient to rewrite the equation by using 
\beq
\varphi = r^{\frac12 (n-3)} \phi~, \qquad \ell^2 = \Delta_0 + \left( \frac{n-3}{2} \right)^2.
\eeq
Then we get
\beq
S = - \frac12 \int  \d r\, r \left(  \partial_\mu \varphi \partial^\mu \varphi +  (\partial_r \varphi)^2 +  \frac{ \ell^2}{r^2} \varphi^2 \right).
\eeq
and
\beq
-\frac{1}{r}\partial_r ( r \partial_r \varphi) +  \frac{ \ell^2}{r^2}  \varphi = E \varphi~.
\eeq
This is the same eigenvalue equation as the Schr\"odinger equation on a flat two dimensional space with angular momentum $\ell$. 
It is the Bessel equation if we perform the change of variables $ \sqrt{E} r \to r$. 
Therefore, physically we know that negative eigenvalues do not exist if $\ell^2 >  0$. 

More mathematically, we may argue as follows. In the asymptotic regions $r \to \infty$ and $r \to 0$, a solution of the eigenvalue equation behaves as
\beq
\varphi \to \left\{ \begin{array}{ll} 
c_1^{(\infty)} e^{ \sqrt{-E} r} + c_2^{(\infty)} e^{ -\sqrt{-E} r}  \quad&( r \to \infty)~, \vspace{0.3cm} \\
c_1^{(0)} (\sqrt{-E} r)^\ell+ c_2^{(0)} (\sqrt{-E}r)^{-\ell}  \quad & (r \to 0)~. 
\end{array} \right.
\eeq
Suppose that we have an eigenmode with a negative eigenvalue $E<0$. 
We want the solution to be normalizable, so we should choose a solution with $c_1^{(\infty)} =0$. 
After choosing it, there is no freedom except for overall normalization, and we expect that generically both $c_1^{(0)}$ and $ c_2^{(0)} $ are nonzero. (This is indeed confirmed by using the properties of the Bessel function). If $\ell >0$, the term $c_2^{(0)} (\sqrt{-E}r)^{-\ell}$ would grow too rapidly in such a way that
neither $\int \d r r (\partial_r \varphi)^2$ nor $\int \d r r^{-1} \varphi^2$ is convergent in the region $r \to 0$. 
In the context of brane solutions, we do not trust the analysis of this appendix in the region $r \to 0$. However, the divergence of these integrals imply that
these modes (or more precisely their energy densities) are not localized in the large $r$ region; 
they are living in the small $r$ region. Our interest in this appendix is whether or not instability happens in the region $r \to \infty$.
The region $r \to 0$ should be studied by a different method.
Therefore, we do not consider modes which grow like $r^{-\ell} $ in the region $r \to 0$. We conclude that there are no negative eigenvalue modes
which are localized in the large $r$ region if $\ell^2>0$. 

On the other hand, if $\ell^2 <0$ so that $\ell$ is purely imaginary, $(\sqrt{-E} r)^{\pm \ell}$ are oscillatory. 
In this case, a negative $E$ is possible. To be concrete (although a bit artificial), suppose we put a cutoff on $r$ as $r \geq \epsilon$ for some small positive $\epsilon$,
and impose the Dirichlet boundary condition $\varphi|_{r = \epsilon} =0$.  
We seek negative eigenvalues such that $|E| \ll \epsilon^{-2} $ because eigenvalues of order $|E| \sim   \epsilon^{-2}  $ can be an artifact of the cutoff at $r = \epsilon$.
When $\ell$ is positive, the leading term $c_2^{(0)} (\sqrt{-E}r)^{-\ell} $ cannot be zero as mentioned above, and hence there is no solution 
satisfying the boundary condition at $r=\epsilon$.
On the other hand, when $\ell $ is pure imaginary, the leading term in the small $r$ region behaves as 
\beq
\varphi \sim c^{(0)} \sin \left( |\ell | \log (\sqrt{-E} r) +\alpha^{(0)} \right),
\eeq
where $\alpha^{(0)}$ and $c^{(0)}$ are real constants, and we have taken into account the fact that $\varphi$ is real. 
The boundary condition requires $ |\ell | \log (\sqrt{-E}  \epsilon) +\alpha^{(0)} \in \pi \bZ$, and there are infinitely many eigenvalues 
such that $|E| \ll \epsilon^{-2}$. 
We expect that this conclusion holds regardless of the details of the boundary condition in the small $r$ region.\footnote{
More precisely, what is expected to be valid more generally is not the precise eigenvalues, but the qualitative feature of eigenvalue distribution that the density of states per energy, $\rho(E) $,
is given by $\rho(E)\d E \sim -\frac{ |\ell |}{\pi} \d \log \sqrt{-E}$ or $\rho(E) \sim \frac{ |\ell |}{2\pi (-E)}$. 
In the case of the Dirichlet boundary condition, it comes from $ |\ell | \log (\sqrt{-E}  \epsilon) +\alpha^{(0)} \in \pi \bZ$.
At least when $|\ell|$ is large, the density of states can be derived from a semiclassical method.
We compute the phase space volume $\frac{1}{2\pi} \d r \d p$ of the region 
$E \leq H(r,p) \leq E + \d E$, 
where $H(r,p) = p^2 + \ell^2/r^2$ is a classical Hamiltonian, $p $ is the momentum corresponding to $ -\i \partial_r$, 
and we have neglected $\cO(1)$ ambiguities in relating the quantum and classical Hamiltonians which may be justified when $|\ell| \gg 1$.
For the purposes of this appendix, we need to be careful about such $\cO(1)$ terms, and hence the more precise discussions in the main text is required.
}

Therefore, as a condition that there is no tachyonic instability in the region $r \to \infty$, we require
\beq
\ell^2 =  \Delta_0 + \left( \frac{n-3}{2} \right)^2> 0~.  \label{eq:tachyon free}
\eeq

\subsection{Examples}\label{sec:example}
Now we would like to study whether the condition \eqref{eq:tachyon free} is satisfied
in some examples. 

\subsubsection{Embedding the tangent bundle into a gauge group $\SO(N)$}

Let us take a gauge group to be $\SO(N)$, and consider the following gauge field configuration. We assume $m:=N-n>0$, and
then take a subgroup $\SO(n) \times \SO(m) \subset \SO(n+m)$.
The gauge bundle of $\SO(n)$ on $S^n$ is taken to be the same as the tangent bundle $T S^n $ with the explicit connection determined by the Riemannian metric on $S^n$, while  the gauge bundle and connection of $\SO(m)$ is taken to be trivial. 

We denote the indices of $\SO(n)$ and $\SO(m)$ in the fundamental representation by lower and upper case Latin letters $i,j,\cdots$ and $I, J, \cdots$, respectively. If we write gauge indices of $\sa_a$ explicitly, it is denoted as
\beq
\sa_a = \begin{pmatrix} \sa_{aij} & \sa_{a i J} \\ \sa_{aI j} & \sa_{aIJ} \end{pmatrix}, 
\eeq
where
\beq
 \sa_{aij} = -\sa_{aji} ~, \quad \sa_{a i I} = -  \sa_{a I i}  ~,\quad \sa_{a IJ}  = - \sa_{a JI}~.
\eeq
The nonzero components of $F_{ab}$ are
\beq
F_{ab ij} = R_{abij} = \delta_{ai} \delta_{bj} - \delta_{aj}\delta_{bi}~,
\eeq
where we have used the same orthonormal frame for both the tangent bundle and the gauge $\SO(n)$ bundle.
By using it, the operator $ L: \sa_a \mapsto R_{ab}\sa_b -2[F_{ab}, \sa_b]$ defined in \eqref{eq:L-op} can be computed, with the results
\beq
&L:\, \sa_{aij} \mapsto   (n-1) \sa_{aij} -2 \delta_{ai} \sa_{kkj}+2\delta_{aj}\sa_{kki} +2\sa_{iaj} -2\sa_{jai} ~, \\
&L:\, \sa_{aiI} \mapsto   (n-1) \sa_{aiI} -2 \delta_{ai} \sa_{kkI}+2\sa_{iaI}   ~, \\
&L:\, \sa_{aIJ} \mapsto (n-1)\sa_{aIJ} ~.
\eeq

We demonstrate the existence of an instability by using the second equation involving $\sa_{aiI}$. If we take $\sa_{aiI} = \delta_{ai} \sb_I$, then it is an eigenvector of $L$ with a negative eigenvalue,
\beq
L :\, \sb_I \mapsto  -  (n-1) \sb_I~.
\eeq
Moreover, $\sb_I$ is a scalar on $S^n$ because $I$ is an index of a trivial bundle. 
Thus we can minimize the operator $- D_a D_a$ by taking $\sb_I$ to be just a constant on $S^n$. 
Therefore, the minimum eigenvalue of the Laplacian $\Delta = - D_a D_a +L$ acting on $\sb_I$ is $\Delta_0 = - (n-1)$ and hence
\beq
 \Delta_0 + \left( \frac{n-3}{2} \right)^2 = -(n-1) +  \left( \frac{n-3}{2} \right)^2.
\eeq
For instance, if we put $n=8$, one can see that this is negative, implying the existence of a tachyonic mode. 

What happens if $m=0$, i.e. the part $\SO(m)$ is absent?
In this case, we only need to consider $\sa_{aij}$. We will use some of the results for later purposes, so let us also work out this case in detail.
We decompose $\sa_{aij}$ as
\beq
\sa_{aij} = (\delta_{ai} \sb_j - \delta_{aj} \sb_i) + \sc_{aij} + \sd_{aij}~,
\eeq
where $\sd_{aij}$ is an antisymmetric tensor with respect to the three indices $a,i,j$, and $\sc_{aij}$ is such that 
$\sc_{kkj}=0$, $\sc_{aij} = - \sc_{aji}$, and $\sc_{aij} +\sc_{jai} + \sc_{ija}=0$. These conditions for $ \sc_{aij}$ guarantee that it is orthogonal to 
$ (\delta_{ai} \sb_j - \delta_{aj} \sb_i) $ and $ \sd_{aij}$. Now the operator $L$ acts independently on $\sb_i, \sc_{aij}$ and $\sd_{aij}$ as
\beq
&L:\, \sb_{i} \mapsto -(n-3) \sb_{i}~, \\
&L:\, \sc_{aij} \mapsto (n+1) \sc_{aij}~, \\
&L:\, \sd_{aij} \mapsto (n-5) \sd_{aij}~. \label{eq:Lbcd}
\eeq
Let us examine each of them.

First, we immediately see that $\sc_{aij}$ has a positive definite action. Hence there is no tachyonic mode arising from it. 

Next consider $\sb_i$. 
Following harmonic analysis in Hodge theory, we decompose $\sb_i$ as 
\beq
\sb_i = D_i \sb' + \sb''_i~,
\eeq
where $\sb''_i$ represents the modes that are orthogonal to the modes of the form $D_i \sb' $, or more explicitly $D_i \sb''_i=0$. 
For $\sb''_i$, we use
\beq
\int_{S^n} (D_i \sb''_j)(D_i \sb''_j) =  \int_{S^n} \left(\frac12(D_i \sb''_j + D_j \sb''_i)(D_i \sb''_j + D_j \sb''_i) - (D_i \sb''_i)^2 + R_{ij}\sb''_i \sb''_j \right).
\eeq
The second term on the right-hand side, namely $- (D_i \sb''_i)^2$, is zero. The other terms are nonnegative with 
$R_{ij}\sb''_i \sb''_j = (n-1) \sb''_i \sb''_i$. Thus the eigenvalues of $-D_i D_i$ are no less than $(n-1)$, and hence 
\beq
\Delta \geq (n-1) - (n-3) =2~.
\eeq
Thus there are no tachyonic modes arising from $\sb''_i$. 

For $\sb_i = D_i \sb'$, we use
\beq
\int_{S^n} (D_i \sb_j)(D_i \sb_j) 
&=  \int_{S^n} \left(\frac12(D_i \sb_j - D_j \sb_i)(D_i \sb_j - D_j \sb_i) + (D_i \sb_i)^2 - R_{ij}\sb_i \sb_j \right) \nonumber \\
&=  \int_{S^n} \left(  \sb' \Delta_S^2 \sb' - (n-1) \sb' \Delta_S \sb' \right) 
\eeq
where $\Delta_S = -D_i D_i$ acting on scalars. First notice that $\sb'$ cannot be a constant because if it were a constant, then we would have $\sb_i = D_i \sb'=0$. 
Let $\Delta'_1$ be the lowest eigenvalue of nonzero modes of $\Delta_S$. Then, from the above result, we see that the lowest eigenvalue of $-D_iD_i$ acting on 
vectors of the form $\sb_i = D_i \sb'$ is given by $\Delta'_1 - (n-1)$. It is well-known that eigenvalues of $\Delta_S$ are given by 
$\Delta'_k= k (k+n-1)~$ ($k=0,1,2,\cdots$) where $k=0$ is the zero mode.\footnote{
This can be obtained as follows. Consider a flat space $\bR^{n+1}$ with coordinates $X^M~(M=1,\cdots,n+1)$. 
For a constant $k$-th symmetric-traceless tensor $C_{M_1 \cdots M_k}$, we consider a function $\phi(X) = C_{M_1 \cdots M_k} X^{M_1} \cdots X^{M_k}$.
It satisfies $- \partial^2  \phi=0$ on $\bR^{n+1}$ because $C_{M_1 \cdots M_k}$ is constant and traceless. 
In polar coordinates with the radial coordinate $R=\sqrt{X^2}$ and angular coordinates $\Omega^M = X^M/R$, 
we have $\phi(X) =R^{k} \phi(\Omega)$.
Also notice that in polar coordinates the Laplacian is given by $ - \partial^2 \phi = - R^{-n} \partial_R (R^n \partial_R\phi) + R^{-2} \Delta_S \phi$. 
Since $- R^{-n} \partial_R (R^n \partial_R\phi) = -k(k+n-1) R^{-2} \phi$, we get 
$\Delta_S \phi(\Omega) = k(k+n-1)\phi(\Omega)$, as claimed. \label{harmonics}
} 
Therefore, the first nonzero mode has $\Delta'_1=n$ and hence $\Delta'_1 - (n-1) =1$. 
We conclude that the lowest eigenvalue of the original $\Delta$ acting on $\sb_i = D_i \sb'_i$ is\,\footnote{
As a check, when $n=4$, the tangent bundle consists of a single (anti)instanton of each $\SU(2)$ of $\SU(2) \times \SU(2) = \Spin(4)$.
We expect that instanton moduli give zero modes, and indeed $-(n-4)=0$ when $n=4$. The number of instanton moduli of a single $\SU(2)$ on $S^4$ is known to be 5, and hence
we have $10=5+5$ moduli for $\SU(2) \times \SU(2) $. On the other hand, the zero eigenmodes of the form $\sb_i = D_i \sb'$ are in the vector representation of the $\SO(5)$ rotational symmetry of $S^5$ as one can check from the construction of footnote~\ref{harmonics}. Thus $\sb_i$ provides 5 zero modes. The remaining 5 zero modes will be provided by $\sd_{aij}$.
}
\beq
\Delta_0 = 1 -(n-3) = - (n-4).\label{eq:delta0-1}
\eeq
Notice that this is negative for $n=8$.
But to check stability, we must consider
\beq
\Delta_0 +  \left( \frac{n-3}{2} \right)^2 = - (n-4)  + \left( \frac{n-3}{2} \right)^2 = \frac{(n-5)^2}{4}~. \label{eq:delta0-2}
\eeq
This is nonnegative, and hence there is no instability. (The case $n=5$ is marginal, and will not be necessary for the present paper.) 

Finally, let us briefly consider $\sd_{aij}$. For this we need a case-by-case analysis. 
(i) When $n=2$, there is no antisymmetric tensor with three indices and hence $\sd_{aij}=0$. (ii) When $n=3$, it is the same as a scalar $\sd_{aij} = \epsilon_{aij} \sd$ 
where $\epsilon_{aij}$ is the totally antisymmetric tensor in $n=3$ dimensions. The lowest eigenvalue comes from constant $\sd$, and from \eqref{eq:Lbcd} one gets a tachyonic mode. (iii) When $n=4$,
it is the same as a vector $\sd_{aij} = \epsilon_{ijab} \sd_b$ where $\epsilon_{ijab}$ is the totally antisymmetric tensor in $n=4$ dimensions.
In this case the analysis is the same as $\sb_i$, and there is no tachyonic mode. (iv) When $n \geq 5$, there is no tachyonic mode.

\subsubsection{Embedding a spin bundle into a gauge group $\SO(N)$}
Here we consider some real spin bundles on $S^n$ associated to the tangent bundle. Let $d_n$ be the dimension of the spinor representation.
For $N = d_n +m$ with $m>0$, we embed the spin bundle into $\SO(d_n) \subset \SO(d_n) \times \SO(m) \subset \SO(d_n+m)$.
In this case, $\sa_a$ is of the form
\beq
\sa_a = \begin{pmatrix} \widehat \sa_{a} & \widetilde \sa_{a J} \\ - ( \widetilde \sa_{aI} )^T& \sa_{aIJ} \end{pmatrix}, 
\eeq
where two spinor indices for $ \widehat \sa_{a}$ and one spinor index for $\widetilde \sa_{aI}$ are implicit, and we regard them as matrices and vectors in the space of spinors, respectively. 

Let $\Gamma_i$ be gamma matrices for the space of spinors, and $\Gamma_{ij} = \frac12 (\Gamma_i \Gamma_j - \Gamma_j \Gamma_i)$. The nonzero components of $F_{ab}$ are
\beq
F_{ab } = \frac{1}{4} \Gamma_{ij} R_{abij} = \frac12 \Gamma_{ab}~.
\eeq
By using this, we obtain
\beq
&L:\, \widehat \sa_{a} \mapsto  (n-1) \widehat\sa_{a}  -[\Gamma_{ab} , \widehat\sa_{b}]   ~, \\
&L:\, \widetilde \sa_{aI} \mapsto    (n-1) \widetilde \sa_{aI} -\Gamma_{ab} \widetilde \sa_{bI}   ~, \\
&L:\, \sa_{aIJ} \mapsto (n-1)\sa_{aIJ} ~.
\eeq

Let us first study the second equation involving $\widetilde \sa_{aI}$. For this purpose, we decompose it into irreducible representations as
\beq
\widetilde \sa_{aI} = \Gamma_{a} \widetilde \sb_I + \widetilde \sc_{aI}
\eeq
where $\widetilde \sc_{aI}$ is such that $\Gamma_a  \widetilde \sc_{aI} =0$. By using $\{\Gamma_a, \Gamma_b\}=2\delta_{ab}$, we get
\beq
\Gamma_{ab} = \Gamma_a \Gamma_b - \delta_{ab}~, \qquad     \Gamma_{ab}  \Gamma_b =  (n-1) \Gamma_a~.
\eeq
Then $L$ acts on $\widetilde \sb_I $ and $\widetilde \sc_{aI}$ as
\beq
&L:\, \widetilde \sb_{I} \mapsto  0~, \\
&L :\, \widetilde \sc_{aI} \mapsto n \widetilde \sc_{aI}~.
\eeq
Thus, the eigenvalues of $L$ are nonnegative. Moreover, $- D_i D_i$ is positive definite on $\widetilde \sb_{I} $ since
there are no covariantly constant spinors on $S^n$.\footnote{If $\widetilde \sb_{I} $ were covariantly constant, i.e., $D_a \widetilde \sb_{I} =0$, we would have $0=[D_a, D_b] \widetilde \sb_{I} = \frac12 \Gamma_{ab} \widetilde \sb_{I} $ which would be impossible unless $\widetilde \sb_{I} =0$. }
We conclude that $\Delta$ is positive definite and there is no tachyonic mode in the direction $\widetilde \sa_{aI}$. 

The stability in the direction $\widehat \sa_{a}$ depends on $n$ and a choice of the real spinor representation. 
For our purposes, the most important case is $n=8$ and the representation is one of the irreducible spinor representations with dimension 8. 
This is the case that is relevant to the 0-brane.
In this case, we can use the following argument.
By the triality of $\so(8)$, the embedding of the tangent bundle and either of the irreducible spinor bundles is equivalent for $\so(8)$.
We have already shown in the previous subsection that $\sa_{aij}$ has no tachyonic mode. Therefore, for $n=8$, $\widehat \sa_a$ also does not have any tachyonic mode.

Let us recapitulate the situation. As we have seen in \eqref{eq:delta0-1} and \eqref{eq:delta0-2}, the lowest eigenvalue $\Delta_0$ has the properties that
\beq
\Delta_0 = -(n-4)=-4<0, \qquad \Delta_0 +  \left( \frac{n-3}{2} \right)^2 =\frac{(n-5)^2}{4}=\frac{9}{4}>0~.
\eeq
Thus, the negative mode on $S^8$ is stabilized by the effect of the radial direction $r$.

We also remark that the triality is not preserved when $\SO(8)$ is realized as a subgroup $\SO(8) \subset \SO(N)$, so the above argument cannot be used for $\widetilde \sa_{aI} $.
In fact, we have seen in the previous section that $\sa_{aiI}$ has a tachyonic mode, while we have shown in this subsection that $\widetilde \sa_{aI} $ does not have a tachyonic mode.

\section{Some properties of chiral CFTs}
\label{app:CFTsection}

In this appendix, we will review basic facts about chiral CFTs, including a derivation of the equivalence 
\begin{equation}
G_1 = [H_k \times \SO(n)_1]/(-1)^{\sF_\sL}~,\label{eq:equivalence-app}
\end{equation}
where the list of $(G,H,k,n)$ we use is given in Table \ref{table:equivalence-app}.
The same equivalences were known long ago, and recently re-derived using a unified construction in \cite{BoyleSmith:2023xkd}. 
Here we provide a slightly different perspective; we will also be more careful about the global form of the groups involved.

\begin{table}
  \centering
  \begin{tabular}{c||c|c|c|c}
  $G$ & $\hete$ & $\hets$ & $\hete$ & $\hets$   \\ 
  \hline
  \hline
  $H$ & $E_8$ & $\SU(16)/\bZ_4$ & $(E_7 \times E_7)/\bZ_2$ & $\Spin(24)/\bZ_2$ \\
  \hline
  $k$ & $2$ & $1$ & $1$ & $1$  \\
  \hline
  $n$ & $1$ & $2$ & $4$ & $8$  \\
  \end{tabular}
  \caption{The combinations of $(G, H, k, n)$ for which the equivalence \eqref{eq:equivalence-app} holds. 
  }
  \label{table:equivalence-app}
  \end{table}

\subsection{Some chiral CFTs}\label{sec:someCFT}

We first recall some basic facts about Narain CFTs. For our purposes, we will only be interested in chiral CFTs. Consider an $N$-dimensional lattice $\bL \subset \bR^N$ with the inner product $(\bullet, \bullet)$ defined by the one induced from $\bR^N$. Let $\bL^*$ be the dual lattice, consisting of elements $ w \in \mathbb{R}^N$ such that for any $v \in \bL$ the inner product satisfies $(v, w) \in \bZ$. A (chiral) Narain CFT consists of $N$ periodic left-moving bosons $\phi=(\phi_i)_{i=1,\cdots,N}$ with periodicity
\beq
\phi \sim \phi + 2\pi  w~, \qquad w \in \bL^*~.
\eeq
In other words, operators $\exp( \i v \cdot \phi)$ for $v \in \bL$ are allowed, where $v \cdot \phi :=(v,\phi)$. If we combined left- and right-moving bosons, the action would be
\beq
S = \frac{1}{8\pi} \int \d^2 \sigma \partial_\alpha \phi \partial^\alpha \phi~.
\eeq
We say that a lattice $\bL$ is self-dual or unimodular if $\bL^* = \bL$. 
Furthermore, a lattice is said to be even when  we have $v^2:=(v,v) \in 2\bZ$ for any $v \in \bL$. 

If we include spin structures in the data of Riemann surfaces, a Narain CFT is modular invariant (modulo gravitational anomalies) if the lattice $\bL$ is self-dual. For instance, the operator $\exp( \i v \cdot \phi)$ has left and right scaling dimensions $(h_L, h_R) = (\frac{1}{2} v^2, 0)$, and hence its spin is $h_L - h_R=\frac{1}{2} v^2$. This is half-integer or integer, i.e. $\frac12 v^2 \in \frac12 \bZ$, because $v \in \bL = \bL^*$, and hence $v^2 \in \bZ$. We refer to CFTs that depend on spin structures as spin-CFTs. If we further assume that $\bL$ is even, then there is no dependence on spin structures. For instance, for even $\bL$ the spin of the operator $\exp( \i v \cdot \phi)$ is $\frac{1}{2} v^2 \in \bZ$. We refer to CFTs that do not depend on spin structures as bosonic CFTs. 

Suppose we are given a Lie group $H$. Except for $\hete$, we will consider only connected groups (i.e. $\pi_0(H)=0$), and for $\hete$ we use the connected component containing the identity in the following discussion. The rank of $H$ will be denoted by $N$. All possible weight vectors of $H$ form a lattice $\bL \subset \bR^N$, where the inner product is normalized such that root vectors have squared length 2. Here, for weight vectors we only consider genuine representations of $H$, rather than representations of its Lie algebra $\mathfrak{h}$. For instance, in $\SO(3) = \SU(2)/\bZ_2$, half-integer spin representations are not allowed because they change sign under a ``$2\pi$-rotation''. If the weight lattice is self-dual, it gives a Narain CFT with a global symmetry (at least) $H$. This is a current algebra theory at level 1, and we denote it by $H_1$. Physical states of the CFT on $S^1$ with Neveu-Schwarz (NS) spin structure are genuine representations of $H$. For instance, these states include the ones corresponding to operators $\exp( \i v \cdot \phi)$ for $v \in \bL$. On the other hand, there is possibly a mixed anomaly between $H$ and worldsheet Lorentz symmetry, and in that case, states in the Ramond (R) spin structure are in projective representations of $H$.

Let us give examples which are relevant to the present paper. In general, it is known (e.g. from a consideration of gravitational anomalies)\footnote{The anomaly polynomial is given by $\frac{c_L-c_R}{24} p_1$, where $p_1$ is the first Pontryagin class. On 4-manifolds without assuming spin structure, only the integral of $p_1/3$ is integer-valued (which is shown by signature index theorem).  
We conclude that, for bosonic theories, we need $(c_L-c_R)/8 \in \bZ$
for the integrality of the anomaly polynomial.} that a CFT with $c_L-c_R \notin 8\bZ$ is a spin-CFT. The following examples are all such spin-CFTs.  

\paragraph{{$H=\SU(16)/\bZ_4$} :} Let $N$ and $M$ be positive integers. In general, one can check that the weight lattice of $\SU(N M)/\bZ_N$ and the weight lattice of $\SU(NM)/\bZ_M$ are dual to each other. Applying this fact to the case $N=M=4$, we see that the weight lattice of $\SU(16)/\bZ_4$ is self-dual. This lattice is not even, as can be seen explicitly as follows. By using $\su(NM) \subset \u(NM)$, we can represent weights by using the subspace of $\bR^{NM}$ consisting of $(\sx_1,\cdots,\sx_{NM}) \in \bR^{NM}$ with $\sum_{i=1}^{NM} \sx_i =0$. In $\SU(NM)/\bZ_N$, the $N$-th antisymmetric tensor representation is allowed, and has highest weight $v_N$ given by
\beq
v_N=(1, \cdots, 1,0,\cdots,0) - \frac{1}{M}(1,\cdots,1)~,
\eeq
where in the first term  $1$ is repeated $N$ times, and the second term is required by the condition $\sum_{i=1}^{NM} \sx_i =0$ mentioned above. We have $v_N^2=\frac{N(M-1) }{M}$, and when $N=M=4$ this is an odd integer, $v_3^2=3$.

\paragraph{$H=(E_7 \times E_7)/\bZ_2$ :}
It is known that the center of $E_7$ is $\bZ_2 \subset E_7$, and this $\bZ_2$ acts nontrivially on the $56$-dimensional representation ${\bf 56}$ of $E_7$. Let $\bL_r$ and $\bL_w$ be the root and weight lattices of $E_7$, and let $v_{\bf 56}$ be the highest weight of the ${\bf 56}$. The weight lattice $\bL_w$ is obtained from the root lattice $\bL_r$ by adding $v_{\bf 56}$ as an additional generator, and the two are dual to each other, i.e. $\bL_w^* = \bL_r$. The length squared of $v_{\bf 56}$ is a half-integer, $v_{\bf 56}^2 \in \frac12 + \bZ$ (otherwise we would have $\bL_w=\bL_r$, which is not true). 
More explicitly, one can compute $v_{\bf 56}^2 = \frac32$.\footnote{
For instance, consider the subalgebra $\su(2) \times \so(12) \subset \e_7$ under which ${\bf 56}$ is decomposed as $({\bf 1} \otimes {\bf 32}) \oplus ({\bf 2} \otimes {\bf 12})$. The spinor representation ${\bf 32}$ of $\so(12)$ has weight vectors whose length squared is $3/2$. The representations ${\bf 2}$ of $\su(2)$ and ${\bf 12}$ of $\so(12)$ have weight vectors with length squared $1/2$ and $1$, respectively, and hence $({\bf 2} \otimes {\bf 12})$ has weight vectors whose length squared is $1/2 +1 =3/2$.
} 
The weight lattice $\bL$ of $(E_7 \times E_7)/\bZ_2$ is obtained by adding to $\bL_r \oplus \bL_r$ an additional generator $v_{\bf 56} \oplus v_{\bf 56}$, upon which one can see that $\bL$ is self-dual. It is not even since $(v_{\bf 56} \oplus v_{\bf 56})^2=2v_{\bf 56}^2 =3$ is an odd integer.

\paragraph{$H=\Spin(24)/\bZ_2$ :} We take the $\bZ_2 \subset \Spin(24)$ appearing in the quotient to be the one which acts trivially on one of the spinor representations, $s$, and nontrivially on the other one, $c$. Let $N$ be a positive integer. In general, the fact that the weight lattice of $\Spin(8N)/\bZ_2$ is self-dual can be shown in the same way as the standard fact of string theory that the weight lattice of $\hets$ is self-dual. Let $v_s$ be the highest weight of the spinor representation $s$ of $\Spin(8N)/\bZ_2$,
\beq
v_s = \left(\frac12, \cdots, \frac12\right)~.
\eeq
Its length is given by $v_s^2=N$, and it is an odd integer for $N=3$, namely $v_s^2=3$. Hence the weight lattice of $\Spin(24)/\bZ_2$ is not even. 

Incidentally, as is well-known, the cases $N=1$ and $2$ also play important roles in string theory. Let $\bZ'_2 \subset \Spin(8N)$ be another subgroup which acts trivially on $c$ and nontrivially on $s$. For $N=1$, the lattices of $\Spin(8)/\bZ_2$, $\Spin(8)/\bZ'_2$, and $\SO(8)$ are all equivalent, and it is the triality of $\so(8)$. For $N=2$, the theory based on $\Spin(16)/\bZ_2$ has the enhanced symmetry $E_8$.
\\

All of the above examples are current algebra theories at level 1 realized as Narain CFTs.
Given a group $H$, we denote its current algebra theory at level $k$ by $H_k$. The chiral central charge $c_L$ of the theory $H_k$ is given by the Sugawara central charge
\beq
c_L = \frac{k\dim H}{h^\vee +k}~,\label{eq:sugawara}
\eeq
where $h^\vee$ is the dual Coxeter number of $H$. 
Note that for a simply laced $\h$ at level $k=1$, the Sugawara central charge \eqref{eq:sugawara} is just equal to the rank, as is evident from the Narain construction.\footnote{%
This leads to a curious equality $(h^\vee +1)\mathop{\mathrm{rank}}G=\dim G$.}

It is not always true that we have a purely left-moving spin-CFT for a given $H$ and $k$.
Still, for some choices of $H$ and $k$, we do get a spin-CFT. In the examples above, we have $k=1$ and $H=\SU(16)/\bZ_4,~(E_7 \times E_7)/\bZ_2,$ and $\Spin(24)/\bZ_2$.

We would like to mention two other examples:

\paragraph{$(E_8)_2$ :} That this is a spin-CFT may be seen as follows. First note that the current algebra $(E_8)_1 \times (E_8)_1$ has a subalgebra $(E_8)_2$ associated to the diagonal subgroup $E_8 \subset E_8 \times E_8$. By using the fact that $\dim E_8 = 248$ and $h^\vee=30$ for $E_8$, the chiral central charge for $(E_8)_2$ is given by $\frac{31}{2}$. Then the coset $[ (E_8)_1 \times (E_8)_1]/(E_8)_2$ has chiral central charge $8+8 - \frac{31}{2} = \frac12$. This must be a single Majorana-Weyl fermion (by the classification of minimal models), which is a spin-CFT. Therefore, $(E_8)_2$ should also be a spin-CFT, since it should be possible to combine the Majorana-Weyl fermion and $(E_8)_2$ to give the bosonic CFT $(E_8)_1 \times (E_8)_1$. We discuss their relation later.

\paragraph{$\SO(n)_1$ :} This current algebra theory is realized by $n$ Majorana-Weyl fermions. We define $\SO(1)_1$ to be a single Majorana-Weyl fermion. These are obviously spin-CFTs.

\subsection{Equivalences of chiral CFTs}
\label{sec:equivalenceintro}

We now discuss the current algebra CFTs $G_1$ for $G=\hets$ and $\hete$
and establish the equivalence \eqref{eq:equivalence-app}, which we repeat here for convenience:
\begin{equation}
G_1 = [H_k \times \SO(n)_1]/(-1)^{\sF_\sL}~.
\end{equation}
The meaning of the notation $/(-1)^{\sF_\sL}$ is to gauge the $\bZ_2$ symmetry associated to the fermion parity of the spin-CFT, as discussed below. More concretely, these equivalences can be understood as follows.

Let us first consider the equivalences at the level of the algebra, neglecting various global structures. For the algebra $\g$ of $G$, we can find a subalgebra 
\beq
\h \times \so(n) \subset \g
\eeq
where $\h$ and $\so(n)$ are defined as follows:
\begin{itemize}
\item $n=1$: We have the diagonal subalgebra $\e_8 \subset \e_8 \times \e_8$. Thus in this case $\h = \e_8$, and instead of $\so(1)$ we consider a single Majorana-Weyl fermion as mentioned above.
\item $n=2$: We have $\su(16) \times \u(1) = \u(16) \subset \so(32)$. Thus in this case $\h = \su(16)$ and we identify $\so(2) =\u(1)$. 
\item $n=4$: We have $\e_7 \times \su(2) \subset \e_8$ and hence $\e_7 \times \e_7 \times \su(2) \times \su(2) \subset \e_8 \times \e_8$. Thus in this case $\h = \e_7 \times \e_7$ and we identify $\so(4) =\su(2) \times \su(2)$.
\item $n=8$: We have $\so(24) \times \so(8) \subset \so(32)$. Let us mention a well-known fact that $\so(8)$ has an automorphism $\so(8) \xrightarrow{\sim} \so(8)$ under which a vector representation is mapped to one of the spinor representations (say $s$). This is the triality of $\so(8)$.
\end{itemize}
One can compute the Sugawara central charge of the current subalgebra $\h \times \so(n)$ given by the formula \eqref{eq:sugawara}.
The sum of the central charges of the $\h$ part and the $\so(n)$ part saturates the central charge of the full theory $G_1$. 
This saturation means that the current algebra theories for $\g$ can be decomposed in terms of a finite number of irreducible representations of the current subalgebra $\h \times \so(n)$.

Including global structure, we argue as follows. In the previous subsection, we discussed the theories $H_k$ and $\SO(n)_1$ which appear in Table~\ref{table:equivalence}. There, we found that all of them are spin-CFTs. Thus the theory $H_k \times \SO(n)_1$ is also a spin-CFT.

In general, suppose that we have a spin-CFT whose chiral central charge $c_L - c_R$ is a multiple of $8$. Then we can obtain a bosonic theory by gauging the $\bZ_2$ symmetry generated by $(-1)^{\sF_\sL}$, where $(-1)^{\sF_\sL}$ is fermion parity, which is roughly the ``$2\pi$-rotation''. The condition that the chiral central charge must be a multiple of $8$ can be understood as follows. Naively, one might at first think that from any spin-CFT we can obtain a bosonic CFT by gauging the fermion parity $\bZ_2$ symmetry. However, there is a subtlety about gravitational anomalies. When gravitational anomalies exist, partition functions on Riemann surfaces are ill-defined. This ill-definedness forbids the simple argument that gauging the $\bZ_2$ symmetry gives a bosonic theory.\footnote{ Incidentally, it is known that the $\bZ_2$ symmetry is anomaly-free under a less restrictive condition that $c_L -c_R \in 4\bZ$. When $c_L - c_R \in 4+8\bZ$, the theory obtained after gauging the $\bZ_2$ is again a spin-CFT rather than a bosonic one.  
For more details, we refer the readers to \cite{BoyleSmith:2024qgx}.} Recall that the gravitational anomaly of a CFT is determined by the chiral central charge $c_L-c_R$. To avoid gravitational anomalies, we add a ``spectator CFT'' given by some copies of the bosonic CFT $(E_8)_1$ or its orientation reversal. If $c_L -c_R$ is a multiple of 8, we can cancel the gravitational anomaly by such a spectator bosonic CFT. Then the subtlety about gravitational anomalies disappears, and we can now just gauge the $\bZ_2$ symmetry to obtain a bosonic CFT. The bosonic spectator CFT is not involved in this $\bZ_2$ gauging, so we conclude that we can gauge the $\bZ_2$ symmetry of the original CFT to get a bosonic CFT.

In our case, the chiral central charge of $H_k \times \SO(n)_1$ is $16$, and hence it is possible to gauge the fermion parity to get a bosonic CFT. Such a bosonic chiral CFT with $c_L=16$ must be either $[\hets]_1$ or $[\hete]_1$, and which one it is is just determined by the algebra that we have already discussed. This concludes our derivation of the equivalences in \eqref{eq:equivalence}.

\section{A mod-2 index}
\label{app:mod2}
In this appendix, we develop a mod-2 index used in Sec.~\ref{sec:vacuumstructure-7brane}.
The theory considered there was a theory with a single left-moving fermion $\psi$ 
and a single \Nequals{(0,1)} chiral multiplet $\Phi=\{\phi,\tilde\psi\}$ consisting of a scalar $\phi$ and a right-moving fermion $\tilde\psi$.
Let $(-1)^\sF$ and $(-1)^{\sF_\sL}$ be the total and left-moving fermion parity symmetries, respectively. 
We are going to show that
\begin{itemize}
\item vacua always come in pairs exchanged by $(-1)^{\sF_\sL}$, and
\item the number of pairs of vacua modulo 2 is invariant under any continuous deformation.
\end{itemize}
In fact, the following mod-2 index can be defined not only for the above theory, but also for any theory under the following assumptions.
\begin{enumerate}
\item The theory has ${\cal N}=(0,1)$ supersymmetry without pure gravitational anomalies.
\item The theory also has an additional $\bZ_2$ symmetry whose generator is denoted by $(-1)^{\sF_\sL}$.
\item The $\bZ_2=\{1,(-1)^{\sF_\sL} \}$ has an anomaly such that the $(-1)^{\sF_\sL}$ is part of the algebra given by \eqref{eq:Calgebra}, \eqref{eq:Calgebra2} and \eqref{eq:Calgebra3} discussed below.
\end{enumerate}
One class of examples is the $\cN=(1,1)$ sigma models with the target space $S^n$ regarded as $\cN=(0,1)$ theory, when $n \equiv 1 \mod 4$. The reason that $n \mod 4$ is relevant is as follows. In the following analysis, we only use the reduction of the theory to one-dimensional quantum mechanics with a time-reversal symmetry which comes from the two-dimensional CPT symmetry. The relevant anomalies in this case are known to be classified by the reduced ${\rm pin}^-$ bordism group for the classifying space $B\bZ_2$ of the group $\bZ_2$, which is given by $\widetilde \Omega^{{\rm pin}^-}_2(B\bZ_2) \simeq \bZ_4$~\cite[Theorem~16]{Guo:2018vij}. The anomaly of the $S^n$ sigma model is determined by $n \mod 4$.\footnote{More explicitly, the relevance of $n \mod 4$ for the $S^n$ sigma models may be seen as follows. In footnote~\ref{footnote:harmonic}, we have seen that $(-1)^{\sF_\sL}$ may be identified as $(-1)^{\sF_\sL}=\i^{\frac12 n(n+1)} \star (-1)^{\frac12 \sP(\sP+1)}$ in the space of differential forms. Later we will also consider CPT symmetry $\mathsf{CPT}$ which may be identified with just complex conjugation in the space of differential forms. Because of the imaginary factor $\i^{\frac12 n(n+1)} $, we get $\mathsf{CPT}(-1)^{\sF_\sL} = (-1)^{\frac12 n(n+1)}(-1)^{\sF_\sL}  \mathsf{CPT}$. The sign factor $(-1)^{\frac12 n(n+1)}$ depends on $n \mod 4$. }

Now we go to the definition of the mod-2 index.
The analysis below is similar to the one in the Appendix of \cite{Tachikawa:2023nne}.


Consider the theory on $S^1_W$ wth R spin structure. 
Let $Q$ be the supercharge for the ${\cal N}=(0,1)$ supersymmetry, and let $(-1)^{\sF_\sR}$ be the right-moving fermion parity which is obtained by the composition of $(-1)^\sF$ and $(-1)^{\sF_\sL}$.
The algebra generated by $Q, (-1)^\sF$, $(-1)^{\sF_\sL}$ and $(-1)^{\sF_\sR}$ is as follows,
\begin{equation}
  \begin{aligned}
&   (-1)^\sF  = \i (-1)^{\sF_\sL} \cdot (-1)^{\sF_\sR}~, \\
&    \{Q , (-1)^\sF\} = \{Q, (-1)^{\sF_\sR}\}=0~, \quad [Q, (-1)^{\sF_\sL}]=0~,\\
&     \{(-1)^{\sF_\sL}, (-1)^{\sF_\sR}\}  = \{(-1)^\sF, (-1)^{\sF_\sR}\} =  \{(-1)^\sF, (-1)^{\sF_\sL}\} =  0~,\\
 &   ((-1)^\sF)^2= ((-1)^{\sF_\sR})^2= ((-1)^{\sF_\sL})^2=1~.
    \end{aligned}\label{eq:Calgebra}      
\end{equation}
Among these relations, the third relation that $(-1)^{\sF_\sL} $, $ (-1)^{\sF_\sR}$ and $(-1)^\sF$ anti-commute (rather than commute) requires explanation. (The imaginary unit $\i$ in the first relation is necessary to make $(-1)^{\sF_\sL} $, $ (-1)^{\sF_\sR}$ and $(-1)^\sF$ hermitian when they anti-commute with each other.)
This is due to an anomaly, and we may check it in the simplest case of the theory $\psi, \Phi$ without any superpotential as follows. In this case, the fermions $\psi$ and $\tilde \psi$ are free and can be quantized straightforwardly. The only subtlety comes from the zero modes $\psi_0$ and $\tilde \psi_0$ of these fermions on $S^1_W$. After normalizing them appropriately, quantization of the zero modes gives the algebra 
\beq
\psi_0^2=\tilde \psi_0^2 =1~, \qquad \{\psi_0, \tilde \psi_0\} =0~,
\eeq
which is just a Clifford algebra.
The irreducible representation of this algebra has two states $\ket{0}$ and $\ket{1}$ characterized by
\beq
(\psi_0 + \si \tilde \psi_1)\ket{0}=0~, \qquad \ket{1} = \frac{1}{2}(\psi_0 - \si \tilde \psi_1)\ket{0}~.
\eeq
The symmetry $(-1)^{\sF_\sR}$ changes the sign of $\tilde \psi_0$, and hence it must exchange these two states (possibly up to a phase factor which is irrelevant for the current discussion),
\beq
(-1)^{\sF_\sR} \ket{0} \propto \ket{1}~, \qquad (-1)^{\sF_\sR} \ket{1} \propto \ket{0}~.
\eeq
On the other hand, the state $ \ket{1}$ is obtained by acting with the fermionic operator $ (\psi_0 - \si \tilde \psi_1)$ on $\ket{0}$, so if $\ket{0}$ is (say) a bosonic state, then $\ket{1}$ is fermionic,
\beq
(-1)^\sF \ket{0} = \ket{0}~, \qquad (-1)^\sF \ket{1} = -\ket{1}~.
\eeq
Thus we see that $(-1)^{\sF_\sR}$ sends a bosonic state to a fermionic state and vise versa. Even more explicitly, if we only take into account the zero modes and completely neglect nonzero modes, we can just take
\beq
(-1)^{\sF_\sR} = \psi_0~, \qquad (-1)^{\sF_\sL} = \tilde \psi_0~, \qquad (-1)^\sF = - \i  \psi_0 \tilde \psi_0~. \label{eq:explicitgenerator}
\eeq
Therefore, we conclude that $(-1)^{\sF_\sL} $, $(-1)^{\sF_\sR} $ and $(-1)^\sF$ anti-commute. This property can be interpreted as a mixed anomaly between the worldsheet Lorentz symmetry (in particular $(-1)^\sF$) and the symmetry $\bZ_2=\{1, (-1)^{\sF_\sL} \}$, and hence it should be robust under deformations (e.g. by $F(\widehat \phi  )$).\footnote{The anomalies of two-dimensional theories are elements of $\Omega^\text{spin}_3(B\bZ_2) \simeq \bZ_8$ where $\bZ_2$ is the one generated by $(-1)^{\sF_\sL}$. However, the discussions of this appendix only use the reduction to one-dimensional quantum mechanics and the anomalies there are elements of $\widetilde \Omega^{{\rm pin}^-}_2(B\bZ_2) \simeq \bZ_4$ as mentioned before. Analogous modifications of symmetry algebras by anomalies are studied in e.g. \cite{Delmastro:2021xox}.}

In addition to the symmetries mentioned above, we also have the CPT symmetry $\mathsf{CPT}$ on the worldsheet. It acts on Majorana fermions as
\beq
\mathsf{CPT}\psi(\sigma) \mathsf{CPT}^\dagger = \psi(-\sigma)~, \qquad \mathsf{CPT} \tilde \psi(\sigma) \mathsf{CPT}^\dagger  = - \tilde \psi(-\sigma)~.
\eeq
The relative sign difference is required by the fact that $\si \psi \tilde \psi$ is a real scalar operator, while the absolute sign is our choice of convention, since it can be changed by a redefinition $\mathsf{CPT} \to \mathsf{CPT}(-1)^\sF$. From \eqref{eq:explicitgenerator}, we see that
\beq
[\mathsf{CPT}, (-1)^\sF] = [\mathsf{CPT}, (-1)^{\sF_\sR} ]=0, \quad \{\mathsf{CPT}, (-1)^{\sF_\sL}\}=0~. \label{eq:Calgebra2}
\eeq
The supercharge is given by $Q =  \dot{\phi} \tilde\psi + \cdots$, where 
$\dot{\phi}$ is the time derivative of $\phi$ and hence changes sign under $\mathsf{CPT}$. The field $\tilde \psi$ also changes sign, and we find
\beq
[\mathsf{CPT} , Q]=0~, \qquad \mathsf{CPT}^2=1~, \label{eq:Calgebra3}
\eeq
where the second equation $\mathsf{CPT}^2=1$ is just a general fact in two dimensional theories without gravitational anomalies (see e.g. \cite{Delmastro:2021xox,Tachikawa:2023nne} for how it is modified when there are pure gravitational anomalies.).

We have demonstrated the symmetry algebra relations \eqref{eq:Calgebra}, \eqref{eq:Calgebra2} and \eqref{eq:Calgebra3} by using the theory $\psi, \Phi$ as an explicit example. The mod-2 index is defined as long as the theory has anomalies which lead to this symmetry algebra. 

 $\mathsf{CPT}$ is an antilinear operator whose square is 1. Therefore, it gives the Hilbert space the structure of a real vector space. More precisely, we define a real vector space $\cH_\bR$ to be the subspace of the Hilbert space $\cH$ that is invariant under $\mathsf{CPT}$, 
\beq
\cH_\bR = \{ \ket{\psi} \in \cH~|~ \mathsf{CPT} \ket{\psi} = \ket{\psi} \}~. 
\eeq
Then, $\cH$ is the complexification of $\cH_\bR$, i.e. $\cH = \bC \otimes \cH_\bR$.

Note that the square $Q^2$ is the right-moving Hamiltonian $H_R = (H-P)/2$, where $H$ and $P$ are the Hamiltonian and the momentum. Let us consider  the eigenspace of $Q^2$ with eigenvalue $E$. When $E>0$, we denote
\beq
  \se_0 :=(-1)^\sF~, \qquad \se_1 := (-1)^{\sF_\sR}~, \qquad \se_2:= \frac{Q}{\sqrt{E}}~. \label{eq:se}
\eeq
They satisfy the Clifford algebra with three generators,
\beq
\{\se_i, \se_j\} = 2 \delta_{ij}~, \qquad i,j = 0,1,2~. \label{eq:realclifford}
\eeq
When $E=0$, we only consider the two generators $\se_i$ for $i=0,1$, since $Q=0$ when $E=0$. 
The $\se_i$ commute with $\mathsf{CPT}$,
\beq
[\mathsf{CPT}, \se_i]=0~.
\eeq
Then the Clifford algebra generators $\se_i$ must be represented on $\cH_{\bR}$ by real matrices. 

We now use the algebra above to define the mod-2 Witten index:
\begin{itemize}
\item 
When $E=0$, the irreducible representation of the algebra \eqref{eq:realclifford} is given by $\se_0=\sigma_3$ and $\se_1=\sigma_1$, where $\sigma_1, \sigma_2, \sigma_3$ are the usual Pauli matrices,
\beq
\sigma_1 = \begin{pmatrix}0&1\\1&0\end{pmatrix}~, \qquad  \sigma_2 = \begin{pmatrix}0&-\si\\ \si&0\end{pmatrix}~, \qquad \sigma_3 = \begin{pmatrix}1&0\\0&-1\end{pmatrix}~.
\eeq
Notice that $\sigma_3$ and $\sigma_1$ are real matrices. If we denote the two states as $\ket{0}$ and $\ket{1}$, then by recalling the definitions \eqref{eq:se} we have
\beq
(-1)^\sF \ket{0}=\ket{0}, \quad (-1)^\sF\ket{1} = - \ket{1}, \quad  (-1)^{\sF_\sR}\ket{0}=\ket{1}, \quad  (-1)^{\sF_\sR}\ket{1} = \ket{0}. \label{eq:vacstructure}
\eeq
Therefore, the number of supersymmetric vacua $N_\text{vacua}$ is a multiple of 2, and these vacua consist of pairs satisfying \eqref{eq:vacstructure}. 
\item
In contrast, when $E>0$, the irreducible representation of the Clifford algebra is given by $\se_0=\sigma_3$, $\se_1=\sigma_1$, and $\se_2=\sigma_2$ if we forget the reality condition. However, $\sigma_2$ is a complex matrix and hence the representation space is complex 2-dimensional. To obtain a real representation, we need to take the underlying real vector space with dimension $4 = 2 \times 2$. 
\end{itemize}
Now note that
\beq
\frac{N_\text{vacua}}{2} \mod 2 \label{eq:mod2Witten}
\eeq
is invariant under continuous deformations of the theory. Indeed, some states with $E>0$ may become $E=0$ during continuous deformations, but the change of $N_\text{vacua}$ from such states is always a multiple of 4. Thus the quantity \eqref{eq:mod2Witten} is invariant, and this is the mod-2 Witten index. 

\section{Elliptic genera of the internal theories}\label{app:elliptic}

In this appendix we compute the elliptic genera of the worldsheet theories
$X_1$, $X_2$, $X_4$, and $X_8$,
corresponding to the internal part of the 7-, 6-, 4-, and 0-branes, respectively,
introduced in Sec.~\ref{sec:names}.
We will see that the elliptic genera for the 7-, 6-, 4-branes are all trivial,
while the elliptic genus $Z^\text{ell,phys}(q)$ for the 0-brane is $24$, which is independent of $q$.

Let $H$ and $P$ be the worldsheet Hamiltonian and momentum operators, respectively, and let $H_L = \frac12 (H+P)$ and $H_R=\frac12 (H-P)$. For CFTs, we have 
\beq
H_L=  L_0-\frac{c_L}{24}~, \qquad H_R  = \overline L_0-\frac{c_R}{24}~.
\eeq
Elliptic genera are deformation invariants for general SQFTs without assuming conformal invariance. We let $\cH_m$ be the eigenspace of $P$ in the R-sector Hilbert space with
eigenvalue $P =m$. Due to the worldsheet gravitational anomalies, eigenvalues $m$ in the R-sector take values in
\beq
P=m \in -\frac{\nu}{24} +\bZ~,
\eeq
as one can check in the case of free fermions, where $\nu$ is the degree of $\sqft_\nu$. In the case of CFTs, we have $\nu=-2(c_L-c_R)$.

The ordinary elliptic genus is simply the partition function on $T^2$ with periodic spin structure,
\beq
Z^\text{ell,phys} = \tr (-1)^\sF q^{H_L} \overline q^{H_R}~.
\eeq
It is independent of $\overline q$ by the standard argument for the Witten index and the superalgebra $Q^2=H_R$, where $Q$ is the supercharge. Only the states with $H_R=0$ and $H_L=P$ contribute.
We will also be interested in the mod-2 elliptic genus, defined in \cite{Tachikawa:2023nne}. 

\subsection{The 7-brane}
We begin with the angular theory for the 7-brane, for which gravitational anomaly $ -2(c_L-c_R)=-31$ is $+1$ mod $8$,
 so we can consider the mod-2 elliptic genus of \cite{Tachikawa:2023nne}.\footnote{
When gravitational anomaly $\nu$ is odd, we take the convention that $(-1)^\sF$ is not defined on the R-sector; see e.g. the discussions around Eq.~(6.3) of \cite{Witten:1998cd}.}
Then $\dim \cH_m$ mod 2 is a deformation invariant.

The R-sector of the $E_8$ level-2 theory consists of the irreducible representation of the level 2 current algebra whose highest weight vectors transform as $\mathbf{248}$ of $E_8$.
Therefore \begin{equation}
\tr q^{H_L}=\sum q^{m} \dim \cH_m= \chi_{\mathbf{248}} = q^{7/24}(248+34504q + 1022752 q^2+ \cdots)~.
\end{equation}
Clearly the first three coefficients are all even.

We will now show that all of the $q$-expansion coefficients are even, so that the mod-2 elliptic genus is trivial.
To do so, it is useful to rewrite $\chi_{\mathbf{248}}$ using the equivalence of theories in \eqref{eq:equivn=1}. One can show that \begin{equation}
[q^{1/24}\prod_n (1+q^n)  ] \chi_{\mathbf{248}} 
= \frac12 \left( \left(\frac{E_4(q)}{\eta(q)^8} \right)^2 -\frac{E_4(q^2)}{\eta(q^2)^8} \right) \label{E8-level2-rewrite}
\end{equation} where \begin{equation}
E_4(q) = \sum_{v\in \Lambda_{E_8}} q^{v^2/2}
\end{equation} is the Eisenstein series of degree 4,
which is also the theta function of the root lattice of $E_8$.

One way to understand the equality \eqref{E8-level2-rewrite} is as follows. Consider the theory $(E_8)_2 \times \SO(1)_1$. Let $\sa$ be the $\bZ_2$ gauge field for the $\bZ_2^{(0)}$ symmetry of this theory which is basically the fermion parity symmetry. Let $\sb$ be the $\bZ_2$ gauge field for the dual $\bZ_2^{(1)}$ symmetry, which corresponds to the $\bZ_2$ symmetry exchanging the two $E_8$ factors in the theory $(\hete)_1$. As in \eqref{eq:dualcoupling2}, they are coupled as $\pi \i \int \sa \wedge \sb$. Let $z$ be the coordinate of $T^2$ with $z \sim z+1$ and $z \sim z+\tau$. We call $z \to z+1$ as the A-cycle and $z \to z+\tau$ as the B-cycle. Then $\pi \i \int \sa \wedge \sb = \pi \i (\sa_A \sb_B - \sa_B \sb_A)$, where $\sa_A~(\sb_A)$ and $\sa_B~(\sb_B)$ are components of $\sa~(\sb)$ in the A-cycle and the B-cycle, respectively. If we perform integration of $\sa$ in both the A-cycle and the B-cycle, we get the theory $(\hete)_1$. However, it is possible to integrate only $\sa_B$. We can think that $\sa_A$ is just set to zero, or we can also think that both $(\sa_A, \sa_B)$ are integrated and we also integrate $\sb_B$ which acts as a Lagrange multiplier setting $\sa_A=0$.  We conclude that the partition function of the theory $(E_8)_2 \times \SO(1)_1$ with $\sa_B$ integrated is equivalent to the partition function of the theory $(\hete)_1$ with $\sb_B$ integrated. They give the left- and the right-hand sides of \eqref{E8-level2-rewrite}, respectively. The minus sign in the second term of the right-hand side is due to the discrete theta angle discussed in Sec.~\ref{sec:dual7gauge}.

Then, $\chi_{248}\equiv 0$ mod 2 is equivalent to the condition that 
the right-hand side of \eqref{E8-level2-rewrite} is zero mod 2,
which is further equivalent to \begin{equation}
1-\frac{E_4(q^2)}{E_4(q)^2} \left(\frac{\eta(q)^2}{\eta(q^2)}\right)^8 \equiv 0 \mod 4~.
\end{equation} 
This can be shown using the following two identities. One is given in \eqref{eq:Eisenstein},
\begin{align}
E_4(q)&= 1 + 240 \sum_{ n=1}^{\infty} \frac{n^3 q^n}{1-q^n} \,\,\equiv\,\, 1 \mod 240~.  
\end{align}
The other is
\beq
\left(\frac{\eta(q)^2}{\eta(q^2)} \right)^8  &= \prod_{n=1}^\infty \left( \frac{(1-q^n)^2}{1-q^{2n}} \right)^8 
= \prod_{n=1}^\infty \left(1 - \frac{2q^n}{1+q^{n}} \right)^8 \,\,\equiv\,\, 1 \mod 16~.
\eeq
This concludes our derivation of $\chi_\mathbf{248}\equiv 0$ mod 2.

\subsection{The 6-brane}
We next consider the angular part of the 6-brane, for which the gravitational anomaly $2(c_R-c_L)=-30$ is $+2$ mod $8$,
and so we can consider the mod-2 elliptic genus of \cite{Tachikawa:2023nne}.
In this case, $\dim \cH_m$ is always even, and $(\dim\cH_m)/2$ mod 2 is a deformation-invariant.

As can be determined by discussions as in Sec.~\ref{sec:fermion-spectrum} (see also \cite{BoyleSmith:2023xkd}),
the R-sector of the $\SU(16)/\bZ_4$ level-1 theory consists of irreducible representations
whose highest weights transform as $\wedge^2$, $\wedge^6$, $\wedge^{10}$, $\wedge^{14}$ respectively,
where $\wedge^n$ is the $n$-index antisymmetric tensor representation of $\mathfrak{su}(16)$.\footnote{One may check it by starting from $\widetilde \U(16)/\bZ_2 \subset \hets$ at level 1, where $\widetilde \U(16)$ is the spin cover of $ \U(16)$. This theory is realized by 16 complex Weyl fermions gauged by $(-1)^{\sF}$. (It is just the theory $(\hets)_1$, with the emphasis of the subgroup $\widetilde \U(16)/\bZ_2 $.) Then we split the $\u(1)$ current algebra and the $\su(16)$ current algebra.
We will also use a reverse procedure to this from \eqref{eq:embedding} and onwards.}
We then have \begin{equation}
\sum q^{H_L} \dim \cH_m = \chi_{\wedge^2}+\chi_{\wedge^6}+\chi_{\wedge^{10}}+\chi_{\wedge^{16}}~,
\end{equation}
where $\chi_R$ is the character of the affine algebra whose highest weight is $R$.

The assignment of fermion parity is such that $\wedge^2$ and $\wedge^{10}$ have $(-1)^\sF=+1$,
and $\wedge^{6}$ and $\wedge^{14}$ have $(-1)^\sF=-1$.
As the complex conjugation exchanges the representations as $\wedge^2\leftrightarrow \wedge^{14}$ 
and $\wedge^6\leftrightarrow \wedge^{10}$,
we see $\chi_{\wedge^2}=\chi_{\wedge^{14}}$ and $\chi_{\wedge^6}=\chi_{\wedge^{10}}$
when all the flavor symmetry holonomies are turned off, and $\dim \cH_m$ is indeed even for all $m$. This is a consequence of the general fact that the $\text{CPT}$ symmetry anticommutes (rather than commutes) with $(-1)^{\sF}$ when the gravitational anomaly $2(c_R-c_L)$ is $2 \mod 8$ (see e.g. the Appendix of \cite{Tachikawa:2023nne}).
We  then have\begin{equation}
\frac12 \tr q^{H_L} = \sum q^{m} \frac{\dim \cH_m}2  =  \chi_{\wedge^2} + \chi_{\wedge^{10}}~. \label{6braneEG}
\end{equation}

Let us now show that all the coefficients of the expression \eqref{6braneEG} are even.
For this, let us show a stronger claim that all the coefficients of $\chi_{\wedge^{2}}$ and $\chi_{\wedge^{10}}$ are individually even.
Using the description of the $\mathfrak{su}(16)$ level-1 affine algebra using 15 free bosons, it is clear that we have \begin{equation}
\chi_{\wedge^{k}} = \eta(q)^{-15} \sum_{v\in W_k} q^{v^2/2}
\end{equation} where $W_k$ on the right hand side is the subset of weight vectors of $\mathfrak{su}(16)$ such that the center $e^{2\pi \i /16} \in \SU(16)$ acts on the corresponding representation as $e^{2\pi \i k/16}$.

To compute these sublattice sums, let us recall that we can embed $W_k$ in $\bR^{16}$. Let 
\beq
a=(1,1,\ldots,1)\in \bZ^{16}~. \label{eq:embedding}
\eeq
Then $v \in W_k$ is embedded in $\bR^{16}$ as
\beq
v=w - \frac{a \cdot w}{16} a , \quad w=(w_1,\cdots, w_{16}) \in \bZ^{16}, \quad a \cdot w=\sum_{i=1}^{16} w_i \equiv k \mod 16~.
\eeq
By noticing that $w^2 = v^2 + (a \cdot w)^2/16$,
we have \begin{equation}
\left( \sum_{ n \in \bZ} t^n q^{n^2/2} \right)^{16} 
=\sum_{w \in \bZ^{16}} t^{w \cdot a} q^{w^2/2} = \sum_{k \in \bZ} t^k q^{k^2/32} \sum_{v\in W_{k}} q^{v^2/2}~.
\end{equation} 
We now evaluate the left hand side modulo 2. We have \begin{align}
\left( \sum_{ n \in \bZ} t^n q^{n^2/2} \right)^{16}  \equiv \sum_{n\in \bZ} t^{16n} q^{16 n^2/2} \mod 2~,
\end{align} where we used $(x+y)^2 \equiv x^2+y^2$ modulo $2$ repeatedly.
This shows $\sum_{v\in W_k} q^{v^2/2}$ to have even coefficients unless $k\equiv 0$ mod $16$,
implying that $\chi_{\wedge^k}$ to have all even coefficients unless $k\equiv 0$ mod $16$.
This is what we wanted to show.

\subsection{The 4-brane}
\label{sec:4brane-elliptic}
Let us turn to the theory $[X_4]$ for the 4-brane.
As tabulated in \cite{BoyleSmith:2023xkd},
the R-sector of the $(E_7\times E_7)/\bZ_2$ level-1 theory has the feature that
exchanging two factors of $E_7$ can be compensated by exchanging the $(-1)^\sF=+1$ sector and the $(-1)^\sF=-1$ sector.
This fact can also be understood from our construction of the 4-brane,
where we embedded the $\mathfrak{so}(4)  = \su(2) \times \su(2)$ curvature into $\mathfrak{e}_8\times \mathfrak{e}_8$, such that $\mathfrak{e}_7\times \mathfrak{e}_7$ is preserved. The two $\su(2)$ factors of the $\so(4)$ tangent bundle are exchanged by the parity flip of the $S^4$.
Then the exchange of two $E_7$ factors comes in the UV from the exchange of the two $E_8$ factors and the parity flip of the $S^4$. The parity flip exchanges two chiralities $(-1)^\sF=+1$ and $(-1)^\sF=-1$. 

In summary, there is an anomalous symmetry (of exchanging the two $E_7$ factors) such that the anomaly has the effect that $(-1)^\sF \to -(-1)^\sF$.
This guarantees that the elliptic genus is zero due to the complete cancellation between the $(-1)^\sF=+1$ sector and the $(-1)^\sF=-1$ sector, when the $\mathfrak{e}_7\times \mathfrak{e}_7$  holonomies are turned off.
Therefore we have \begin{equation}
Z^\text{ell,phys}([X_4]) = 
Z^\text{ell,phys}([  ((E_7\times E_7)/\bZ_2)_1   ]) = 0.
\end{equation}
This is consistent with our claim in Sec.~\ref{sec:4braneprelims} that $a$ and $b$ in (\ref{eq:4branecaseeg}) are both zero---in fact, we now see that the elliptic genus is exactly zero.

\subsection{The 0-brane}
Finally, we consider the case of the 0-brane. The $\Spin(24)/\bZ_2$ level-1 theory is known to have a left-moving supercharge \cite{DuncanMackCrane}.\footnote{%
The subgroup of $\Spin(24)/\bZ_2$ commuting with a chosen supercharge is one of the sporadic finite simple groups, $\mathrm{Co}_1$.
But this curious fact does not play any direct role in this paper.}
Therefore the physicists' elliptic genus only gets contributions from the lowest allowed states with $H_L=L_0- \frac12=0$ in the R-sector.

Let us see a little more detail without using the left-moving supersymmetry. The theory $(\Spin(24)/\bZ_2)_1$ is realized by 12 chiral bosons with the lattice given by the weight lattice of $\Spin(24)/\bZ_2$. The theory is not bosonic since its gravitational anomaly is $2(c_R-c_L)=-24 \equiv 8 \mod 16$, so the R-sector states should be described by lattice points in the weight lattice of $\so(24)$ that are not the genuine weight lattice points for $\Spin(24)/\bZ_2$. 
(Otherwise the NS and R sector states would be completely identical and the theory would be bosonic.)
 The minimum value of $L_0$ in the R-sector comes from the lattice points for the vector representation of $\so(24)$. They have $L_0=\frac12$, and hence we get
\beq
Z^\text{ell,phys}([  (    \Spin(24)/\bZ_2   )_1  ]) = \tr (-1)^\sF q^{H_L} = 24 + \cdots.
\label{eq:D.18}
\eeq
On the other hand, since $2(c_R-c_L)= -24 \equiv 0 \mod 24$, the elliptic genus \eqref{eq:D.18} is a modular invariant function.
The only modular invariant functions that start from the term $q^0$ are constant as follows from the fact mentioned around \eqref{eq:modularforms}. Therefore we conclude \begin{equation}
Z^\text{ell,phys}([X_8]) = 
Z^\text{ell,phys}([(\Spin(24)/\bZ_2)_1]) = 24~. \label{eq:0branegenus}
\end{equation} 
In this case the more subtle definition of the Green-Schwarz coupling given in Sec.~\ref{sec:generalGS} is not applicable, but in the first place
the non-zero elliptic genus already tells us that the class $[X_8]$ of the internal theory is non-trivial. 

\bibliographystyle{ytamsalpha}
\baselineskip=.95\baselineskip
\def\arxivfont{\rm}
\bibliography{ref}

\end{document}